\documentclass[lettersize,journal]{IEEEtran}

\usepackage{xargs}      
\usepackage{soul}       
\usepackage{color}      
\usepackage{xspace}     
\usepackage{xpunctuate} 
\usepackage{booktabs, array}
\usepackage{amsmath, amsfonts}
\usepackage{multirow}
\usepackage{algorithmic}
\usepackage{algorithm}
\usepackage[caption=false,font=normalsize,labelfont=sf,textfont=sf]{subfig}
\usepackage{textcomp}
\usepackage{stfloats}
\usepackage{url}
\usepackage{verbatim}
\usepackage{graphicx}
\usepackage{wrapfig}
\usepackage[x11names, table]{xcolor}
\usepackage{hyperref}
\usepackage[most]{tcolorbox}
\usepackage[breakwords]{truncate}
\usepackage[nocompress]{cite}

\graphicspath{{figs/}{figures/}{pictures/}{images/}{./}} 

\newcommand{\ie}{{i.e.,}\xspace}
\newcommand{\eg}{{e.g.,}\xspace}
\newcommand{\etal}{{et~al\xperiod}\xspace}

\newcommand{\etc}{{etc\xperiod}\xspace}

\newcommand{\bpstart}[1]{\vspace{0px}\noindent\textbf{#1}}

\definecolor{lightpink}{RGB}{237,157,202}
\definecolor{lightred}{RGB}{210,121,121}
\definecolor{lightorange}{RGB}{230,170,50}
\definecolor{lightgold}{RGB}{210,194,121}
\definecolor{lightgreen}{RGB}{121,210,121}
\definecolor{lightaqua}{RGB}{121,206,210}
\definecolor{lightblue}{RGB}{121,124,210}
\definecolor{lightpurple}{RGB}{153,102,255}
\definecolor{darkblue}{RGB}{0,0,140}
\definecolor{red}{RGB}{178,34,34}
\definecolor{gray}{RGB}{166,166,166}

\newcommand{\add}[1]{\textcolor{black}{#1}}

\newcommandx{\guest}[3][1=]
    {\setulcolor{lightorange}{\ul{#1}} \textcolor{lightorange} 
    {[\textbf{#2:} #3]}}
\newcommandx{\luke}[2][1=] 
    {\setulcolor{lightred}{\ul{#1}} \textcolor{lightred} 
    {[\textbf{Luke:} #2]}}  
\newcommandx{\jeff}[2][1=] 
    {\setulcolor{lightblue}{\ul{#1}} \textcolor{lightblue}  
    {[\textbf{Jeff:} #2]}}

\definecolor{tactic}{RGB}{255, 212, 205}
\definecolor{strategy}{RGB}{255, 223, 153}
\definecolor{objective}{RGB}{198, 230, 255}
\newtcbox{\xmybox}[1][red]{
    on line,
    arc=4pt,
    colback=#1!90,
    colframe=#1!255!black,
    before upper={
        \rule[-2pt]{0pt}{10pt}
    },
    boxrule=0.5pt,
    boxsep=0pt,
    left=2pt,
    right=2pt,
    top=0.5pt,
    bottom=0.5pt
}

\newcommand{\rot}[1]{\rotatebox{90}{#1}}
\newcommand{\shorten}[1]{\truncate{3.5cm}{#1}}

\newcommand{\figurecholera}{
    \begin{wrapfigure}{l}{0.5\linewidth}
        \centering
        \vspace{1mm}
        \includegraphics[width=1\linewidth]{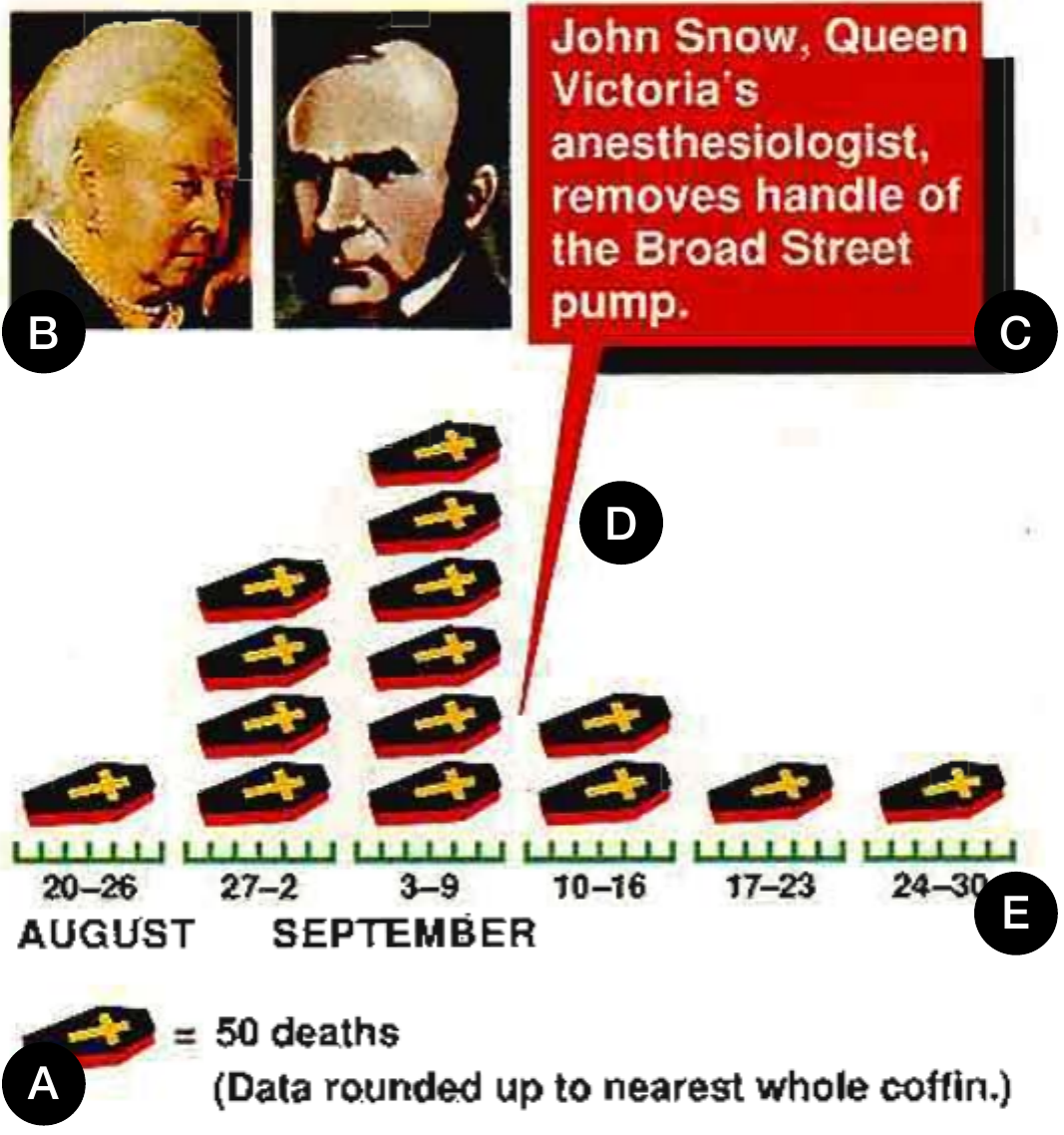}
        \vspace{-13pt}
        \caption{
            Augmented visualization by Tufte~\cite{tufte1997visual} that shows weekly cholera deaths during the 1854 London cholera outbreak. (A) Casket symbol that represents 50 deaths; (B) Pictures of Queen Victoria and John Snow; (C) Annotated text that contextualizes when the handle from the Broad street pump was removed and by whom; (D) Annotated connector that identifies the when the pump was removed; (E) Aggregation to display weekly deaths.
        }
        \label{fig:cholera}
        \vspace{-1mm}
    \end{wrapfigure}
}

\newcommand{\figuredemocrats}{
    \begin{figure}[!t]
        \centering
        \resizebox{1\linewidth}{!}{\includegraphics{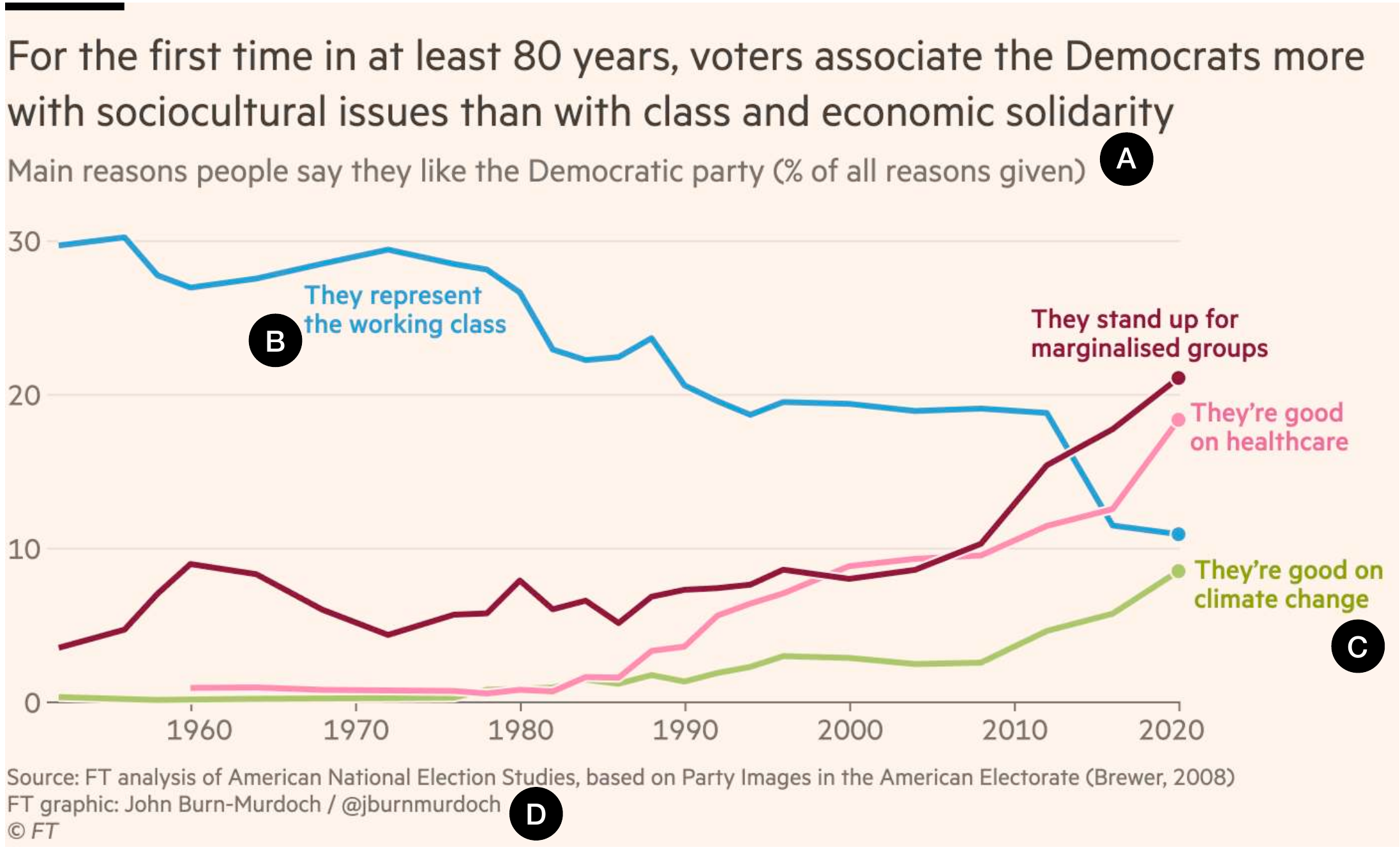}}
        \vspace{-15pt}
        \caption{
            Augmented visualization by Burn-Murdoch~\cite{ft}, which illustrates the change in perceptions of Democratic Party values over time (A). (B) The placement of annotated text subtly encourages left-to-right reading; (C) Color hue is used to relate annotated text with line marks and connote subject matter (\eg blue for blue-collar workers or green for climate); (D) Footnotes provide context for how the data was sourced.
        }
        \label{fig:democrats}
        \vspace{-3mm}
    \end{figure}
}

\newcommand{\figureexplorable}{
    \begin{figure}[!t]
        \centering
        \resizebox{1\linewidth}{!}{\includegraphics{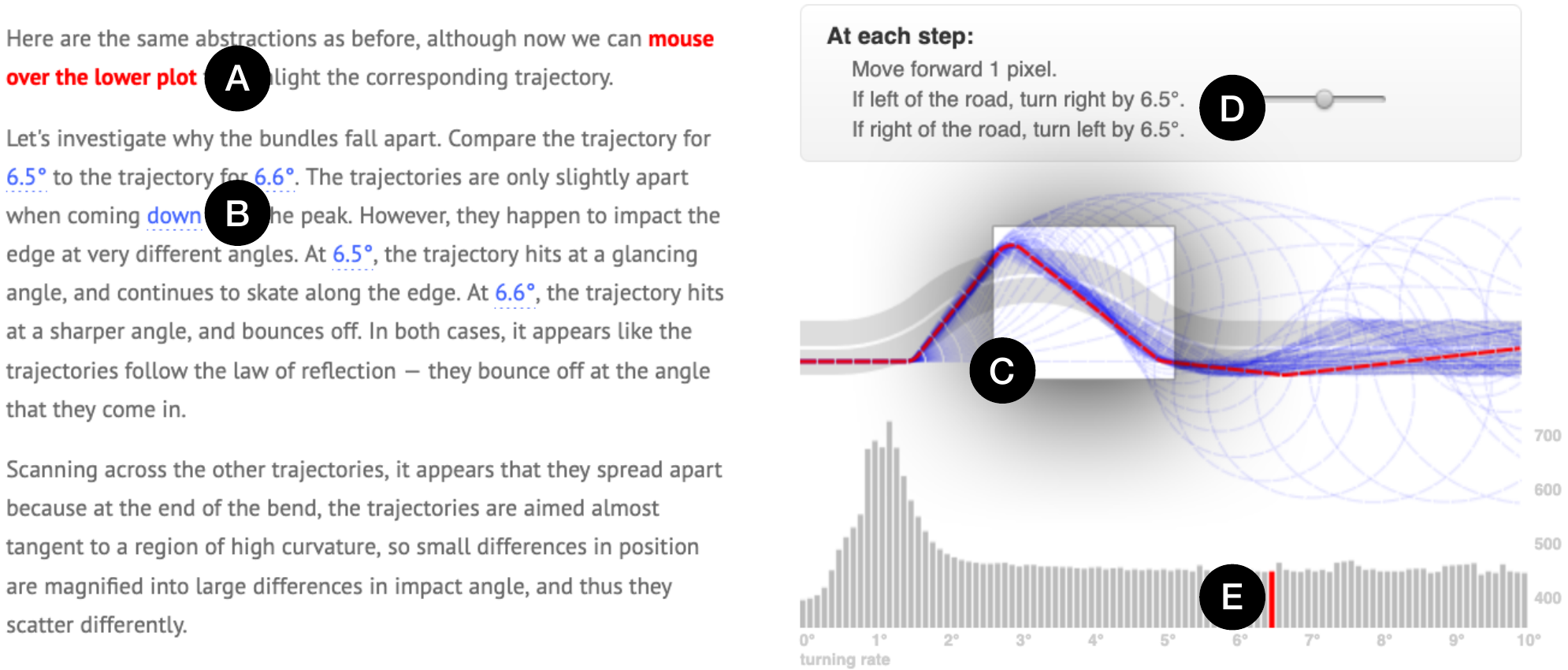}}
        \vspace{-15pt}
        \caption{
            Augmented visualization by Victor~\cite{victor2011ladder}, demonstrating interactive and explorable explanations of a car simulation algorithm. (A) Linked color highlighting to indicate mouse hovering, which sets the algorithm parameter (degrees of turn) in views (D) and (E); (B) Linked color highlighting to indicate mouse clicking, which sets a specific parameter value in views (D) and (E) and focuses the region of interest (C); (D) Slider to explore different parameter values.
        }
        \label{fig:explorable}
        \vspace{-3mm}
    \end{figure}
}

\newcommand{\figuredesignspace}{
    \begin{figure*}[!t]
        \centering
        \resizebox{1\linewidth}{!}{\includegraphics{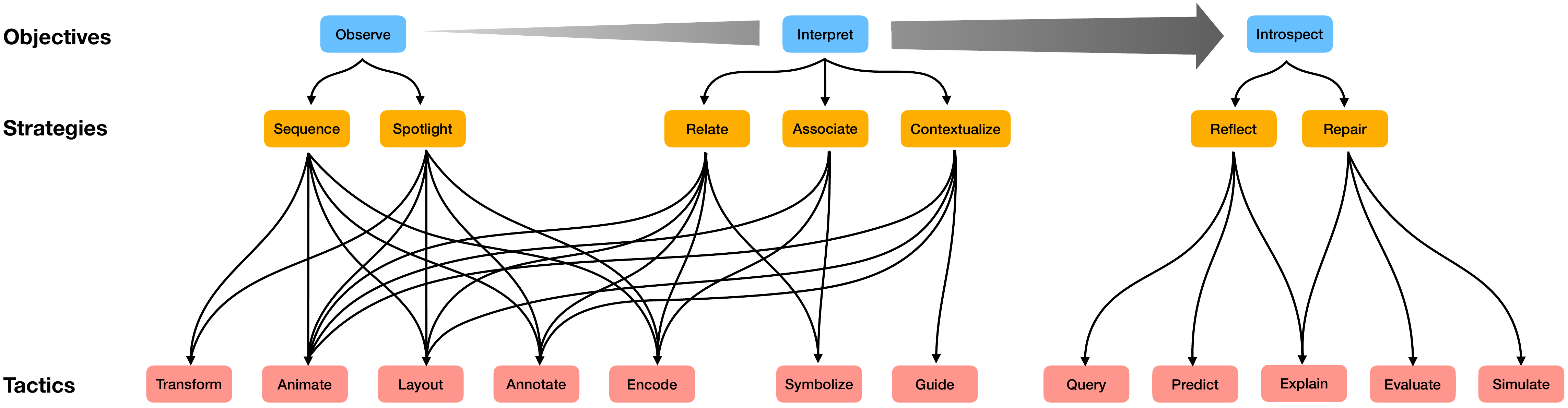}}
        \vspace{-15pt}
        \caption{ 
            Overview of augmentations by Objective (Observe, Interpret, Introspect), Strategy (Sequence, Spotlight, Relate, Associate, Contextualize, Reflect, Repair), and Tactic (Transform, Animate, Layout, Annotate, Encode, Symbolize, Guide, Query, Predict, Explain, Evaluate, Simulate).
            The objectives span from lower-level perceptual aims to higher-level cognitive aims: observation goals seek to direct attention by emphasizing and sequencing content; interpretation aids add supplemental information affecting inference; and introspection helps readers assess and update their understanding.
            }
        \label{fig:design-space}
        \vspace{-4mm}
    \end{figure*}
}

\newcommand{\figurecarbon}{
    \begin{figure*}[!h]
        \centering
        \resizebox{1\linewidth}{!}{\includegraphics{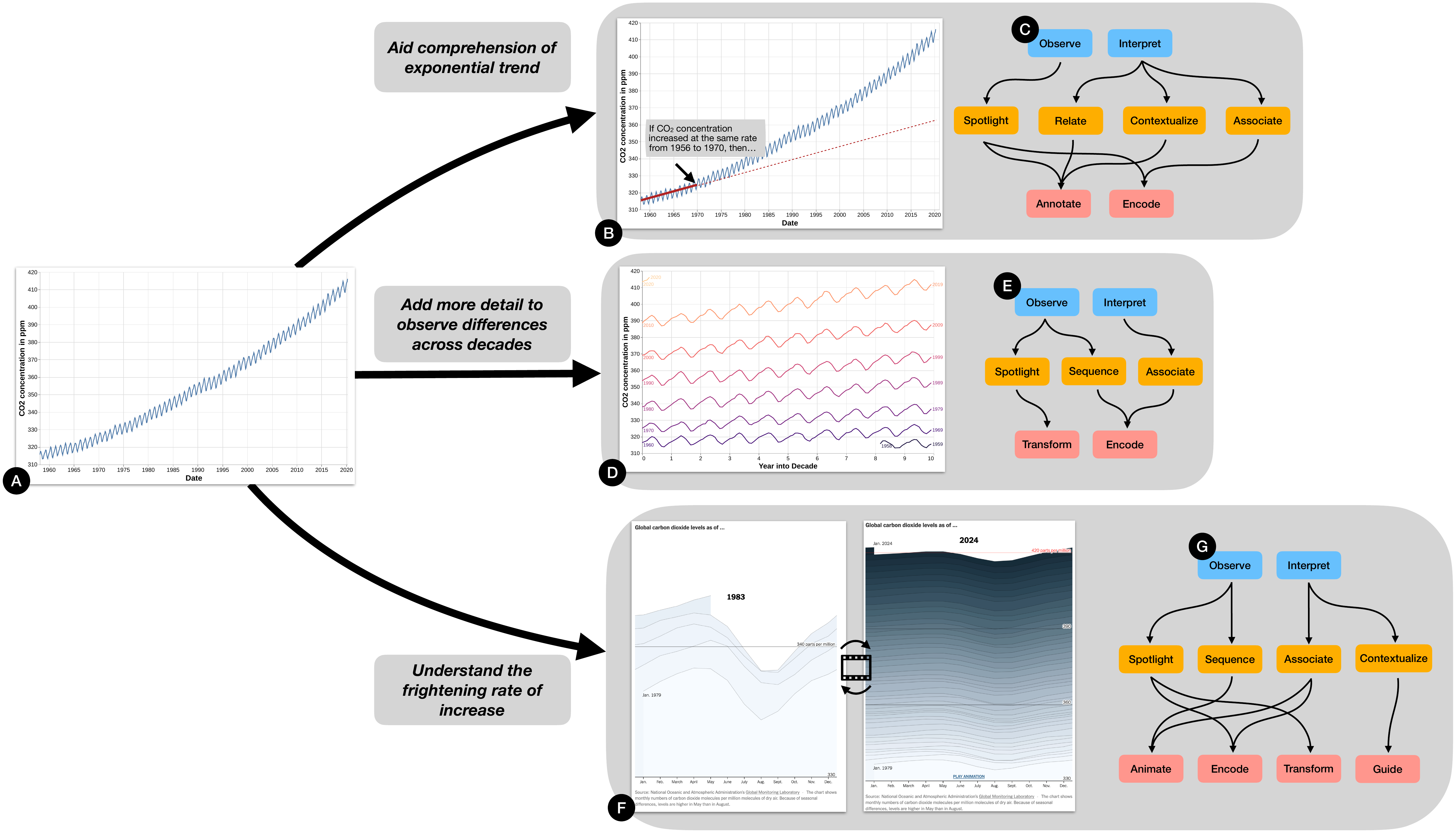}}
        \vspace{-15pt}
        \caption{ 
            Application of our design space to a CO$_2$ emissions chart to generate new designs. (A) Original, non-augmented chart. (B, D, F) A variety of possible augmented versions with (C, E, G) associated augmentation graphs.
            Designers can map high-level intents to a set of actionable tactics via strategies of their choice, enabling search and exploration of augmented communicative visualizations.
        }
        \label{fig:co2-emissions}
        \vspace{-3mm}
    \end{figure*}
}

\newcommand{\figureincome}{
    \begin{figure*}[!h]
        \centering
        \resizebox{1\linewidth}{!}{\includegraphics{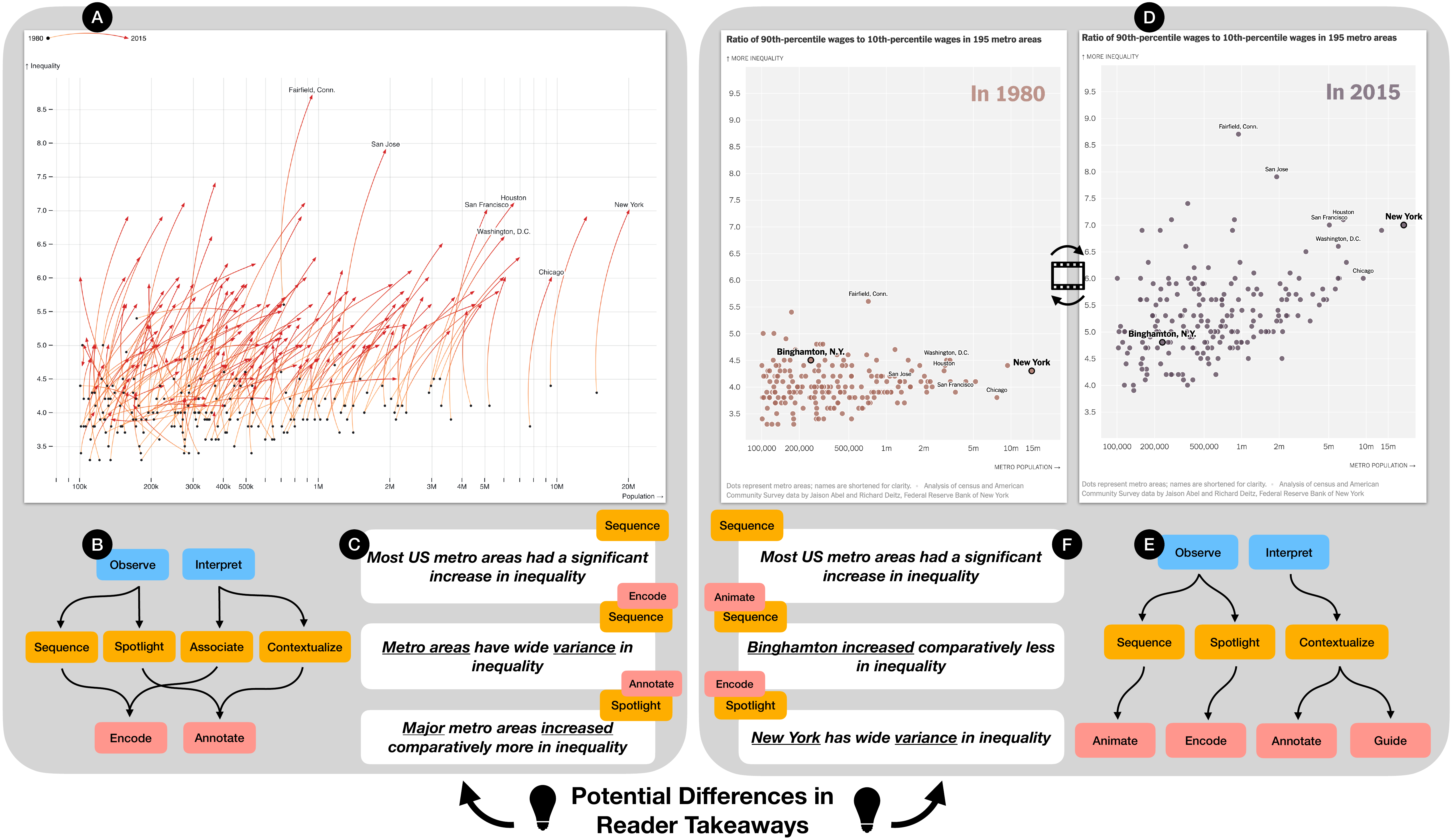}}
        \vspace{-15pt}
        \caption{ 
            Application of our design space to analyze augmentation differences between designs of the same income inequality data across US metro areas. (A) First augmented chart with (B) associated augmentation graph that prioritizes recognizing global trends; (D) second augmented chart with (E) associated augmentation graph that prioritizes recognizing change over time between two metro areas. Potential differences between reader takeaways (C, F) can be linked to augmentation differences, aiding design reasoning to align designer intent with reader takeaways.
        }
        \label{fig:income-inequality}
        \vspace{-3mm}
    \end{figure*}
}

\newcommand{\figureselfpredict}{
    \begin{figure*}[!t]
        \centering
        \resizebox{1\linewidth}{!}{\includegraphics{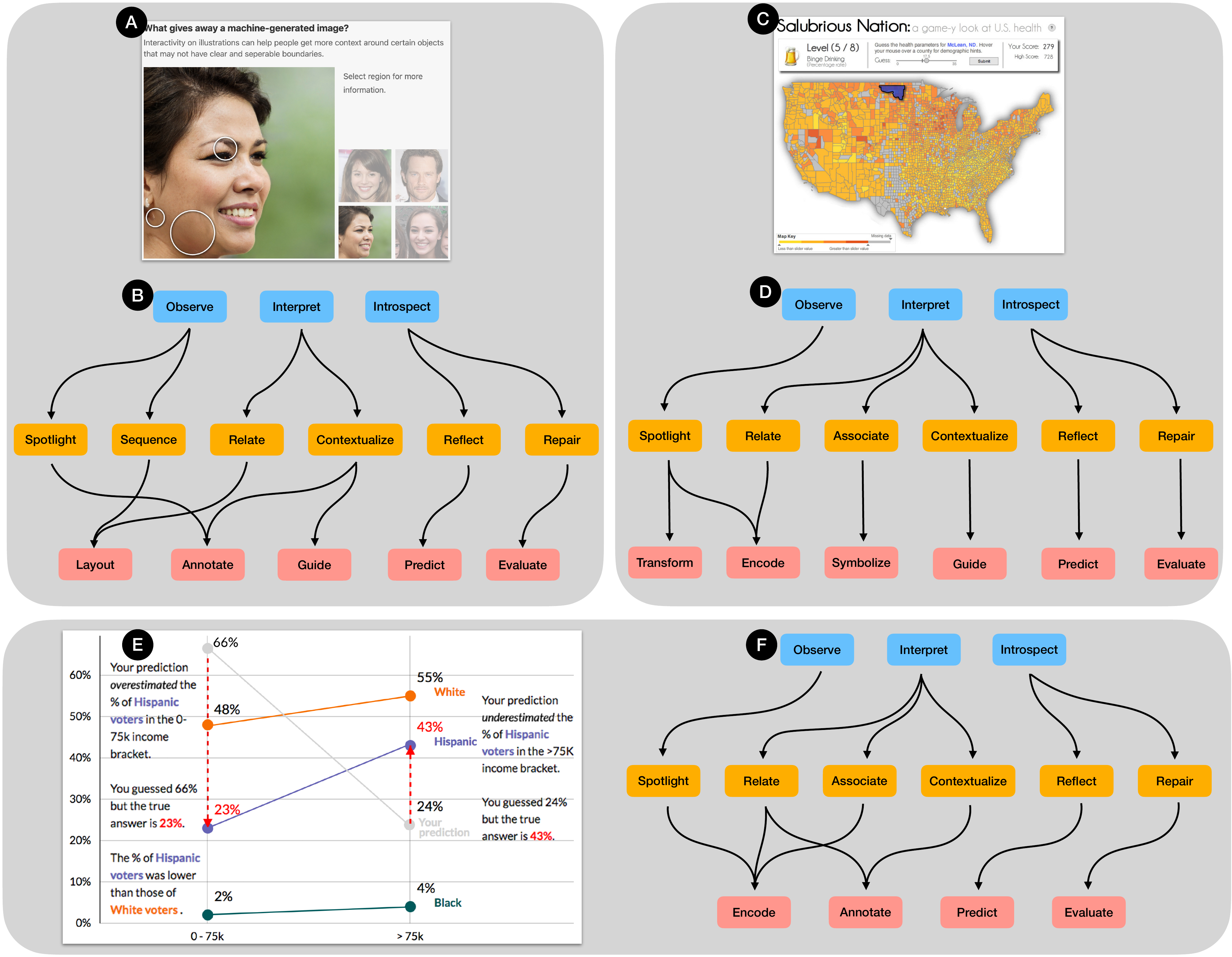}}
        \vspace{-15pt}
        \caption{ 
            Application of our design space to different visualizations that encourage active learning. Augmented visualizations (A, C, E) are shown with associated augmentation graphs (B, D, F). 
            Each visualization uses the same strategies for introspection (\emph{reflect} and \emph{repair} via self-prediction), but with varying tactics for other objectives (details-on-demand (A), game-playing (C) and graphical elicitation (E)), suggesting that learning differences may be attributed to supporting tactics and strategies which aid observation and interpretation, or how readers are prompted to engage in learning.
        }
        \label{fig:selfpredict}
        \vspace{-3mm}
    \end{figure*}
}

\newcommand{\figurerecognizetactics}{
    \begin{figure*}[!t]
        \centering
        \resizebox{1\linewidth}{!}{\includegraphics{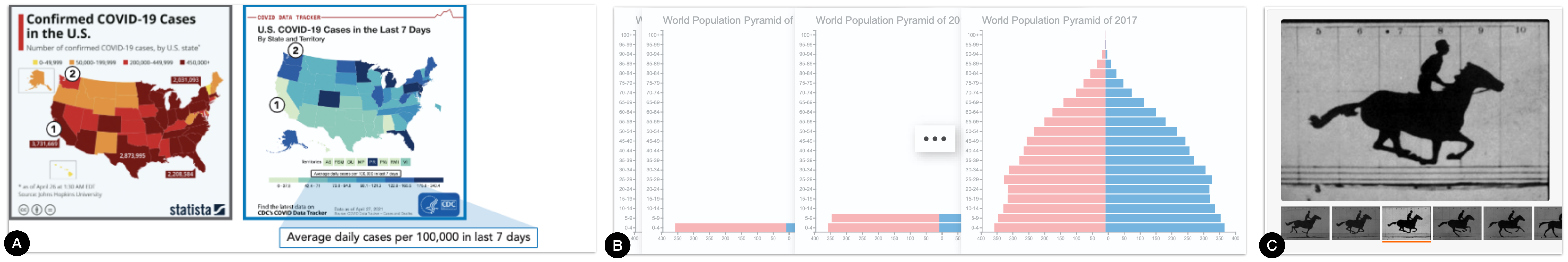}}
        \vspace{-15pt}
        \caption{
            Example augmentation tactics to \textbf{observe}. (A) Applying a normalization transform to COVID-19 data, which alters the sequence of data~\cite{ge2024v}; (B) Animated sequence of world population demographics from young to old to draw attention to the gradual changes between each age group~\cite{ge2021cast}; (C) overview+detail with faceted layout to sequence a horse's trot, revealing that all four hooves break contact with the ground~\cite{hohman2020communicating}.
        }
        \label{fig:recognize-tactics}
        \vspace{-3mm}
    \end{figure*}
}

\newcommand{\figureinterprettactics}{
    \begin{figure*}[!t]
        \centering
        \resizebox{1\linewidth}{!}{\includegraphics{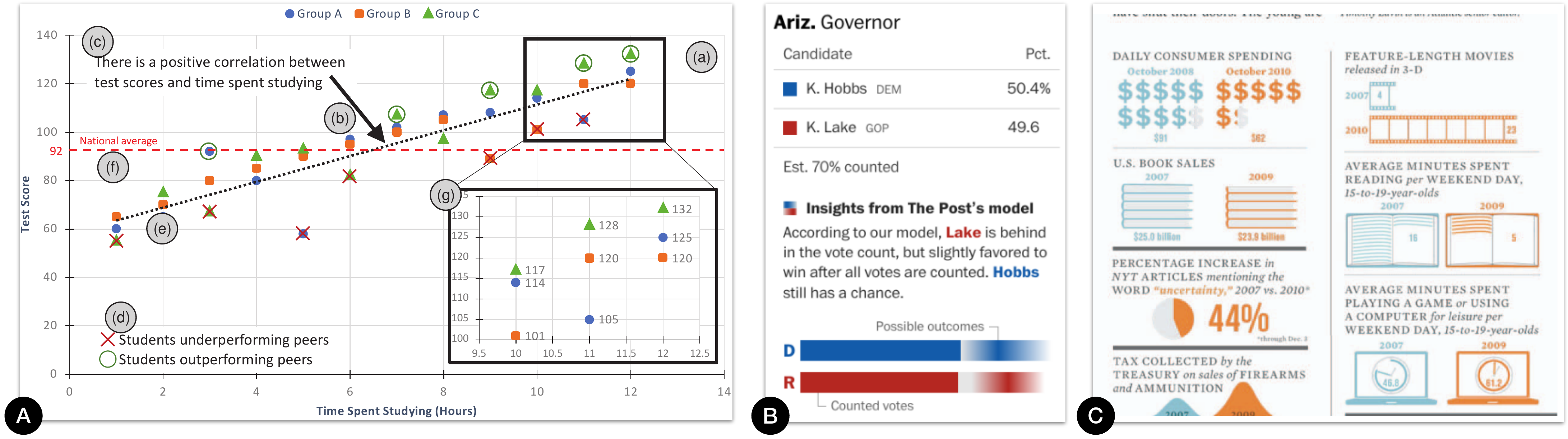}}
        \vspace{-15pt}
        \caption{
            Example augmentation tactics to \textbf{interpret}. (A) Annotated trend line (dotted blue line), national average (dashed red line), and student performance relative to peers (green circle and red strike-through glyphs, with a legend provided in the bottom-left) to provide context about student's test scores~\cite{rahman2024qualitative}; (B) linked red / blue hue encoding to relate visualizations and text during an Arizona gubernatorial race and associate with political parties~\cite{cai2024watching}; (C) use of repeated symbols to both relate consumer metrics after the 2008 recession (each row) and help associate the subject matter~\cite{hullman2011visualization}.
        }
        \label{fig:interpret-tactics}
        \vspace{-3mm}
    \end{figure*}
}

\newcommand{\figureintervenetactics}{
    \begin{figure*}[!t]
        \centering
        \resizebox{1\linewidth}{!}{\includegraphics{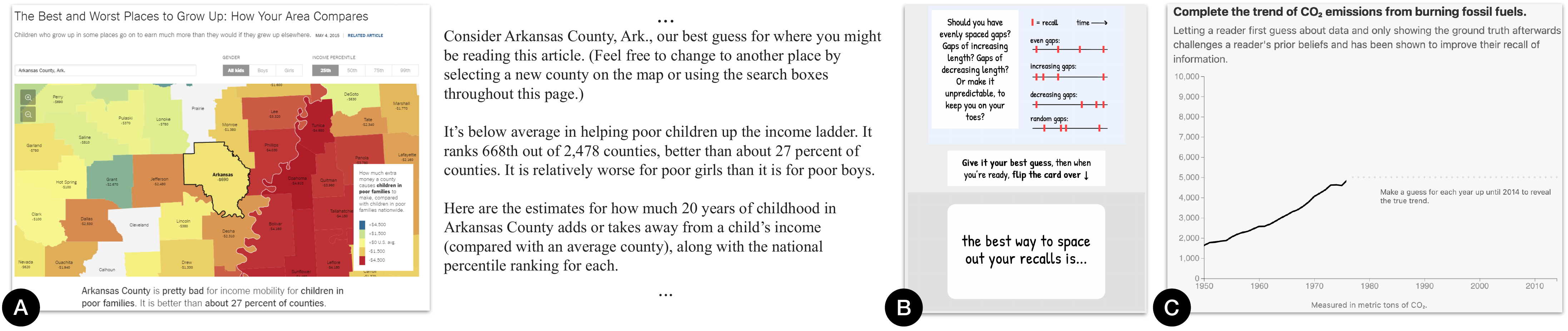}}
        \vspace{-15pt}
        \caption{
            Example augmentation tactics to \textbf{interpret} (A) and \textbf{introspect} (B,C). (A) Personalized content as guides, which displays income mobility data for the state that the reader accesses the article from~\cite{adar2017persalog}; (B) use of quizzes (within an article about spaced repetition) to query reader knowledge and promote reflection~\cite{hohman2020communicating}; (C) self-predicting the trend of CO$_2$ emissions to encourage reflection~\cite{hohman2020communicating}.
        }
        \label{fig:intervene-tactics}
        \vspace{-3mm}
    \end{figure*}
}

\newcommand{\tableaugmentations}{
    \begin{table}[!t]
    \centering
    \setlength\tabcolsep{0pt} 
    \smallskip 
    
    \def\arraystretch{1.0}%
    \scalebox{1.0}{
        \begin{tabular*}{\linewidth}{@{\extracolsep{\fill}} cll}
        \toprule
        Type & Augmentation & Description \\
        \toprule 

        \multirow{3}{*}{\xmybox[objective]{Objective}} & Observe & Direct attention to one or more elements \\
        & Interpret & Manipulate semantic or pragmatic intent \\
        & Introspect & Promote learning and reflection \\

        \midrule

        \multirow{7}{*}{\xmybox[strategy]{Strategy}} & Sequence & Arrange content in a particular order \\
        & Spotlight & Focus viewer attention on specific content \\
        & Relate & Conceptually link disparate data or views \\
        & Associate & Ground representations in known referents \\
        & Contextualize & Supplement chart with additional information \\
        & Reflect & Self-examine knowledge \\
        & Repair & Self-correct knowledge \\

        \midrule
        
        \multirow{12}{*}{\xmybox[tactic]{Tactic}} & Transform & Apply transformations to data \\
        & Animate & Temporally sequence static frames \\
        & Layout & Spatially arrange two or more views \\
        & Annotate & Introduce notational graphical elements \\
        & Encode & Map data attribute(s) to visual channel \\
        & Symbolize & Symbolically represent object or idea \\
        & Guide & Describe data semantics or thematic content \\
        & Query & Probe reader's knowledge directly \\
        & Predict & Self-predict unseen data \\
        & Explain & Self-explain observed data\\
        & Evaluate & Critique elicited prior knowledge \\
        & Simulate & Update visualized data, driven by reader input \\
    
        \bottomrule
        \end{tabular*}
    }
    \vspace{4pt}
    \caption{ 
        Overview of visual augmentations.
    }
    \label{table:augmentations}
    \vspace{-8mm}
    \end{table}
}

\newcommand{\tableaugmentationmap}{
    \begin{table*}[!t]
    \centering
    \setlength\tabcolsep{3.5pt} 
    \smallskip 
    
    \def\arraystretch{1.0}%
    \scalebox{1.0}{
        \begin{tabular*}{\linewidth}{@{\extracolsep{\fill}}llllcccccccccccccccccccccc}

        \multicolumn{4}{c}{} & \multicolumn{3}{l}{\multirow{2}{*}{Objectives}} & \multicolumn{7}{c}{\multirow{2}{*}{Strategies}} & \multicolumn{12}{c}{\multirow{2}{*}{Tactics}} \\
        & & & & & & & & & & & & & & & & & & & & & & & & & \\
        \cmidrule{5-7} \cmidrule(rl){8-14} \cmidrule{15-26} \vspace{-3mm} \\

        \rot{Reference} & \rot{Title} & \rot{First Author} & \rot{Year} & \rot{Observe} & \rot{Interpret} & \rot{Introspect} & \rot{Sequence} & \rot{Spotlight} & \rot{Relate} & \rot{Associate} & \rot{Contextualize} & \rot{Reflect} & \rot{Repair} & \rot{Animate} & \rot{Layout} & \rot{Transform} & \rot{Annotate} & \rot{Encode} & \rot{Symbolize} & \rot{Guide} & \rot{Explain} & \rot{Predict} & \rot{Query} & \rot{Evaluate} & \rot{Simulate} \\
        \midrule

        \cite{victor2011explorable} & \shorten{Explorable Explanations} & Victor & 2011 & \cellcolor{objective!255} & \cellcolor{objective!255} & \cellcolor{objective!255} & \cellcolor{strategy!250} & \cellcolor{strategy!250} & \cellcolor{strategy!250} & & & \cellcolor{strategy!250} & \cellcolor{strategy!250} & & \cellcolor{tactic!255} & & & \cellcolor{tactic!255} & & & & \cellcolor{tactic!255} & & & \cellcolor{tactic!255} \\ \addlinespace[.25em]

        \cite{victor2006magic} & \shorten{Magic Ink} & Victor & 2006 & \cellcolor{objective!255} & \cellcolor{objective!255} & \cellcolor{objective!255} & \cellcolor{strategy!250} & \cellcolor{strategy!250} & \cellcolor{strategy!250} & & \cellcolor{strategy!250} & \cellcolor{strategy!250} & \cellcolor{strategy!250} & & \cellcolor{tactic!255} & & & \cellcolor{tactic!255} & & \cellcolor{tactic!255} & & \cellcolor{tactic!255} & & & \cellcolor{tactic!255} \\ \addlinespace[.25em]

        \cite{hohman2020communicating} & \shorten{Communicating with Interactive Articles} & Hohman & 2020 & \cellcolor{objective!255} & \cellcolor{objective!255} & \cellcolor{objective!255} & \cellcolor{strategy!250} & \cellcolor{strategy!250} & \cellcolor{strategy!250} & \cellcolor{strategy!250} & \cellcolor{strategy!250} & \cellcolor{strategy!250} & \cellcolor{strategy!250} & & \cellcolor{tactic!255} & & \cellcolor{tactic!255} & & \cellcolor{tactic!255} & \cellcolor{tactic!255} & & \cellcolor{tactic!255} & \cellcolor{tactic!255} & \cellcolor{tactic!255} & \cellcolor{tactic!255} \\ \addlinespace[.25em]
        
        \cite{crichton2021new} & \shorten{A New Medium for Communicating Research on Programming Languages} & Crichton & 2021 & \cellcolor{objective!255} & \cellcolor{objective!255} & \cellcolor{objective!255} & \cellcolor{strategy!250} & \cellcolor{strategy!250} & \cellcolor{strategy!250} & & \cellcolor{strategy!250} & \cellcolor{strategy!250} & \cellcolor{strategy!250}  & & \cellcolor{tactic!255} & & \cellcolor{tactic!255} & \cellcolor{tactic!255} & & & & \cellcolor{tactic!255} & & \cellcolor{tactic!255} & \\ \addlinespace[.25em]

        \cite{gunturu2024augmented} & \shorten{Augmented Physics: Creating Interactive and Embedded Physics Simulations from Static Textbook Diagrams} & Gunturu & 2024 & \cellcolor{objective!255} & \cellcolor{objective!255} & \cellcolor{objective!255} & \cellcolor{strategy!250} & & \cellcolor{strategy!250} & & & \cellcolor{strategy!250} & \cellcolor{strategy!250} & \cellcolor{tactic!255} & & & \cellcolor{tactic!255} & \cellcolor{tactic!255} & & & & \cellcolor{tactic!255} & & & \cellcolor{tactic!255} \\ \addlinespace[.25em]

        \cite{hullman2011benefitting} & \shorten{Benefitting InfoVis with Visual Difficulties} & Hullman & 2011 & \cellcolor{objective!255} & \cellcolor{objective!255} & \cellcolor{objective!255} & \cellcolor{strategy!250} & & \cellcolor{strategy!250} & \cellcolor{strategy!250} & \cellcolor{strategy!250} & \cellcolor{strategy!250} & \cellcolor{strategy!250} & \cellcolor{tactic!255} & \cellcolor{tactic!255} & & & & \cellcolor{tactic!255} & \cellcolor{tactic!255} & \cellcolor{tactic!255} & & & & \\ \addlinespace[.25em]

        \cite{diakopoulos2011playable} & \shorten{Playable Data: Characterizing the Design Space of Game-y Infographics} & Diakopoulos & 2011 & \cellcolor{objective!255} & \cellcolor{objective!255} & \cellcolor{objective!255} & & \cellcolor{strategy!250} & \cellcolor{strategy!250} & \cellcolor{strategy!250} & \cellcolor{strategy!250} & \cellcolor{strategy!250} & \cellcolor{strategy!250} & & & \cellcolor{tactic!255} & & \cellcolor{tactic!255} & \cellcolor{tactic!255} & \cellcolor{tactic!255} & & \cellcolor{tactic!255} & & \cellcolor{tactic!255} & \\ \addlinespace[.25em]

        \cite{ajani2021declutter} & \shorten{Declutter and Focus: Empirically Evaluating Design Guidelines for Effective Data Communication} & Ajani & 2021 & \cellcolor{objective!255} & \cellcolor{objective!255} & & & \cellcolor{strategy!250} & \cellcolor{strategy!250} & & \cellcolor{strategy!250} & & & & & & \cellcolor{tactic!255} & \cellcolor{tactic!255} & & & & & & & \\ \addlinespace[.25em]
        
        \cite{chevalier2016animations} & \shorten{Animations 25 Years Later: New Roles and Opportunities} & Chevalier & 2016 & \cellcolor{objective!255} & \cellcolor{objective!255} & & \cellcolor{strategy!250} & \cellcolor{strategy!250} & \cellcolor{strategy!250} & \cellcolor{strategy!250} & \cellcolor{strategy!250} & & & \cellcolor{tactic!255} & & & & & & & & & & & \\ \addlinespace[.25em]

        \cite{ge2021cast} & \shorten{CAST: Authoring Data-Driven Chart Animations} & Ge & 2021 & \cellcolor{objective!255} & \cellcolor{objective!255} & & \cellcolor{strategy!250} & \cellcolor{strategy!250} & \cellcolor{strategy!250} & & & & & \cellcolor{tactic!255} & & & & & & & & & & & \\ \addlinespace[.25em]

        \cite{hullman2013deeper} & \shorten{A Deeper Understanding of Sequence in Narrative Visualization} & Hullman & 2013 & \cellcolor{objective!255} & \cellcolor{objective!255} & & \cellcolor{strategy!250} & & \cellcolor{strategy!250} & & & & & & \cellcolor{tactic!255} & & & & & & & & & & \\ \addlinespace[.25em]

        \cite{ploetzner2020review} & \shorten{A Review of Learning Demands in Instructional Animations: The Educational Effectiveness of Animations Unfolds if the Features of Change Need to be Learned} & Ploetzner & 2020 & \cellcolor{objective!255} & \cellcolor{objective!255} & & \cellcolor{strategy!250} & & \cellcolor{strategy!250} & & & & & \cellcolor{tactic!255} & & & & & & & & & & & \\ \addlinespace[.25em]
        
        \cite{satyanarayan2014authoring} & \shorten{Authoring Narrative Visualizations with Ellipsis} & Satyanarayan & 2014 & \cellcolor{objective!255} & \cellcolor{objective!255} & & \cellcolor{strategy!250} & \cellcolor{strategy!250} & \cellcolor{strategy!250} & & \cellcolor{strategy!250} & & & \cellcolor{tactic!255} & \cellcolor{tactic!255} & & \cellcolor{tactic!255} & & & & & & & & \\ \addlinespace[.25em]
        
        \cite{segel2010narrative} & \shorten{Narrative Visualization: Telling Stories with Data} & Segel & 2010 & \cellcolor{objective!255} & \cellcolor{objective!255} & & \cellcolor{strategy!250} & \cellcolor{strategy!250} & \cellcolor{strategy!250} & \cellcolor{strategy!250} & \cellcolor{strategy!250} & & & \cellcolor{tactic!255} & \cellcolor{tactic!255} & & \cellcolor{tactic!255} & & & \cellcolor{tactic!255} & & & & & \\ \addlinespace[.25em]
        
        \cite{srinivasan2018augmenting} & \shorten{Augmenting Visualizations with Interactive Data Facts to Facilitate Interpretation and Communication} & Srinivasan & 2018 & \cellcolor{objective!255} & \cellcolor{objective!255} & & & \cellcolor{strategy!250} & & & \cellcolor{strategy!250} & & & & & \cellcolor{tactic!255} & \cellcolor{tactic!255} & \cellcolor{tactic!255} & & & & & & & \\ \addlinespace[.25em]

        \cite{heer2012interactive} & \shorten{A Taxonomy of Tools that Support the Fluent and Flexible Use of Visualizations} & Heer & 2012 & \cellcolor{objective!255} & \cellcolor{objective!255} & & \cellcolor{strategy!250} & \cellcolor{strategy!250} & \cellcolor{strategy!250} & \cellcolor{strategy!250} & \cellcolor{strategy!250} & & & \cellcolor{tactic!255} & \cellcolor{tactic!255} & & \cellcolor{tactic!255} & \cellcolor{tactic!255} & \cellcolor{tactic!255} & \cellcolor{tactic!255} & & & & & \\ \addlinespace[.25em]

        \cite{bateman2010useful} & \shorten{Useful Junk? The Effects of Visual Embellishment on Comprehension and Memorability of Charts} & Bateman & 2010 & \cellcolor{objective!255} & \cellcolor{objective!255} & & & & & \cellcolor{strategy!250} & & & & & & & & & \cellcolor{tactic!255} & & & & & & \\ \addlinespace[.25em]

        \cite{shah1999graphs} & \shorten{Graphs as Aids to Knowledge Construction: Signaling Techniques for Guiding the Process of Graph Comprehension} & Shah & 1999 & \cellcolor{objective!255} & \cellcolor{objective!255} & & \cellcolor{strategy!250} & & \cellcolor{strategy!250} & & & & & & & \cellcolor{tactic!255} & & \cellcolor{tactic!255} & & & & & & & \\ \addlinespace[.25em]

        \cite{xiong2021visual} & \shorten{Visual Arrangements of Bar Charts Influence Comparisons in Viewer Takeaways} & Xiong & 2021 & \cellcolor{objective!255} & \cellcolor{objective!255} & & \cellcolor{strategy!250} & & \cellcolor{strategy!250} & & & & & & \cellcolor{tactic!255} & & & & & & & & & & \\ \addlinespace[.25em]

        \cite{rahman2024qualitative} & \shorten{A Qualitative Analysis of Common Practices in Annotations: A Taxonomy and Design Space} & Rahman & 2024 & \cellcolor{objective!255} & \cellcolor{objective!255} & & & \cellcolor{strategy!250} & \cellcolor{strategy!250} & & \cellcolor{strategy!250} & & & & & \cellcolor{tactic!255} & \cellcolor{tactic!255} & & & & & & & & \\ \addlinespace[.25em]

        \cite{bach2018design} & \shorten{Design Patterns for Data Comics} & Bach & 2018 & \cellcolor{objective!255} & \cellcolor{objective!255} & & \cellcolor{strategy!250} & \cellcolor{strategy!250} & \cellcolor{strategy!250} & & \cellcolor{strategy!250} & & & & \cellcolor{tactic!255} & & & & & & & & & & \\ \addlinespace[.25em]

        \cite{clark2010graphics} & \shorten{Graphics for Learning: Proven Guidelines for Planning, Desinging, and Evaluating Visuals in Training Materials} & Clark & 2010 & \cellcolor{objective!255} & \cellcolor{objective!255} & & \cellcolor{strategy!250} & \cellcolor{strategy!250} & \cellcolor{strategy!250} & \cellcolor{strategy!250} & & & & & \cellcolor{tactic!255} & & \cellcolor{tactic!255} & \cellcolor{tactic!255} & \cellcolor{tactic!255} & & & & & & \\ \addlinespace[.25em]

        \cite{sultanum2021leveraging} & \shorten{Leveraging Text-Chart Links to Support Authoring of Data-Driven Articles with VizFlow} & Sultanum & 2021 & \cellcolor{objective!255} & \cellcolor{objective!255} & & \cellcolor{strategy!250} & \cellcolor{strategy!250} & \cellcolor{strategy!250} & & \cellcolor{strategy!250} & & & & \cellcolor{tactic!255} & & \cellcolor{tactic!255} & \cellcolor{tactic!255} & & & & & & & \\ \addlinespace[.25em]

        \cite{mayer2005cognitive} & \shorten{Cognitive Theory of Multimedia Learning} & Mayer & 2005 & \cellcolor{objective!255} & \cellcolor{objective!255} & & \cellcolor{strategy!250} & \cellcolor{strategy!250} & \cellcolor{strategy!250} & & \cellcolor{strategy!250} & & & & \cellcolor{tactic!255} & & & & & \cellcolor{tactic!255} & & & & & \\ \addlinespace[.25em]

        \cite{boy2015storytelling} & \shorten{Storytelling in Information Visualizations: Does it Engage Users to Explore Data?} & Boy & 2015 & \cellcolor{objective!255} & \cellcolor{objective!255} & & \cellcolor{strategy!250} & \cellcolor{strategy!250} & \cellcolor{strategy!250} & & \cellcolor{strategy!250} & & & & \cellcolor{tactic!255} & \cellcolor{tactic!255} & & \cellcolor{tactic!255} & & \cellcolor{tactic!255} & & & & & \\ \addlinespace[.25em]

        \cite{shi2021communicating} & \shorten{Communicating with Motion: A Design Space for Animated Visual Narratives in Data Videos} & Shi & 2021 & \cellcolor{objective!255} & \cellcolor{objective!255} & & \cellcolor{strategy!250} & \cellcolor{strategy!250} & \cellcolor{strategy!250} & \cellcolor{strategy!250} & & & & \cellcolor{tactic!255} & & & & & & & & & & & \\ \addlinespace[.25em]

        \cite{tufte1997visual} & \shorten{Visual Explanations} & Tufte & 1997 & \cellcolor{objective!255} & \cellcolor{objective!255} & & \cellcolor{strategy!250} & \cellcolor{strategy!250} & \cellcolor{strategy!250} & \cellcolor{strategy!250} & \cellcolor{strategy!250} & & & & \cellcolor{tactic!255} & \cellcolor{tactic!255} & \cellcolor{tactic!255} & \cellcolor{tactic!255} & \cellcolor{tactic!255} & & & & & & \\ \addlinespace[.25em]

        \cite{hill1991deixis} & \shorten{Deixis and the Future of Visualization Excellence} & Hill & 1991 & \cellcolor{objective!255} & \cellcolor{objective!255} & & \cellcolor{strategy!250} & \cellcolor{strategy!250} & \cellcolor{strategy!250} & & & & & \cellcolor{tactic!255} & & & & \cellcolor{tactic!255} & & & & & & & \\ \addlinespace[.25em]
        
        \cite{tversky2002animation} & \shorten{Animation: Can it Facilitate?} & Tversky & 2002 & \cellcolor{objective!255} & \cellcolor{objective!255} & & \cellcolor{strategy!250} & & \cellcolor{strategy!250} & & & & & \cellcolor{tactic!255} & & & & & & & & & & & \\ \addlinespace[.25em]
        
        \cite{guo2024we} & \shorten{What We Augment When We Augment Visualizations: A Design Elicitation Study of How We Visually Express Data Relationships} & Guo & 2024 & \cellcolor{objective!255} & \cellcolor{objective!255} & & \cellcolor{strategy!250} & \cellcolor{strategy!250} & \cellcolor{strategy!250} & & & & & & & \cellcolor{tactic!255} & \cellcolor{tactic!255} & \cellcolor{tactic!255} & & & & & & & \\ \addlinespace[.25em]

        \cite{ancker2024you} & \shorten{Do You Want to Promote Recall, Perceptions, or Behavior? The Best Data Visualization Depends on the Communication Goal} & Ancker & 2024 & \cellcolor{objective!255} & \cellcolor{objective!255} & & & & & & & & & & & & & & & & & & & & \\ \addlinespace[.25em]

        \cite{ge2024v} & \shorten{V-FRAMER: Visualization Framework for Mitigating Reasoning Errors in Public Policy} & Ge & 2024 & \cellcolor{objective!255} & \cellcolor{objective!255} & & \cellcolor{strategy!250} & & & \cellcolor{strategy!250} & & & & & & \cellcolor{tactic!255}  & & \cellcolor{tactic!255} & \cellcolor{tactic!255} & & & & & & \\ \addlinespace[.25em]

        \cite{cai2024watching} & \shorten{Watching the Election Sausage Get made: How Data Journalists Visualize the Vote Counting Process in U.S. Elections} & Cai & 2024 & \cellcolor{objective!255} & \cellcolor{objective!255} & & \cellcolor{strategy!250} & & \cellcolor{strategy!250} & \cellcolor{strategy!250} & \cellcolor{strategy!250} & & & \cellcolor{tactic!255} & & & \cellcolor{tactic!255} & \cellcolor{tactic!255} & \cellcolor{tactic!255} & \cellcolor{tactic!255} & & & & & \\ \addlinespace[.25em]

        \cite{conlen2019capture} & \shorten{Capture \& Analysis of Active Reading Behaviors for Interactive Articles on the Web} & Conlen & 2019 & & \cellcolor{objective!255} & \cellcolor{objective!255} & & & \cellcolor{strategy!250} & \cellcolor{strategy!250} & \cellcolor{strategy!250} & \cellcolor{strategy!250} & \cellcolor{strategy!250} & & \cellcolor{tactic!255} & & \cellcolor{tactic!255} & \cellcolor{tactic!255} & \cellcolor{tactic!255} & & & \cellcolor{tactic!255} & & & \cellcolor{tactic!255} \\ \addlinespace[.25em]

        \cite{conlen2018idyll} & \shorten{Idyll: A Markup Language for Authroing and Publishing Interactive Articles on the Web} & Conlen & 2018 & & \cellcolor{objective!255} & \cellcolor{objective!255} & \cellcolor{strategy!250} & & \cellcolor{strategy!250} & & & \cellcolor{strategy!250} & \cellcolor{strategy!250} & & \cellcolor{tactic!255} & & & & & & & \cellcolor{tactic!255} & & & \cellcolor{tactic!255} \\ \addlinespace[.25em]

        \cite{victor2011scientific} & \shorten{Scientific Communication as Sequential Art} & Victor & 2011 & & \cellcolor{objective!255} & \cellcolor{objective!255} & \cellcolor{strategy!250} & & \cellcolor{strategy!250} & & & \cellcolor{strategy!250} & \cellcolor{strategy!250} & & \cellcolor{tactic!255} & & & \cellcolor{tactic!255} & & & & \cellcolor{tactic!255} & & & \cellcolor{tactic!255} \\ \addlinespace[.25em]
        
        \cite{quadri2024you} & \shorten{Do You See What I See? A Qualitative Study Eliciting High-Level Visualization Comprehension} & Quadri & 2024 & \cellcolor{objective!255} & & & \cellcolor{strategy!250} & \cellcolor{strategy!250} & & & & & & & \cellcolor{tactic!255} & & \cellcolor{tactic!255} & & & & & & & & \\ \addlinespace[.25em]
        
        \cite{brehmer2013multi} & \shorten{A Multi-Level Typology of Abstract Visualization Tasks} & Brehmer & 2013 & \cellcolor{objective!255} & & & \cellcolor{strategy!250} & \cellcolor{strategy!250} & & & & & & & \cellcolor{tactic!255} & \cellcolor{tactic!255} & \cellcolor{tactic!255} & \cellcolor{tactic!255} & & & & & & & \\ \addlinespace[.25em]
        
        \cite{quadri2021survey} & \shorten{A Survey of Perception-Based Visualization Studies by Task} & Quadri & 2021 & \cellcolor{objective!255} & & & \cellcolor{strategy!250} & \cellcolor{strategy!250} & & & & & & & & \cellcolor{tactic!255} & & \cellcolor{tactic!255} & & & & & & & \\ \addlinespace[.25em]
        
        \cite{mackinlay1986automating} & \shorten{Automating the Design of Graphical Presentations of Relational Information} & Mackinlay & 1986 & \cellcolor{objective!255} & & & \cellcolor{strategy!250} & & & & & & & & \cellcolor{tactic!255} & & & & & & & & & & \\ \addlinespace[.25em]
        
        \cite{fygenson2023arrangement} & \shorten{The Arrangement of Marks Impacts Afforded Messages: Ordering, Partitioning, Spacing, and Coloring in Bar Charts} & Fygenson & 2023 & \cellcolor{objective!255} & & & \cellcolor{strategy!250} & & & & & & & & \cellcolor{tactic!255} & \cellcolor{tactic!255} & & \cellcolor{tactic!255} & & & & & & & \\ \addlinespace[.25em]
                
        \cite{heer2007animated} & \shorten{Animated Transitions in Statistical Data Graphics} & Heer & 2007 & & \cellcolor{objective!255} & & \cellcolor{strategy!250} & & \cellcolor{strategy!250} & & & & & \cellcolor{tactic!255} & & & & & & & & & & & \\ \addlinespace[.25em]
        
        \cite{kim2017graphscape} & \shorten{Graphscape: A Model for Automated Reasoning about Visualization Similarity and Sequencing} & Kim & 2017 & & \cellcolor{objective!255} & & \cellcolor{strategy!250} & & \cellcolor{strategy!250} & & & & & & \cellcolor{tactic!255} & & & & & & & & & & \\ \addlinespace[.25em]

        \cite{kim2020gemini} & \shorten{Gemini: A Grammar and Recommender System for Animated Transitions in Statistical Graphics} & Kim & 2020 & & \cellcolor{objective!255} & & \cellcolor{strategy!250} & & \cellcolor{strategy!250} & & & & & \cellcolor{tactic!255} & & & & & & & & & & & \\ \addlinespace[.25em]
        
        \cite{mckenna2017visual} & \shorten{Visual Narrative Flow: Exploring Factors Shaping Data Visualization Story Reading Experiences} & McKenna & 2017 & & \cellcolor{objective!255} & & \cellcolor{strategy!250} & & \cellcolor{strategy!250} & & \cellcolor{strategy!250} & & & \cellcolor{tactic!255} & \cellcolor{tactic!255} & & & & & \cellcolor{tactic!255} & & & & & \\ \addlinespace[.25em]
        
        \cite{YangLeni2023EWEL} & \shorten{Explaining with Examples: Lessons Learned from Crowdsourced Introductory Description of Information Visualizations} & Yang & 2023 & & \cellcolor{objective!255} & & & & & & \cellcolor{strategy!250} & & & & & & & & & \cellcolor{tactic!255} & & & & & \\ \addlinespace[.25em]

        \cite{adar2017persalog} & \shorten{PersaLog: Personalization of News Article Content} & Adar & 2017 & & \cellcolor{objective!255} & & & & & & \cellcolor{strategy!250} & & & & & & & & & & \cellcolor{tactic!255} & & & & \\ \addlinespace[.25em]

        \cite{padilla2020uncertainty} & \shorten{Uncertainty Visualization} & Padilla & 2020 & & \cellcolor{objective!255} & & & & & \cellcolor{strategy!250} & & & & & \cellcolor{tactic!255} & & & \cellcolor{tactic!255} & & & & & & & \\ \addlinespace[.25em]

        \cite{maceachren2012visual} & \shorten{Visual Semiotics \& Uncertainty Visualization: An Empirical Study} & MacEachren & 2012 & & \cellcolor{objective!255} & & & & & \cellcolor{strategy!250} & & & & & \cellcolor{tactic!255} & & & \cellcolor{tactic!255} & & & & & & & \\ \addlinespace[.25em]

        \cite{alma99126810900001452} & \shorten{Understanding Comics} & McCloud & 1993 & & \cellcolor{objective!255} & & \cellcolor{strategy!250} & & \cellcolor{strategy!250} & \cellcolor{strategy!250} & \cellcolor{strategy!250} & & & & \cellcolor{tactic!255} & & \cellcolor{tactic!255} & & \cellcolor{tactic!255} & & & & & & \\ \addlinespace[.25em]

        \cite{kim2017explaining} & \shorten{Explaining the Gap: Visualizing One's Predictions Improves Recall and Comprehension of Data} & Kim & 2017 & & & \cellcolor{objective!255} & & & & & & \cellcolor{strategy!250} & \cellcolor{strategy!250} & & & & & & & & \cellcolor{tactic!255} & \cellcolor{tactic!255} & & \cellcolor{tactic!255} & \\ \addlinespace[.25em]

        \cite{levie1982effects} & \shorten{Effects of Text Illustrations: A Review of Research} & Levie & 1982 & & & \cellcolor{objective!255} & & & & & & \cellcolor{strategy!250} & \cellcolor{strategy!250} & & & & & & & & \cellcolor{tactic!255} & & & & \\ \addlinespace[.25em]

        \midrule
        \end{tabular*}
    }
    \vspace{2pt}
    \caption{ 
        Overview of papers that were used to code our design space, categorized by use or discussion of Objectives, Strategies, and Tactics. 
        Papers are sorted according to the number of objectives present.
    }
    \label{table:augmentation-map}
    \vspace{0mm}
    \end{table*}
}

\begin{document}

\title{\texorpdfstring{Crossing the Chasm:\\~Bridging Visual Augmentations and Designer Intent}{Crossing the Chasm: Bridging Visual Augmentations and Designer Intent}}

\author{Luke S. Snyder, Maureen C. Stone, Jeffrey Heer
\thanks{This is the author’s version of the article that has been published in IEEE Transactions on Visualization and
Computer Graphics. The final version of this record is available at: https://doi.org/10.1109/TVCG.2026.3671259.}
\thanks{This research was supported in part by an NSF IIS Award (2402718).}%
\thanks{The authors are with the University of Washington, Seattle, WA, USA (e-mail: snyderl@cs.washington.edu; mcstone@cs.washington.edu; jheer@cs.washington.edu).}}%




\maketitle

\begin{abstract}
To direct attention and communicate context, visualization designers often employ augmentations such as annotations, animated transitions, and stylistic layouts.
While graphical perception principles can helpfully prescribe effective low-level representations of data, they lack guidance for augmentations that service higher-level communication goals.
To bridge this gap, we contribute a design space that \add{frames designer intent as viewer-oriented cognitive behaviors, grounding communicative aims} in actionable visualization design techniques, including both visual encodings and augmentations.
This design space consists of common augmentation tactics (annotation, animation, stylized encodings, \etc) that implement design strategies (spotlighting, sequencing, association, \etc) to achieve higher-level design goals (observe, interpret, introspect), often simultaneously.
We demonstrate the analytic and generative value of our design space with examples across varied designer objectives.
We discuss how our contributions help align designer intent with reader takeaways, pave the way for readers to learn more effectively, and enable future (semi-)automated systems to support visualization designers in achieving their communication goals.
\end{abstract}





\section{Introducing the Chasm}
\label{sec:introduction}

\IEEEPARstart{V}{isualizations} are used as communicative aids in contexts such as teaching, business, news, and scientific papers.
During the design process, designers must consider their audience, what message they want to convey, and how they want to convey it.
Designers typically convey this intent by augmenting their visualizations (\eg annotations, animation, semantic color encoding, \etc), but how can they ensure that their intent was correctly decoded by the reader?
Graphical perception literature provides grounding for effective low-level representation of data~\cite{10.5555/1095597, mackinlay1986automating}, but we lack similar guidelines to help designers express communicative intent. 

\add{
We use ``communicative'' and ``designer'' intent interchangeably throughout the paper. This can be defined at multiple levels, each increasingly difficult to assess.
For instance, the designer's intent may be rhetorical~\cite{hullman2011visualization} or affective~\cite{lee2022affective} in nature: to persuade or evoke an emotional response.
On the other hand, the goal may be for readers to learn something important~\cite{adar2020communicative} or modify their future behavior based on new knowledge.
Each of these goals, however, may be secondary to less savory intents such as deceit or profit.
We frame communicative intent as explicit properties of the design: \textbf{\textit{general cognitive actions} that the designer wishes to induce on behalf of the reader}, such as \textit{observing} an important fact about the data or \textit{reflecting} on their own knowledge.
Though these behaviors may still be in service of higher-level goals, this more abstract framing has two primary advantages: (1) we are able to more concretely reason about design objectives from a varied set of augmentations used throughout the literature, in contrast to more specific and complex objectives (\eg learning, emotional, \etc) that are difficult to relate to each other and explicate without discussions or assessments with designers and readers; (2) designers are encouraged to reflect on the viewer's behavior, using our objectives as actionable guides to help achieve their ultimate intents.
}


Consider the political issues that US voters associate with the Democratic Party across past decades (\autoref{fig:democrats}).
After the data has been encoded, a number of decisions were made that affect communicative intent~\cite{adar2020communicative, lee2021learning}.
For instance, color not only helps relate annotated text with the data, it also exploits cultural associations to ground meaning (\ie visual metaphors~\cite{hullman2011visualization}): green for climate change, blue for working class (``blue-collar''), and pink for healthcare (breast cancer ribbon).
The placement of the annotated text also orders the content left-to-right to convey the temporal shift from working class to other issues more clearly.
However, if additional design constraints were present (\eg more data, heavy overlap of lines, \etc), other sequencing techniques might be used instead, such as animation (\eg fading to layer in each line), annotated labels (to enumerate content), faceting, a stepper, \etc
Additional annotations with historical context could help \textit{explain} voters' shift, but would alter the communicative intent.
Each of these decisions---or lack thereof---make use of \textbf{visual augmentations} (defined in~\autoref{sec:related-work}) to direct readers' attention and convey context. 
These objectives and supporting augmentations move beyond standard graphical perception concerns to influence the message that the designer conveys and what readers may learn.

A gap remains between design intent and the specific visual augmentations: how might a designer's communication goals be achieved within a vast space of varied augmentations?
To bridge this gap, we contribute a design space that characterizes connections between \textit{how} designers augment their charts and \textit{why} they employ these augmentations.
We survey literature in narrative visualization, graphical perception, communication, and learning sciences to distill a design space with three tiers: a common set of \emph{augmentation tactics} (annotation, animation, stylized encodings, \etc) that implement \emph{design strategies} (spotlighting, sequencing, association, \etc) to achieve higher-level \emph{design objectives} (observe, interpret, introspect).
We find that many tactics and strategies are reused across other tiers, often achieving multiple goals simultaneously.
In addition, our design space can be aligned with prior work on learning objectives~\cite{lee2021learning,adar2020communicative} to support more systematic design and assessment in the future.

Our design space serves as both an analytic tool to evaluate and compare communication aims across different visualizations, as well as a generative aid for exploring new designs with competing augmentation strategies and objectives.
We unpack examples in communicative visualization to demonstrate these contributions, noting trade-offs and opportunities.

By grounding communicative intent in visual augmentations, designers gain more actionable techniques---including visual encodings and augmentations---that they can use as communicative ``levers'' to convey different messages.
These ``levers'' may also encourage reflection about what readers will take away, better aligning designers and readers.
We discuss this alignment in greater depth, along with how our contributions can help pave the way for readers to learn more effectively, enable future (semi-)automated systems to support visualization designers in achieving their communication goals, and help scaffold and evaluate visual augmentations.

In summary, we make the following contributions:
\begin{enumerate}
    \item We identify and characterize the gap between visual augmentations and design intent as a disconnect between \textit{how} designers augment their charts and \textit{why}.
    \item We formally bridge this gap by surveying the visualization literature to distill a design space of \textit{what} augmentations designers use (tactics), \textit{how} they use them (strategies), and \textit{why} (to what aim) they use them (objectives). 
    \item We demonstrate the analytic and generative value of our design space across examples with varied designer intents, identifying both trade-offs and opportunities.
\end{enumerate}

\section{Characterizing the Chasm: Background}
\label{sec:related-work}

To illustrate the gap between visual techniques and designer intent, we first define visual augmentations, using motivating scenarios (\autoref{fig:democrats}, \autoref{fig:cholera}, \autoref{fig:explorable}) to surface potential types of designer objectives and supporting aids.

\vspace{1mm}
\bpstart{Defining Visual Augmentation.}
We define visual augmentation as \emph{any use or modification of existing or supplemental graphical elements designed to communicate information beyond denotative graphical decoding}.


\noindent 
Our definition broadens the scope of augmentations explored by prior work, such as annotations~\cite{rahman2024qualitative} (which constitute a subset of visual aids) and augmentations used to highlight data relationships~\cite{srinivasan2018augmenting, quadri2021survey} (which represent one kind of design goal).
In addition, we recognize that non-supplemental, standard encoding choices (not only new ones~\cite{guo2024we}) made during the initial data mapping may also serve communicative goals (\eg in addition to categorizing data, red and blue colors connote the Republican and Democratic parties in the United States).
We fully operationalize our definition in~\autoref{sec:design-space}, identifying augmentation tactics, design strategies, and higher-level goals to form a design space.

\figuredemocrats

\vspace{1mm}
\bpstart{Example: Democratic Party Trends.}
Perceptions of political party values have shifted over time.
This shift within the Democratic Party is demonstrated in \autoref{fig:democrats}, with voters' increasingly associating the party with issues related to marginalized groups, healthcare, and climate regulation. 
The title (\autoref{fig:democrats}(A)) and footnote (\autoref{fig:democrats}(D)) serve as \textbf{guides} for additional \textbf{context} on what the data represents and how it was sourced, aiding \textbf{interpretation}.
\textbf{Annotated text} (\autoref{fig:democrats}(B,C)) provides additional \textbf{context} for each of the line marks.
The placement of these annotations \textbf{sequence} the reader left-to-right, prompting them to \textbf{observe} the shift in associations from working class to other topical issues.
Finally, color hue (\autoref{fig:democrats}(B,C)) assists \textbf{interpretation} by \textbf{relating} annotated text with each line mark and connoting higher-level \textbf{associations} (\eg blue for blue-collar workers, pink for breast cancer awareness, and green for the climate). \figurecholera

\bpstart{Example: Cholera Outbreak.}
London suffered a cholera epidemic in 1854. 
Physician John Snow surmised that the source was contaminated water, removing the Broad Street pump to test his hypothesis. 
\autoref{fig:cholera} visualizes Snow's actions within the context of the epidemic, with several visual augmentations employed to support communication goals. 
The visualization makes use of \textbf{textual} and \textbf{connector} annotations (\autoref{fig:cholera}(C, D)) that provide \textbf{context} (\autoref{fig:cholera}(C)) as well as \textbf{spotlight} (\autoref{fig:cholera}(D)) when the Broad Street pump was removed.

Rather than use standard bar marks, the designer encoded the number of deaths with symbolic caskets (\autoref{fig:cholera}(A)) and includes pictures of historical figures (\autoref{fig:cholera}(B)).
These symbolic representations may help the reader \textbf{associate} the meaning of the data with known referents, which can aid \textbf{interpretation} of the subject matter.
It is worth noting that the decision to include these associations was strongly criticized by Tufte~\cite{tufte1997visual}.
Indeed, this decision may have little effect on the reader's ability to accurately decode the information presented in the chart, a topic of some debate~\cite{hullman2011benefitting, bateman2010useful}.
However, with respect to higher-level communication objectives, symbols may more directly convey a visualization's thematic content by engaging a reader's prior associations.

Finally, the chart \textbf{aggregates} data by week (\autoref{fig:cholera}(E)) to help the reader \textbf{observe} the reduction in deaths after the handle was removed.
This tactic must be used with care: depending on the context, aggregation achieves rhetorical aims or it may mislead the reader by hiding data that conflicts with a chosen narrative~\cite{hullman2011visualization, tufte1997visual}. 
Here, the number of \textit{daily} deaths were already decreasing before the handle was removed. 


\vspace{1mm}
\bpstart{Example: Exploring Algorithms.}
\figureexplorable
Explorable Explanations surface underlying models in the form of interactive content for readers to investigate, test, and refine both the assumptions that the model and themselves make~\cite{victor2011explorable}. 
One example of this type of content is demonstrated in \autoref{fig:explorable}, which communicates different layers of abstraction in a car simulation algorithm~\cite{victor2011ladder}.
As in previous examples, repeated use of linked colors help \textbf{relate} different content and functionality, aiding \textbf{interpretation}.
Red coloring (\autoref{fig:explorable}(A,C,E)) is used consistently throughout the article to help the reader \textbf{observe} specific interactive abilities (\eg mouse hovering in \autoref{fig:explorable}(A)), as well as \textbf{spotlight} selected content in \textbf{faceted} views (\autoref{fig:explorable}(C,E)).
Underlined blue text (\autoref{fig:explorable}(B)) represents snapshots of different model states, which the reader can click to \textbf{spotlight} corresponding content (\autoref{fig:explorable}(C)).

Finally, an interactive slider (\autoref{fig:explorable}(D)) enables the reader to explore and test the algorithm across faceted views.
Critically, this functionality encourages the reader to \textbf{introspect} by \textbf{self-reflecting} and \textbf{repairing} their knowledge. 
Specifically, the reader engages in a cycle in which they make tacit \textbf{predictions} about the algorithm, use the interactive slider to test these predictions, and revise their assumptions if the \textbf{simulated} content differs from the expected outcome.



\vspace{1mm}
\bpstart{Summary of Examples.}
These examples illustrate varied design tactics (encodings, annotations,
\add{interactive simulations}), strategies (spotlight, contextualize, relate), and objectives (observe, interpret) to scaffold \textit{how} and \textit{why} visual augmentations are employed.
Crucially, a single tactic can serve different strategies and objectives, sometimes simultaneously.
For example, as in \autoref{fig:democrats}(A), annotations may \textbf{sequence} data to guide readers in \textbf{observing} important data insights or phenomena, as well as provide \textbf{context} to connect and explain these insights, aiding \textbf{interpretation}.
Consistent use of colors can both \textbf{relate} and semantically \textbf{associate} content for \textbf{interpretability} (\autoref{fig:democrats}(B,C)).

During the design process, authors must consider (1) their communicative objectives, (2) what encoding and augmentation techniques are available to them, (3) which techniques are appropriate for their objectives, and (4) trade-offs and interactions between different techniques.
Currently, the literature provides guidance on points (1) and (2), but stops short of relating them to achieve points (3) and (4).

\section{Examining the Chasm: Related Work}
We review literature on graphical perception, visualization taxonomies, communicative visualization, and learning sciences to note current gaps in scaffolding communicative aids and providing designers with actionable design guidance. 

\subsection{Graphical Perception}
While the creative and effective use of visualization to leverage human cognition can be traced back as early as William Playfair's \textit{The Commercial and Political Atlas}~\cite{playfair1801commercial}, Bertin's \textit{Semiology of Graphics}~\cite{10.5555/1095597} arguably contributes the first formalism for visual encodings and the appropriate mappings from different types of data (\eg categorical vs. quantitative).
These principles have been empirically studied~\cite{pinker1990theory, kosslyn1989understanding} and mapped to effectiveness rankings for automatic systems~\cite{cleveland1984graphical}, smart defaults in visualization systems~\cite{mackinlay1986automating, satyanarayan2016vega, bostock2011d3}, and general criteria for appropriate chart design (\eg Mackinlay's expressiveness and effectiveness principles~\cite{mackinlay1986automating}).
Recent work has shown that perceptual principles which govern our visual system (\eg Gestalt principles~\cite{pinker1990theory}) can even affect higher-level comprehension, as readers focus on certain data relationships (\eg trends) which affects their insights and interpretation of a chart's intended message~\cite{fygenson2023arrangement, xiong2021visual}.
In part, this work empirically validates early efforts of psychologists to understand how humans comprehend graphics, as well as why various chart types and encodings are better suited for certain tasks than others (\eg using bar chart to compare overall trends vs. a line chart to comprehend changes in slopes)~\cite{pinker1990theory}.

Graphical perception principles helpfully prescribe effective low-level representations of data, such that readers can (on average) accurately decode visualized data. 
However, visualizations within a communicative context may be viewed more appropriately as a learning problem~\cite{adar2017persalog}.
As such, while perceptual principles might help readers \textit{quickly} and \textit{accurately} comprehend a chart (applying a cognitive-efficiency perspective), they fail to guide designers in reasoning about and achieving higher-level communication and learning goals.
Within the context of visual augmentations, graphical perception work may guide \textit{how} the augmentation should be implemented (\eg use color to encode categorical data or animation to convey change~\cite{tversky2002animation}), but not \textit{why} in the context of communication (\eg use red / blue color to culturally associate data with Republican / Democratic parties, or an interactive stepper that sequences the order of content to improve the reader's interpretation of the narrative).

\subsection{Visualization Tasks \& Evaluation}
Most visualization taxonomies consider the role of the data analyst, providing useful guidance for appropriate types of charts~\cite{quadri2021survey} or interactions~\cite{brehmer2013multi} based on that context.
Low-level task taxonomies operationalize graphical perception principles as different analytic tasks that the user might perform (\eg compare distributions, retrieve / identify values, \etc) and the appropriate representations for these tasks (\eg scatter plot for detecting outliers)~\cite{quadri2021survey}.
Brehmer and Munzner's typology of abstract visualization tasks~\cite{brehmer2013multi} mirror our efforts in the analytic space, bridging the \textit{why}, \textit{what}, and \textit{how} for analysis tasks.
However, this typology relegates all communicative objectives to a single abstract category: \textit{present}.
As Adar \etal~\cite{adar2020communicative} note, although this decision can reflect the constant cycle between analysis and communication~\cite{heer2007voyagers, nobarany2012facilitating}, it fails to delineate nuanced interactions among varied communication aims.
Naturally, this design process involves a different, possibly more complex, set of questions beyond the scope of analysis -- namely, \textit{what} to communicate, \textit{how} to communicate it, \textit{why} it should be communicated and \textit{to whom} it will be communicated (\ie the audience).
By bridging design intent with augmentation tactics (\ie how designers may realize their intents), our design space provides an initial foundation to help answer these questions, including reasoning about tradeoffs and interactions between visual augmentations which can be formally assessed in future studies.

Commonly, the effectiveness of a visualization is evaluated with a suite of cognitive-efficiency metrics (\eg accuracy, speed, size) that focus on data insights~\cite{fygenson2023arrangement, xiong2021visual}.
However, in isolation these metrics are often not the most appropriate for communicative visualization.
Rather, the primary concern is whether the reader \textit{actually} learned what the author intended~\cite{adar2020communicative}, as opposed to how quickly or accurately they could decode insights, some of which might be tangential to the original intent.
Although the use of visual augmentations outlined in our design space might require additional effort and time to decode (\eg annotations, explorable explanations~\cite{victor2011explorable}, \etc), we expect that what the reader \textit{learned} will be closer to the designer's intent.
We leave this evaluation as future work.

\subsection{Communicative Visualization \& Design}
A large body of work helps scaffold the design dimensions of communicative visualization across multiple contexts, including data storytelling and narration~\cite{segel2010narrative}, business dashboards~\cite{sarikaya2018we}, interactive articles~\cite{conlen2018idyll, conlen2019capture, hohman2020communicating}, and research papers~\cite{heer2023living}.
Although these design spaces address \textit{how} designers might author a visualization, they remain disconnected from \textit{why} (to what aim).
As a result, designers lack guidance when reasoning from high-level design goals, including how their final design might be improved to better align with reader takeaways.
For instance, when are \textit{explorable explanations} appropriate (which have been used incorrectly, much to Bret Victor's chagrin~\cite{victor2011explorable}), and what are alternative techniques depending on the viewing context?

We take direct inspiration from this line of work, asking, \textit{``How do varied techniques relate to designers' communicative goals?''}, and use literature in this area to scaffold our design space, noting similarities in how divergent tactics can be used to accomplish similar goals, often simultaneously.
A central benefit to bridging visual augmentations with design intents is the ability to identify potential mismatches with what readers actually learn.
For example, under what contexts (\eg background knowledge, domain, learning goals \etc) do narrative systems (\eg Idyll~\cite{conlen2018idyll}, explorable explanations~\cite{victor2011explorable}, playable guessing games~\cite{diakopoulos2011playable}) provide learning benefits to readers?
To answer this question empirically, we must first unpack the intended uses and interactions behind such augmentations, including \textit{when} and \textit{why} an augmentation outperforms others.

There exists a long-standing debate as to under which circumstances visual augmentations are beneficial to readers and to what the extent they are mere ``chart-junk''~\cite{bateman2010useful}.
Tufte's well-known prescription to maximize the data-to-ink ratio of a chart~\cite{tufte1990data} (\ie representing the most amount of (necessary) data with the smallest amount of ``ink'' in a chart) has been shown to not always equate with maximal comprehension or learning~\cite{hullman2011benefitting}, a central aim of communicative graphics.
Hullman \etal~\cite{hullman2011benefitting} characterize these differences as visual difficulties, such as charts with lower data-to-ink ratios that are more likely to engage the reader.
As an alternative to cognitive-efficiency evaluations, these benefits may arise (in certain contexts) by inducing active processing and self-explanation to better manipulate internal representations and expectations of data~\cite{chi2013self}. 
Visual augmentations may also provide novelty, aesthetics, personalization, challenge, and reflection, promoting engagement~\cite{hullman2011benefitting}.
By supporting learning outcomes at the expense of cognitive efficiency, we expect future assessments will help formalize the opportunity cost of visual augmentations, including diminishing returns (\eg too many augmentations, additional augmentations needed for more complex types of content or domains, \etc).

\subsection{Visualization Objectives \& Extant Design Spaces}
\add{
Prior work classifies visualization objectives at multiple levels: perceptual, descriptive, and cognitive.
Perceptual goals are often insight-driven~\cite{card1999readings, saraiya2005insight, saraiya2006insight, plaisant2007promoting}, ranging from low-level tasks such as summarization, comparison, and identification~\cite{quadri2021survey, rahman2024qualitative, brehmer2013multi, stokes2023role} to high-level comprehension (\eg reader takeaways)~\cite{pinker1990theory, kosslyn1989understanding, quadri2024you, ajani2021declutter} or aesthetics such as legibility~\cite{tufte1990data, ajani2021declutter}.
Most design spaces, however, often only relate these objectives to a single \textit{tactic} in our design space, such as annotations~\cite{rahman2024qualitative}, text~\cite{stokes2023role, head2022math}, or encodings~\cite{mackinlay1986automating, padilla2020uncertainty}.
}

\add{
Descriptive objectives delineate \textit{what} a visualization or augmentation is intended to accomplish, which roughly correspond to \textit{strategies} in our design space.
For example, prior work in narrative visualization~\cite{segel2010narrative, hullman2013deeper, bach2018design} relates specific implementation features (\eg zooming, progress bar) to structural devices affecting narrative flow (\eg highlighting, transitions, sequential ordering).
We organize implementation features into tactics to more clearly delineate important differences in how divergent tactics are used (\eg zooming, animated transitions, and annotations can each be used to highlight or sequence content, though annotations may be used in other ways, such as to provide external context).
}

\add{
Cognitive goals are the highest level and often the most difficult to assess.
These objectives may span the following contexts: learning~\cite{adar2020communicative, lee2021learning, adams2015bloom, ploetzner2020review}, affective or emotional~\cite{lee2022affective}, rhetorical~\cite{hullman2011visualization}, engagement or reflection~\cite{hohman2020communicating, kim2017explaining, boy2015storytelling}, efficiency~\cite{hullman2011benefitting}, and reasoning or decision-making~\cite{ge2024v}.
However, these objectives may comprise sub-goals themselves.
In the context of learning, ``recall'' / ``remember'' may be viewed as the first level (though this remains controversial~\cite{haladyna2004developing}) before progressing to higher levels such as ``understanding'', ``application'', and ``analysis''.
Though ``recall'' may be used as a proxy for comprehension~\cite{hullman2011benefitting, bateman2010useful, kim2017data}, this begs the question: \textit{why} does the designer want to the viewer to remember certain information?
Cognitive goals (affective, rheotrical, learning) may, for example, be in service of specific \textit{outcomes}, such as modifying the viewer's behavior, improving trust, or profit.
}

\add{
A central issue is that prior work often discusses perceptual and cognitive goals interchangeably at different levels of abstraction that are difficult to define and relate.
For instance, graphical perception work discusses ``interpretation'' in the context of visual encodings and semantics~\cite{kosslyn1989understanding, pinker1990theory} (\ie which marks are better at denoting different types of data and relationships?), while ``interpretation'' as a learning goal~\cite{lee2021learning} is often higher-level and depends on both the encodings and visual augmentations (\eg using annotations to help readers' interpret a correlation or summarize key temporal stages~\cite{adar2020communicative}).
Certain learning verbs (\eg summarize, compare) also overlap with analytic visualization tasks in the context of interaction~\cite{brehmer2013multi} or chart encodings~\cite{quadri2021survey}, which obfuscates what the viewer does \textbf{cognitively} versus what the viewer \textit{can} do \textbf{analytically} (\eg via interaction).
}

\add{
Conceptually, our visualization objectives (observe, interpret, and introspect) abstract descriptive goals into explicit properties of the chart that reflect what the viewer will \textit{do} as a result of a visual augmentation.
These objectives are partitioned in a general manner to help relate perceptual and cognitive aims through our objective hierarchy: observation augmentations serve primarily as attention aids (\eg to identify, compare, \etc), while interpretation and introspection aids modify what the viewer interprets from this attention and how they may assess their interpretation.
Designers can then use these objectives as guides to reflect on the viewer's behavior and reason about how to achieve higher-level cognitive goals (\eg learning, rhetorical, \etc).
}

\add{
Our design space reflects general-to-specific structures between low-level verbs and high-level goals from prior taxonomies~\cite{brehmer2013multi, segel2010narrative, bach2018design}.
However, several gaps remain in extant taxonomies:
\begin{enumerate}
    \item[G1] Communication goals aimed at viewers are often relegated to a single category~\cite{brehmer2013multi} or not abstract enough to relate varied augmentations~\cite{segel2010narrative, bach2018design}.
    \item[G2] Communication goals are described without connection to the full range of tactics~\cite{adar2020communicative, hullman2011benefitting}.
    \item[G3] Augmentations are only discussed for a single tactic~\cite{rahman2024qualitative, chevalier2016animations} or context~\cite{segel2010narrative, bach2018design}.
    \item[G4] Roles and relationships among augmentations at different levels are not delineated, including (1) similar uses of divergent tactics, (2) different uses of the same tactic, (3) or using a single tactic to achieve multiple goals simultaneously.
\end{enumerate}
These gaps pose two primary risks: (1) the visualization community cannot adequately analyze and evaluate diverse communication visualizations across a variety of presentation contexts due to a lack of shared goals and language for describing visual augmentations; (2) without a sufficiently nuanced vocabulary, designers lack tools that both encourage reflection about what the viewer does and guide them with different strategies and tactics for achieving these goals, resulting in potentially sub-optimal designs that limit reader comprehension and learning.
}

\add{
Our design space defines augmentations at multiple levels of specificity (tactics, strategies, objectives), enabling us to capture nuanced roles, relationships, and dependencies (G2-4).
To provide scaffolding, we abstract perceptual and cognitive goals into more general categories that describe viewers' cognitive behaviors (G1).
Since the designer cannot ``force'' their ultimate goal (\eg that the viewer will have learned exactly what they intended), our objectives can also be interpreted as ``guides'' to help designers achieve such higher-level intents. 
}



\subsection{Communication \& Learning Sciences}
Our interest in the relationship between the \textit{use} and \textit{intent} of visual augmentations was inspired heavily by Adar and Lee-Robbins~\cite{adar2020communicative}.
They frame communicative visualization as a learning problem, building off \textit{process}-~\cite{shannon1948mathematical} and \textit{meaning}~\cite{10.5555/1095597}-focused communication theories and extending Bloom's taxonomy of educational learning objectives~\cite{adams2015bloom}.
This framing provides important scaffolding for designers to both describe what they want readers to actually learn and use assessment to highlight gaps in the design.
We view our design space as a next step in this direction, seeking to provide more actionable design techniques to control different learning ``levers''.
While we reason about how our designer objectives, strategies, and tactics (\autoref{sec:design-space}) might relate to Adar's learning taxonomy~\cite{adar2020communicative} (\autoref{sec:discussion}), we leave a more exact mapping to future work given the need for further empirical grounding.
One benefit of keeping an arm's length between our notion of \emph{design intent} and learning objectives is that our intent goals might be related more cleanly to additional objectives, including affective~\cite{lee2022affective} and rhetorical framing~\cite{hullman2011visualization, stokes2023role} goals.

Our design space builds on theoretical foundations in communication theory, cognitive psychology, and semiotics~\cite{10.5555/1095597, pinker1990theory, kosslyn1989understanding}.
Central to this foundation is the notion of syntax (\ie how marks are visually interpreted based on spatial grouping and organization), semantics (\ie what the configurations of marks represent, such as data values, axes, legends, labels, \etc), and pragmatics (\ie how indirect information is conveyed beyond semantic interpretation, which may be affected by the context of viewing, how much information is included, \etc)~\cite{kosslyn1989understanding}.
Graphical perception literature helpfully distills principles grounded in syntactic and semantic notions~\cite{10.5555/1095597, mackinlay1986automating, cleveland1984graphical}, elucidating how humans comprehend visual information (\eg use of shape encoding conveys categorical semantics effectively, but not quantitative).
However, we find that visual augmentations often extend beyond denotative semantics to drive connotative meaning and pragmatic intent.
For example, the use of caskets instead of bars (\autoref{fig:cholera}) may evoke emotion or serve rhetorical aims~\cite{hullman2011visualization} (\eg shifting the focus to John Snow's heroic efforts, or commenting on the religious influences in medicine at that time).

We consider the manner and context in which augmentation tactics might be used, reflecting both denotative semantics and varied connotative and pragmatic goals.
For example, color encodings may be used to spotlight specific content (denotative semantics), associate data with cultural concepts (\autoref{fig:democrats}(B,C)) (connotative semantics), or relate groups of content together due to spatial proximity (\autoref{fig:democrats}(B,C)) (pragmatics).
The distinction between denotative and connotative semantics and pragmatics helped us partition and reason about different uses of tactics more effectively: observation aids often affect denotative semantics, while interpretation and introspection tactics support more complex connotative and pragmatic aims that consider the background of the reader, the complexity of the data, and in what context the visualization will be viewed.

Visual augmentations in communicative contexts other than data visualization have been explored, such as mathematical notation~\cite{head2022math}, constrained highlighting for text~\cite{joshi2024constrained}, and code debugging~\cite{chandra2024watchat}. 
We find that our design space shares certain tactics with these domains (\eg annotation, use of color to link concepts), suggesting that some augmentations might share communicative grounding and learning benefits more broadly.
Our work considers additional tactics central to spatiovisual data representations (\eg animation, sequencing, \& layout), including their relationship with varied design goals.

\section{\texorpdfstring{Bridging the Chasm:\\~A Design Space of Visual Augmentations}{Bridging the Chasm: A Design Space of Visual Augmentations}}
\label{sec:design-space}

To bridge visual augmentations and designer intent, we sought to (1) identify instances of visual augmentations throughout the visualization and perception literature, (2) characterize higher-level design goals, and (3) ground augmentations in these goals.
This survey resulted in a design space with three tiers: objectives, strategies, and tactics (\autoref{table:augmentations}).
\emph{Objectives} encompass core viewer-oriented goals by which authors seek to communicate the message of their visualization, such as directing attention to important observations or adding context to aid a reader's interpretation.
\emph{Tactics} are lower-level implemented design techniques, such as animated transitions or stylized encoding choices, that can influence viewer attention, interpretation, and consideration.
\emph{Strategies} are approaches the author selects to connect the two, applying one or more tactics to achieve an objective (\eg both animation and encoding may be used to emphasize an observation).

\tableaugmentations

\subsection{Survey Methodology}
We seeded our survey of visual augmentations with several key papers and books in communicative visualization, visual learning, and graphical perception, such as Segel and Heer's design space of narrative visualization techniques~\cite{segel2010narrative}, Tversky's evaluation of the effectiveness of animation for learning~\cite{tversky2002animation}, Adar and Lee-Robbins' extended learning objectives for communicative visualization~\cite{adar2020communicative, lee2021learning}, Tufte's \textit{Visual Explanations}~\cite{tufte1997visual}, and McCloud's \textit{Understanding Comics}~\cite{alma99126810900001452}.
We then snowball-sampled papers across related areas, such as uncertainty visualization~\cite{padilla2020uncertainty}, interactive articles~\cite{victor2011explorable, conlen2018idyll, hohman2020communicating}, augmented visual tools~\cite{srinivasan2018augmenting}, existing design spaces~\cite{bach2018design, shi2021communicating} and visualization taxonomies~\cite{quadri2021survey, brehmer2013multi, heer2007animated}.
Additionally, \add{for completeness}, we widened our search by searching for key terms and phrases on Google Scholar, such as ``communicative visualization'', ``visual augmentations'', ``visual learning'', \etc
In total, we collected an initial set of 70 papers.

\add{
As with prior design spaces~\cite{segel2010narrative}, we do not claim that our sample is exhaustive, though we optimized for a rich set of diverse themes within the scope of communicative visualization, prioritizing a diversity of goals (\eg perceptual~\cite{quadri2021survey}, cognitive~\cite{adar2020communicative}, \etc), media forms and contexts (\eg narrative journalism~\cite{segel2010narrative}, comics~\cite{bach2018design}, dashboards~\cite{sarikaya2018we}, \etc), and augmentation roles (\eg annotation~\cite{rahman2024qualitative}, animation~\cite{tversky2002animation}, self-prediction~\cite{kim2017explaining}, \etc).
We also found that additional relevant papers contained either repetitive ideas or were described by taxonomies already present in our corpus (\eg animation in learning contexts, types of communication devices in visualization, \etc).
Still, we acknowledge some inevitable subjectivity in this process.
}
We removed 18 papers from our augmentation coding process that: described graphical comprehension (2/18), visual learning (3/18), or attention (4/18), but not augmentations therein; supported objectives outside the cognitive domain such as affective (1/18), rhetorical (1/18), or social engineering (2/18); or provided textual augmentations not primarily focused on aiding visualization (5/18).

\add{
We open-coded the resulting set of 52 papers, deriving a shared set of 12 tactics, 7 strategies, and 3 objectives, with the relationships between these categories depicted in~\autoref{fig:design-space}.
We describe this open-coding process in more detail, revisit our paper corpus to categorize each paper along our design space dimensions, and provide a selected set of additional examples in the Appendix.
}

\figuredesignspace

\subsection{Augmentation Tactics}
We first describe tactics, which categorize common augmentation techniques and their implementation \add{(e.g., interactive features, context of use, \etc)}.
Tactics indicate \textit{how} an augmentation achieves its design goal.

\vspace{1mm}
\bpstart{\xmybox[tactic]{Transform}}
Designers employ a variety of data transformations including derivations such as regressions~\cite{guo2024we, srinivasan2018augmenting, rahman2024qualitative}, segmentation (drill-down)~\cite{guo2024we}, or normalization~\cite{ge2024v} (\autoref{fig:recognize-tactics}(A)).
Filtering transforms~\cite{brehmer2013multi, quadri2021survey} may also structure visual information or highlight particular insights.

\vspace{1mm}
\bpstart{\xmybox[tactic]{Animate}}
Animation is widely used across a variety of contexts~\cite{chevalier2016animations, mckenna2017visual, ploetzner2020review, hullman2011benefitting, sultanum2021leveraging, shi2021communicating}, including facilitating transitions between statistical views~\cite{hill1991deixis, kim2020gemini, heer2007animated}, providing narrative structure~\cite{segel2010narrative, satyanarayan2014authoring, mckenna2017visual}, aiding learning of complex processes~\cite{tversky2002animation, ploetzner2020review}, or to convey real-time change~\cite{hill1991deixis, tversky2002animation}.
Clever use of animation can provide tacit tutorials for interactions~\cite{segel2010narrative}.
Judicious use of motion and timing can also help group, filter, or understand data relations~\cite{ge2021cast} (\autoref{fig:recognize-tactics}(B)).

\vspace{1mm}
\bpstart{\xmybox[tactic]{Layout}}
Layout refers to any arrangement of multiple views in space used to guide readers or facilitate examination.
Faceting is heavily employed ~\cite{mackinlay1986automating, quadri2024you, hullman2013deeper, kim2017graphscape, hohman2020communicating, victor2011scientific, tufte1997visual, victor2006magic} to aid visual comparison~\cite{victor2011explorable, heer2012interactive} (\autoref{fig:recognize-tactics}(C)).
Space can also structure content in a particular sequence to guide readers~\cite{hullman2013deeper}, including the use of scrolling~\cite{conlen2019capture, hohman2020communicating}, steppers~\cite{mckenna2017visual, sultanum2021leveraging}, or ordering to fit a particular narrative~\cite{segel2010narrative, bach2018design, alma99126810900001452}.
Content may also be superimposed to facilitate engagement and relate distinct views (\eg before and after images~\cite{tufte1997visual}), or hide supplemental information that can be displayed on-demand via mouse hovering or toggle buttons~\cite{hohman2020communicating}.

\vspace{1mm}
\bpstart{\xmybox[tactic]{Annotate}}
Annotation primarily directs attention or adds useful context~\cite{quadri2024you, conlen2019capture, ajani2021declutter, srinivasan2018augmenting, heer2012interactive, heer2012interactive, brehmer2013multi, rahman2024qualitative, conlen2018idyll, clark2010graphics, sultanum2021leveraging, alma99126810900001452, tufte1997visual} (\autoref{fig:interpret-tactics}(A)).
Annotations help add / layer useful marks such as glyphs, threshold lines, and trends~\cite{guo2024we, hill1991deixis} to focus data characteristics.
Occasionally, annotations may be dynamic for details-on-demand, which can be revealed through click or hover interactions~\cite{satyanarayan2014authoring, crichton2021new, segel2010narrative}.
Designers also annotate for context (\eg using a question mark to indicate missing values~\cite{ge2024v} or uncertainty~\cite{heer2012interactive, padilla2020uncertainty}) or to guide readers through explanations~\cite{heer2012interactive} or sequences~\cite{clark2010graphics}.

\vspace{1mm}
\bpstart{\xmybox[tactic]{Encode}}
Encoding refers to a broad set of augmentations that exploit visual channels or mappings of data to convey higher-level information ~\cite{guo2024we, srinivasan2018augmenting, brehmer2013multi, quadri2021survey, clark2010graphics, sultanum2021leveraging, cai2024watching, ge2024v, padilla2020uncertainty} (\autoref{fig:interpret-tactics}(B,C)).
Common channels to emphasize particular information include brightness and size~\cite{hill1991deixis}, color~\cite{ajani2021declutter, victor2006magic, quadri2021survey, hill1991deixis}, or the shapes of marks themselves~\cite{guo2024we}.
Encodings, such as color, can also help order~\cite{guo2024we} or link~\cite{conlen2019capture, ajani2021declutter, victor2006magic, crichton2021new, victor2011explorable, sultanum2021leveraging, victor2011scientific} items within a view.
We also noted that the layout \textit{within} a single view (\eg by modifying the position encoding via sorting or stacking~\cite{quadri2021survey, fygenson2023arrangement, shah1999graphs, xiong2021visual}) could be used to arrange items in a particular way to emphasize or order content~\cite{fygenson2023arrangement, xiong2021visual}.

\vspace{1mm}
\bpstart{\xmybox[tactic]{Symbolize}}
Symbolize refers to the use of icons, pictures, or other referents to ground and concretize the meaning of a visualization, employed broadly across designs~\cite{clark2010graphics, cai2024watching, ge2024v, shi2021communicating, hullman2011benefitting, tufte1997visual, levie1982effects} (\autoref{fig:interpret-tactics}(C)).
For instance, Conlen \etal~\cite{conlen2019capture} utilized an interactive guitar to contextualize an article on tuning.
Other uses might include annotated question marks to reference uncertainty~\cite{heer2012interactive, padilla2020uncertainty} or unit visualizations~\cite{hohman2020communicating}.
The degree to which symbols are used within a visualization can constitute chart junk~\cite{bateman2010useful, hullman2011benefitting}, although in proper contexts, pictorial illustrations can aid learning~\cite{lee2021learning} and storytelling~\cite{bach2018design, alma99126810900001452, segel2010narrative}.

\vspace{1mm}
\bpstart{\xmybox[tactic]{Guide}}
Guides refer to a set of contextual aids that help communicate the visualization's message, intent, or provenance.
Examples include captions / titles~\cite{segel2010narrative, YangLeni2023EWEL}, footnotes~\cite{victor2006magic}, personalized information such as date, time, location~\cite{victor2006magic, adar2017persalog} (\autoref{fig:intervene-tactics}(A)), the level-of-detail of text~\cite{hohman2020communicating, hullman2011benefitting, cordova1996intrinsic, mayer2005cognitive}, or progress bars to aid navigational context~\cite{segel2010narrative, cai2024watching}.
Designers might also include links to external data or views~\cite{heer2012interactive} to indicate the source or methods of data collection.

\vspace{1mm}
\bpstart{\xmybox[tactic]{Query}}
Query mechanisms incorporate quizzes, potentially spaced out over time with recurrence, to test reader knowledge directly and encourage learning~\cite{hohman2020communicating, matuschak2019can} (\autoref{fig:intervene-tactics}(B)).

\vspace{1mm}
\bpstart{\xmybox[tactic]{Predict}}
Prediction tactics engage readers in ``guessing'' the data, encouraging self-reflection (\autoref{fig:intervene-tactics}(C)).
These include graphical elicitation (\ie drawing)~\cite{kim2017explaining, hohman2020communicating}, interactive widgets and annotations to facilitate active reading / question generation~\cite{conlen2019capture, satyanarayan2014authoring, hohman2020communicating, crichton2021new, victor2011explorable, conlen2018idyll, victor2011scientific}, and 
\add{games}~\cite{diakopoulos2011playable}.

\vspace{1mm}
\bpstart{\xmybox[tactic]{Explain}}
Explanation tactics are mechanisms that motivate readers to self-reflect, potentially using inference to fill in missing gaps in their knowledge~\cite{kim2017explaining}.
Examples include self-explanation summaries~\cite{kim2017explaining, hohman2020communicating} or self-drawing to explain complex processes~\cite{levie1982effects}.



\vspace{1mm}
\bpstart{\xmybox[tactic]{Evaluate}}
Evaluation techniques assess reader predictions, including graphical feedback~\cite{kim2017explaining}, results of quizzes~\cite{matuschak2019can}, and game win / loss~\cite{matuschak2019can}.
Self-evaluation can also take place by hiding contextual information behind interactions, allowing readers to in effect use visualizations as flash cards~\cite{hohman2020communicating}.

\vspace{1mm}
\bpstart{\xmybox[tactic]{Simulate}}
Readers' input with a visualization or interactive article can drive / simulate dynamic system behavior, enabling them to compare system output with tacit predictions about their input:~\cite{conlen2019capture, satyanarayan2014authoring, crichton2021new, victor2011explorable, conlen2018idyll, diakopoulos2011playable, hohman2020communicating, victor2006magic}.

\figurerecognizetactics

\subsection{Designer Objectives}
Augmentation tactics are used to implement broader design objectives.
From our survey, we identified three broad categories of goals that guide design decisions, signaling \textit{why} designers use varied augmentations. 
These objectives operationalize intent as core viewer-oriented mechanisms to effect differences in chart (1) attention and observation (\ie what readers notice and in what order), (2) comprehension and inference (\ie the message and intent that the reader discerns, potentially driven by attention aids), and (3) self-assessment (\ie enabling the reader to explore and refine their comprehension).

We intentionally characterize design intent according to more general cognitive descriptors such as attention and understanding (rather than, for example, specific learning~\cite{adar2020communicative} or rhetorical~\cite{hullman2011visualization} goals) to relax conceptual constraints, enabling us to relate and organize augmentation tactics more meaningfully and encode communication design as actionable techniques.
To connect augmentation tactics with objectives, we surface design strategies that represent the manner in which divergent tactics are instrumented to achieve similar goals, grounding tactics in objectives as a result. 
Concretely, strategies represent \textit{how} a tactic contributes to an objective.

\vspace{1mm}
\bpstart{\xmybox[objective]{Observe}}
A variety of augmentations direct readers' attention~\cite{satyanarayan2014authoring, heer2012interactive, xiong2021visual} to specific areas within a visualization. 
However, directing attention often serves the broader aim of helping the reader \emph{observe} specific or important aspects of the data.
We categorize these augmentations as \emph{Observation} aids, which help the designer more readily communicate insights for the reader to observe and takeaway~\cite{heer2012interactive, rahman2024qualitative}.
The following two strategies support this objective, commonly via tactics that facilitate important lower-level perceptual tasks, such as comparison, identification, or summarization~\cite{brehmer2013multi, quadri2024you, quadri2021survey, rahman2024qualitative, shi2021communicating, ancker2024you}:

\subsubsection*{1) \textbf{\xmybox[strategy]{Sequence}}}
Sequencing refers to the general strategy of arranging content in a particular order.
\xmybox[tactic]{Animation} can be effective tactic for readers to observe important aspects about the \textit{manner} of change~\cite{tversky2002animation} (\eg \textit{how} temporal data changes over time).
For instance, items might be unveiled (\eg fading) in a particular order~\cite{shi2021communicating} to draw attention to specific changes between data values (\autoref{fig:recognize-tactics}(B)).
Spatial \xmybox[tactic]{layout} tactics, such as coordinated multiple views~\cite{heer2012interactive}, faceting~\cite{mackinlay1986automating, victor2006magic, victor2011explorable}, or structured narratives~\cite{hullman2013deeper}, can draw readers' attention to important differences that are more readily apparent between sequential views~\cite{guo2024we, hullman2013deeper, victor2006magic, heer2012interactive}.
A historical example of note is that sequenced frames helped people observe that all of a horse's hooves break contact with the ground during a trot~\cite{hohman2020communicating} (\autoref{fig:recognize-tactics}(C)).

Further, position and color \xmybox[tactic]{encodings} (\eg stacking / sorting~\cite{fygenson2023arrangement} or sequential color hue~\cite{guo2024we}) can affect how readers group visual marks into separate arrangements to make comparisons~\cite{fygenson2023arrangement, shah1999graphs, xiong2021visual} or observe salient trends more easily~\cite{brehmer2013multi} based on Gestalt principles of perception~\cite{pinker1990theory}.
In addition, certain \xmybox[tactic]{transformations} such as normalization can substantially alter a data distribution and, as a result, the order of a trend (\autoref{fig:recognize-tactics}(A)).

While \xmybox[tactic]{annotations} are frequently used as attention or contextual aids~\cite{rahman2024qualitative}, they can also explicitly sequence content.
For example, annotated numbers or arrows can enumerate or reinforce existing sequences~\cite{rahman2024qualitative, clark2010graphics, mckenna2017visual}, or draw attention to less obvious sequences~\cite{adar2020communicative, clark2010graphics} that convey important trends across time.

\figureinterprettactics

\subsubsection*{2) \textbf{\xmybox[strategy]{Spotlight}}}
Spotlighting strategies focus or highlight specific information, most often as observation cues.
Lone \xmybox[tactic]{encoding} variations and \xmybox[tactic]{annotations} emphasize and distinguish particular data~\cite{hill1991deixis, rahman2024qualitative, clark2010graphics}.
This emphasis can help a reader recognize anomalies~\cite{hill1991deixis}, values of significance~\cite{guo2024we, clark2010graphics}, missing values (\eg election data~\cite{ge2024v}), or small / occluded items that orient a reader (\eg mouse cursor~\cite{hill1991deixis}).
For instance, Heer \etal~\cite{heer2012interactive} include an example of a reader using a crowdsourced system who draws a line (annotation) mimicking the data distribution, helping other readers recognize that the data distribution is being referenced.
An annotated question mark is then included to further clarify that the distribution confuses the reader.

Linked or repeated \xmybox[tactic]{annotations} and \xmybox[tactic]{encodings} can emphasize groups of items to more readily make comparisons~\cite{guo2024we, ajani2021declutter, victor2006magic, quadri2021survey} and highlight data trends~\cite{quadri2024you, srinivasan2018augmenting}.
These augmentations also extend to different viewing contexts for the reader.
For interactive content driven by user input, linked annotations (\eg underlined text~\cite{victor2011explorable}) may facilitate observation of dynamic updates within the article.
In addition, humans can dynamically \xmybox[tactic]{annotate} during presentations by gesturing with their hands~\cite{hill1991deixis}, such as lassoing with one's finger to attend to a certain grouping (circle annotation) or using spread fingers to indicate the range of data (interval annotation).

While less common, creative uses of \xmybox[tactic]{animation} can function as a filtering / sliding window~\cite{hill1991deixis}, zoom-in lens~\cite{chevalier2016animations}, or motion separator~\cite{shi2021communicating}, reducing information overload and revealing data outliers more easily~\cite{hill1991deixis}.
\xmybox[tactic]{Transformations} such as filtering~\cite{brehmer2013multi, quadri2021survey} and threshold annotations~\cite{rahman2024qualitative} can graphically aid observation of irregularities with data distributions~\cite{hill1991deixis}.
Aggregation or drill-down \xmybox[tactic]{transformations} may also highlight fine-grained details more clearly, such as how individual data items contribute to the whole~\cite{guo2024we}.
Finally, jutaxposted cut-outs of visualizations regions (\xmybox[tactic]{layout}) can emphasis~\cite{bach2018design} important content.

\vspace{1mm}
\bpstart{\xmybox[objective]{Interpret}}
Even when augmentations are used to facilitate observation, readers may lack the appropriate background knowledge to understand the context and message of a visualization~\cite{clark2010graphics}.
To address this issue, \emph{Interpretation} aids are a class of augmentations designed to help the reader understand what the visualization means (\ie the semantics), such as relationships between data or multiple charts, thematic content, or external information important for context.
The following three design strategies aid reader interpretation:


\subsubsection*{1) \textbf{\xmybox[strategy]{Relate}}}
Relate strategies connect and link content to more easily understand conceptual relationships or convey semantic similarity.
Sequence tactics can directly support the relation strategy for readers to more easily comprehend difficult concepts or transitions.
A primary example is the use of \xmybox[tactic]{animation} to follow difficult transitions or maintain orientation, helping readers understand the relationship and connection between disparate views~\cite{hill1991deixis, tversky2002animation, chevalier2016animations, heer2007animated, kim2020gemini, mckenna2017visual, ploetzner2020review, satyanarayan2014authoring, ge2021cast}.
Further, the sequential \xmybox[tactic]{layout} of content can convey cause-and-effect~\cite{clark2010graphics, hohman2020communicating} relationships or narrative events~\cite{alma99126810900001452, bach2018design, shi2021communicating, satyanarayan2014authoring, victor2006magic} more effectively, providing benefits to learning~\cite{mayer2005cognitive}.
Sequential content can also relate the overall structure and organization of a visualization or story, enabling readers to comprehend the intent or message more easily while also enhancing aesthetics and clarity~\cite{ajani2021declutter}.
For example, parallel layouts~\cite{tufte1997visual}, consistent templates (\eg general-to-specific transitions within a narrative story~\cite{segel2010narrative, hullman2013deeper}), and navigation bars~\cite{segel2010narrative, mckenna2017visual} can orient readers within the narrative and help them form connections between views.

More broadly, relate augmentations can help readers understand exact relationships, classifications, or meanings between parts of a visualization (\eg groups of text or marks), affecting the message that is conveyed.
For instance, \xmybox[tactic]{encodings} (\eg partial coloring~\cite{guo2024we}), \xmybox[tactic]{annotations} (\eg glyphs~\cite{rahman2024qualitative}), motion (\eg marks with different \xmybox[tactic]{animated} movements~\cite{ge2021cast} that exploit the Gestalt Law of Common Fate~\cite{koffka1922perception}), repeated use of \xmybox[tactic]{symbols} (\autoref{fig:interpret-tactics}(C)), or \xmybox[tactic]{layout} (\eg clustering) can link similar groups of items~\cite{conlen2019capture, ajani2021declutter, victor2006magic, victor2011explorable, victor2011scientific}.
In addition, certain annotations like arrows or connectors may not only facilitate observation of specific chart data, but also link contextual descriptions to the data~\cite{clark2010graphics, mayer2005cognitive} (\autoref{fig:interpret-tactics}(A)).

Although linked \xmybox[tactic]{encodings} help delineate categories in standard graphical perception~\cite{mackinlay1986automating}, they may also link the semantics (\eg linked content can indicate reactive updates, interactive behavior~\cite{hohman2020communicating, victor2011explorable, boy2015storytelling, victor2011scientific}, or dynamic annotations for details-on-demand~\cite{hohman2020communicating, crichton2021new}).
For example, judicious use of underlined text (\xmybox[tactic]{annotation}) in interactive documents~\cite{conlen2019capture, conlen2018idyll, victor2006magic, victor2011explorable} not only serves to draw attention to updated content, but also for readers to interpret such text as linked and interactive.
This augmentation can then function as a spotlighting mechanism for the reader to search for similar content later on (as with text highlighting~\cite{joshi2024constrained}).

\subsubsection*{2) \textbf{\xmybox[strategy]{Associate}}}
Association strategies ground content with semantic or connotative referents, facilitating interpretation~\cite{conlen2019capture, hullman2011benefitting, bateman2010useful, shi2021communicating, hohman2020communicating} if the reader has sufficient background knowledge~\cite{clark2010graphics}.
For instance, a guitar \xmybox[tactic]{symbol}~\cite{conlen2019capture} readily associates a narrative with musical content, which can serve as a future memory aid~\cite{bateman2010useful, hullman2011benefitting}.
The timing of \xmybox[tactic]{animated} transitions~\cite{chevalier2016animations} and \xmybox[tactic]{encoded} angles and colors of annotations~\cite{heer2012interactive} can connote emotion~\cite{chevalier2016animations}.
\xmybox[tactic]{Encoding} channels and \xmybox[tactic]{animation} can also communicate uncertainty, such as blurred motion~\cite{shi2021communicating} and disordered stacking~\cite{padilla2020uncertainty}.
Further, designers can employ colors signifying political parties~\cite{cai2024watching} or errors~\cite{kim2017explaining} and use \xmybox[tactic]{symbolic} representations of missing data (\eg question marks or disjointed lines~\cite{ge2024v}).

\subsubsection*{3) \textbf{\xmybox[strategy]{Contextualize}}}
Context strategies provide additional information to deepen the reader's understanding.
\xmybox[tactic]{Annotations} are widely used as contextual aids to provide details-on-demand~\cite{satyanarayan2014authoring, segel2010narrative, hohman2020communicating, crichton2021new, conlen2018idyll} or include external information necessary for interpretation~\cite{rahman2024qualitative}.
Annotations can be used in tandem with spotlighting strategies to contextualize specific data~\cite{heer2012interactive}.
\xmybox[tactic]{Guides} also contribute important context, such as captions and titles~\cite{segel2010narrative, kim2017explaining}, legends~\cite{hullman2011benefitting}, footnotes~\cite{victor2006magic}, personalized text~\cite{victor2006magic, hohman2020communicating, adar2017persalog, cordova1996intrinsic, mayer2005cognitive} (\autoref{fig:intervene-tactics}(A)), progress bars~\cite{segel2010narrative, cai2024watching, diakopoulos2011playable}, and external links~\cite{heer2012interactive}.

Content tutorials provide useful context for the reader prior to engaging with main content.
Initial \xmybox[tactic]{layouts} can provide tacit tutorials~\cite{heer2012interactive, segel2010narrative}, while \xmybox[tactic]{animating} specific interactions may help readers understand their functionality~\cite{chevalier2016animations, heer2012interactive}.
Spatial \xmybox[tactic]{layouts} can also provide overview+detail context~\cite{bach2018design}.
Further, sequential illustrations may be used to provide additional information not readily comprehended from text alone~\cite{levie1982effects, bach2018design, alma99126810900001452}.


\figureintervenetactics

\vspace{1mm}
\bpstart{\xmybox[objective]{Introspect}}
Observation and interpretation help designers tailor the message of their visualization, but do not motivate readers to examine or repair their own assumptions, biases, knowledge, or meta-knowledge (\ie the process by which they acquired that knowledge~\cite{kim2017explaining}).
\emph{Introspection} aids are a class of augmentations that facilitate self-learning~\cite{kim2017explaining}, hypothesis generation and testing to explore one's understanding~\cite{victor2011explorable}, and repair of faulty knowledge~\cite{kim2017explaining, matuschak2019can}.
The following two strategies can refine readers' prior beliefs and knowledge:

\subsubsection*{1) \textbf{\xmybox[strategy]{Reflect}}}
Reflect strategies prompt self-examination of one's assumptions to apply their understanding and generate inferences.
Graphical interventions drive readers to \xmybox[tactic]{predict} partially unseen data, encouraging them to construct external representations of their prior knowledge and engage more deeply with understanding the data~\cite{kim2017explaining, hohman2020communicating} (\autoref{fig:intervene-tactics}(C)).
For instance, Kim \etal~\cite{kim2017explaining} asked readers to predict the change in voting percentage of Hispanic voters between income brackets by drawing a line; percentages for White and Black voters were provided as context.
Readers can also generate textual (self-) \xmybox[tactic]{explanations}~\cite{kim2017explaining}, which may prompt them to reflect on their prior knowledge to generate inferences that fill in gaps~\cite{chi2013self}.
For explaining complex processes, self-illustration may be used instead~\cite{levie1982effects}.

Other types of visual augmentations that encourage active reading, such as explorable explanations~\cite{conlen2019capture, hohman2020communicating, crichton2021new, victor2011explorable, conlen2018idyll, satyanarayan2014authoring} (\autoref{fig:explorable}), can prompt (self-) \xmybox[tactic]{prediction} in which readers generate hypotheses about the data that they test by modifying system parameters via dynamic (interactive) widgets.
Assessment \xmybox[tactic]{queries} (\eg quizzes as in \autoref{fig:interpret-tactics}(B) or flashcards~\cite{hohman2020communicating, matuschak2019can}) can directly prompt reflection, and may use a sequencing strategy (\eg interspersing quizzes throughout data narratives~\cite{hohman2020communicating}) to promote learning~\cite{matuschak2019can}.

\subsubsection*{2) \textbf{\xmybox[strategy]{Repair}}}
Repair strategies provide feedback for readers to observe knowledge failures, encouraging self-learning and potentially critiquing meta-knowledge~\cite{kim2017explaining}.
We observed three primary tactics to facilitate self-repair.
\xmybox[tactic]{Evaluate} tactics \emph{directly} assess the correctness and quality of readers' elicited beliefs / knowledge from the reflection stage.
For example, self-prediction techniques help readers repair their knowledge by providing the correct result~\cite{kim2017explaining}.
Graphical feedback may utilize observation and interpretation aids to draw readers' attention to the literal gap in their knowledge (\ie magnitude of the error in their prediction~\cite{kim2017explaining}).
Visualization games~\cite{diakopoulos2011playable} and quizzes / flashcards~\cite{hohman2020communicating, diakopoulos2011playable} can also evaluate reader input (\eg game win / loss, correct result shown after quiz).

Interactive content, such as explorable explanations (\autoref{fig:explorable}), may provide a similar feedback loop to evaluation by \xmybox[tactic]{simulating} dynamic content with interactive input, rather than assessing the correctness of that input.
Specifically, readers may generate predictions and hypotheses, modify the system state via interactive widgets, and use dynamically executed results to infer relationships and help repair faulty knowledge and assumptions.

Prior work~\cite{kim2017explaining} has shown that (self-) \xmybox[tactic]{explanation}~\cite{kim2017explaining} can improve data recall, which may be explained by the mental model repair hypothesis~\cite{chi2013self}: \textit{quality} self-explanations (which may be dependent on adequate prior knowledge) prompt readers to generate inferences that fill in gaps in their prior knowledge, and \emph{repair} faulty knowledge during this process.

\subsection{Augmentation Properties}

\subsubsection{Concurrency}
\add{We observed that certain tactics could achieve multiple strategies or objectives \textit{simultaneously}, as shown by tactics being shared between strategies in \autoref{fig:design-space}, though only for \xmybox[objective]{observation} and \xmybox[objective]{interpretation}.}
After completing the coding process, we went through each tactic, reasoning about cases in which a tactic could support multiple strategies in tandem.
We describe a few illustrative cases here.

\xmybox[tactic]{Animation} can be used as a moving time window, both \xmybox[strategy]{sequencing} and \xmybox[strategy]{spotlighting} content to more easily \xmybox[objective]{observe} errors~\cite{hill1991deixis}.
Dynamic widgets may be used as \xmybox[objective]{interpretation} aids and (self-) \xmybox[tactic]{prediction} mechanisms (\eg using toggle to display contextual annotations in a math formula to check understanding~\cite{hohman2020communicating}).
An \xmybox[tactic]{annotation} can both provide \xmybox[strategy]{context} (aiding \xmybox[objective]{interpretation}) and \xmybox[strategy]{spotlight} specific portions (\xmybox[objective]{observing} data characteristics)~\cite{rahman2024qualitative} (\eg a single text annotation next to a peak in a distribution can both ``point out'' the peak and provide external context to explain its occurrence). 
In the same scenario, an arrow \xmybox[tactic]{annotation} could both point out the peak and \xmybox[strategy]{relate} it to the text annotation.


\subsubsection{Hierarchy}
\add{
Our survey revealed an implicit hierarchical dependency among the objectives (though not for strategies or tactics within each objective), as illustrated by the gradated arrows spanning from \xmybox[objective]{observe} to \xmybox[objective]{introspect} in~\autoref{fig:design-space}.
The objective hierarchy represents the increasing cognitive complexity on the part of the viewer, with simpler objectives often invoked to satisfy more complex ones.
For example, \xmybox[objective]{observation} aids serve as ``first-order'' perceptual cues to highlight salient aspects of the data.
However, \xmybox[objective]{interpretation} aids may be needed to contextualize such insights (\eg~\autoref{fig:co2-emissions}(B,C)) or communicate the message more viscerally (\eg~\autoref{fig:co2-emissions}(F,G)).
When \xmybox[objective]{introspection} is desired, lower-level objectives can provide essential context and direct the reader's attention to their mistakes (\eg~\autoref{fig:selfpredict}(E,F)).
This dependency is reflected in our design space coding (see Table A1 in Appendix): individual objectives, notably \xmybox[objective]{observe} and \xmybox[objective]{introspect}, are less common, while \xmybox[objective]{observe} and \xmybox[objective]{interpret} co-occur together frequently and with \xmybox[objective]{introspect}.
}

\add{
One explanation for this hierarchy is concurrency: use of certain tactics, such as \xmybox[tactic]{encode} or \xmybox[tactic]{annotate}, may be used primarily as \xmybox[objective]{interpretation} aids, but serve as lower-level \xmybox[objective]{observation} aids by default when used appropriately (\eg using a magma color gradient in~\autoref{fig:co2-emissions}(D,E) to \xmybox[strategy]{associate} with temperature also \xmybox[strategy]{sequences} the data by default, though a different color gradient could have been used).
}


\subsubsection{Containment}
\add{
Increasing complexity within the objective hierarchy results in overlap between certain tactics and strategies from different tiers.
Specifically, augmentations classed within an individual objective (\eg \xmybox[tactic]{symbolize} for \xmybox[objective]{interpret} or \xmybox[strategy]{reflect} for \xmybox[objective]{introspect}) may \textit{contain} a tactic / strategy from a lower-level objective within its implementation.
For instance, a textual \xmybox[tactic]{guide} (\eg captions or titles) also \xmybox[tactic]{annotates} the chart; \xmybox[tactic]{symbols} may \xmybox[tactic]{encode} data (\eg~\autoref{fig:cholera}); and \xmybox[strategy]{sequencing} may be used as a sub-strategy to \xmybox[strategy]{relate} disparate views (\eg parallel \xmybox[tactic]{layout}~\cite{tufte1997visual}).  
}

\add{
Conceptually, containment functions similar to concurrency, such as when a textual description both \xmybox[tactic]{annotates} and \xmybox[tactic]{guides}, though at a level below tactics (\ie implementation).
However, we retain properties of containment in our design space to both ensure consistent levels of abstraction (\ie connections only exist between different tiers) and represent the hierarchical nature of certain tactics and strategies themselves.
For instance, \xmybox[tactic]{encode} and \xmybox[tactic]{symbolize} both connect to \xmybox[objective]{interpret} via \xmybox[strategy]{relate} and \xmybox[strategy]{associate} strategies, though do not share any connections through \xmybox[objective]{observe}.
This relationship accurately reflects the fact that, when containment occurs, any \xmybox[strategy]{sequencing} or \xmybox[strategy]{spotlighting} of data is an artifact of the \xmybox[tactic]{encoding} (\ie \textit{how} data values are mapped to \xmybox[tactic]{symbols}).
}

\section{\texorpdfstring{Crossing the Chasm:\\~Generating \& Analyzing Communicative Graphics}{Crossing the Chasm: Generating \& Analyzing Communicative Graphics}}
\label{sec:evaluation}

\figurecarbon

Our design space 
\add{helps express and ground communicative} intent as a series of design goals, which can then be mapped via strategies to specific, actionable tactics (\ie ``crossing'' the chasm from intent to implementation).
This benefit not only supports formal specification of an augmented chart, but also contributes a principled framework which 
\add{can assist with} generating and searching over a space of different augmentations \add{(\eg by varying tactics / strategies for a chosen goal)}.
Our design space also aids analysis and reasoning about affordances different charts may provide.
For instance, differences in reader takeaways can ostensibly be mapped to structural variations in augmentation graphs (\autoref{fig:co2-emissions}(C,E,G)), surfacing communicative tradeoffs to designers (from the perspective of readers) that can help align design intent and reader learning.
We demonstrate both generative and analytic properties in the following examples, along with other insights and observations that result from using different tactics.

\subsection{Example: CO$_\text{2}$ Concentration}
\label{sec:co2-example}
To illustrate the value of our design space, we selected an initial visualization of atmospheric CO$_2$ concentration from 1958-2020 (\autoref{fig:co2-emissions}(A)) absent augmentations (\eg the trendine and axes are required to decode denotative chart semantics).
We then explored several augmentations to generate the following design variations and reason about their differences: (\autoref{fig:co2-emissions}(B)) adding context to more clearly visualize and call attention to the exponential rate of change; (\autoref{fig:co2-emissions}(D)) cleverly cycling the trend to emphasize finer detail (\ie zooming in), with sequential color hue to indicate warmer temperatures (originally authored by Gregor Aisch~\cite{aisch}); and (\autoref{fig:co2-emissions}(F)) using animation with ramping speed to convey CO$_2$ increase more viscerally (originally authored by The New York Times (NYT)~\cite{nytco2}). 

Although a non-linear increase is apparent in \autoref{fig:co2-emissions}(A), suppose that we want to help readers better \textit{understand} how quickly CO$_2$ concentration is increasing (\autoref{fig:co2-emissions}(B,C)).
We can describe this intent as \textit{showing} the reader what a slower rate (\eg linear) would look like (\ie provide \xmybox[strategy]{context} for \xmybox[objective]{interpretation}).
Of the different tactics for \xmybox[strategy]{context}, \xmybox[tactic]{annotating} with external data (\ie linear regression) best enables visual comparison between the linear and exponential trends to facilitate comprehension.
If we also wanted readers to \xmybox[objective]{observe} the increasing gap, we could \xmybox[tactic]{encode} (\eg shade) the region between the two trend lines as a \xmybox[strategy]{spotlighting} technique.
Given that the linear regression is only fit to an initial subset of data (1956-1970), we add a text \xmybox[tactic]{annotation} with this information as \xmybox[strategy]{context}, though \xmybox[tactic]{guides} (\eg footnote, chart title, or caption) could achieve the same effect.
We use an arrow \xmybox[tactic]{annotation} to \xmybox[strategy]{relate} this \xmybox[strategy]{contextual} description with the regression, as well as \xmybox[strategy]{spotlight} when the fitted data ends.
To further help readers \xmybox[objective]{interpret} the regression's fit, we dot the remaining line as ``imaginary'' to \xmybox[strategy]{associate} it as predicted linear data.
Finally, to direct the reader's attention (\ie \xmybox[objective]{observe}) to the linear trend -- the primary focus of this chart -- we \xmybox[strategy]{spotlight} it with a red hue (\xmybox[tactic]{encoding}). 
It is worth noting that \xmybox[objective]{introspection} tactics, such as (self-) \xmybox[tactic]{prediction} or \xmybox[tactic]{simulating} different regression curves, could be used to encourage the reader to \xmybox[strategy]{reflect} on their prior expectations about CO$_2$ levels, given the importance and global urgency of the climate crisis (\autoref{fig:intervene-tactics}(C)).

Expanding the trend in \autoref{fig:co2-emissions}(A) adds more detail, helping reveal both the exponential rate of change and the seasonal variation in level (\autoref{fig:co2-emissions}(D,E)).
Given the inherent ordering of the temporal data, we can make clever use of \xmybox[tactic]{transformation} to cycle or ``wrap'' the data around the y-axis, helping the reader \xmybox[objective]{observe} the over-decades increase.
Although y-axis labels provide the time periods necessary for context, we can use \xmybox[strategy]{sequential} \xmybox[tactic]{encoding} to make the order of data more transparent.
Many sequential color schemes might be appropriate, but the magma color \xmybox[tactic]{encoding} simultaneously exploits color-concept \xmybox[strategy]{assocations} (\ie bluish hues for colder temperatures and reddish ones for warmer) to convey the relationship (\xmybox[objective]{interpret}) between increasing CO$_2$ levels and global temperature.

The variation in \autoref{fig:co2-emissions}(D) could be augmented further to help readers \xmybox[objective]{interpret} this rate more viscerally as demonstrated in \autoref{fig:co2-emissions}(F,G).
This visualization retains the cyclical \xmybox[tactic]{transform} from \autoref{fig:co2-emissions}(D) to help readers \xmybox[objective]{observe} annually increasing temperatures, while using \xmybox[tactic]{animation} to draw attention to this change over time.
Given that periodicity may be less obvious compared with \autoref{fig:co2-emissions}(B,D), we can add a footnote (\xmybox[tactic]{guide}) to \xmybox[strategy]{contextualize} seasonal differences between May and August.
We can also exploit the \textit{speed} of the \xmybox[tactic]{animation} to both convey the rate of increase more directly and connote urgency (\xmybox[strategy]{association}).
In particular, just as the actual CO$_2$ trend increases exponentially, so does the rate of animated motion that introduces each annual time slice in \xmybox[strategy]{sequence} (\ie the animation starts out slowly, ramps up exponentially, then slows down towards the end to avoid an abrupt halt).
As a result, the animation offers the reader a ``roller coaster'' experience to better \xmybox[objective]{interpret} the alarming rate at which CO$_2$ levels are rising.

\figureincome

While sequential color \xmybox[tactic]{encoding} is also retained to help readers' \xmybox[objective]{interpret} \xmybox[strategy]{sequence}, it is not required given the use of animation and labeled year at the center of the visualization.
Further, the color hue in \autoref{fig:co2-emissions}(F) may negatively affect \xmybox[objective]{interpretation} given the mismatch between color-concept \xmybox[strategy]{associations}.
Specifically, increasing levels of CO$_2$ are encoded with darkening blue hues, despite this scheme's typical \xmybox[strategy]{association} with cooler temperatures (as in \autoref{fig:co2-emissions}(D)).
While this augmentation is in line with standard graphical perception principles, our design space helpfully reveals how this encoding might conflict with higher-level \xmybox[objective]{interpretation} objectives.
One possible explanation for this discrepancy is that a red hue \xmybox[tactic]{encoding} is used at the end of the animation to instead \xmybox[strategy]{spotlight} CO$_2$ concentration surpassing 420 parts per million (enabling \xmybox[strategy]{association} with dangerous levels), which is referenced later in the article to compare with levels before the Industrial Revolution~\cite{nytco2}.
However, other \xmybox[tactic]{encodings} (\eg bolded font) or \xmybox[tactic]{annotations} (\eg arrow call-out) might be used instead.


\subsection{Example: Income Inequality}
\label{sec:income-example}

To demonstrate the analytic affordances of our design space, we consider a visualization authored by Mike Bostock that snapshots the change in income inequality across major United States metros in 1980 versus 2015~\cite{d3incomeinequality} (\autoref{fig:income-inequality}(A)).
This visualization was adapted from a NYT publication~\cite{nytincomeinequality}, which uses animation instead of single static frame (\autoref{fig:income-inequality}(D)).

The static visualization (\autoref{fig:income-inequality}(A, B)) uses arrow and sequential color hue \xmybox[tactic]{encoding} as \xmybox[strategy]{sequencing} strategies for readers to more readily \xmybox[objective]{observe} differences between 1980 and 2015 income inequality.
Selective \xmybox[tactic]{annotations} for certain metro areas serve as both \xmybox[strategy]{spotlighting} emphasis to \xmybox[objective]{observe} the significant rise in inequality among major cities and \xmybox[strategy]{contextual} aids to help readers \xmybox[objective]{interpret} the data.
Finally, the visualization uses red hue (\xmybox[tactic]{encoding}) as an \xmybox[strategy]{association} for more severe inequality, given that the overwhelming majority of metro areas worsened in this respect.
This visualization also makes subtle use of augmentation \textit{concurrency} with both tactics by using color \xmybox[tactic]{encoding} to achieve multiple strategies simultaneously (\xmybox[strategy]{associate} and \xmybox[strategy]{sequence}) across different objectives (\xmybox[objective]{interpret} and \xmybox[objective]{observe}), as well as \xmybox[tactic]{annotations} (to both \xmybox[strategy]{spotlight} and \xmybox[strategy]{contextualize}).

The NYT visualization (\autoref{fig:income-inequality}(D, E)) uses \xmybox[tactic]{animated} transitions to \xmybox[strategy]{sequence} data points between 1980 and 2015 frames and \xmybox[strategy]{spotlights} (bold font) Binghamton and New York to draw readers' attention (\xmybox[objective]{observe}) to the fact that New York's inequality worsened significantly more than Binghamton's, which remained relatively stagnant.
\xmybox[strategy]{Contextual} \xmybox[tactic]{annotations} for other major metro areas are also provided to aid \xmybox[objective]{interpretation}.
\xmybox[tactic]{Guides} (titles, captions, links, footnotes with data sources) also provide \xmybox[strategy]{context}.

When comparing potential differences between reader takeaways from these two visualizations (\autoref{fig:income-inequality}(C, F)), we can reason from structural divergences in the augmentation graphs (\autoref{fig:income-inequality}(B, E)) to expose such differences more concretely and transparently.
The graphical relationships between objectives and strategies are structurally the same, aside from using an \xmybox[strategy]{association} strategy to aid communication in static form (\autoref{fig:income-inequality}(B)).
As a result, we can expect that differences in takeaways likely result from which tactics are employed.
For instance, readers of both visualizations might notice that the majority of US metro areas had a significant increase in inequality due to shared use of \xmybox[strategy]{sequencing}.
The NYT's \xmybox[tactic]{animated} representation reveals that Binghamton's inequality not only increased substantially less than New York's, but also substantially less relative to most metro areas.
However, Bostock's visualization may bias readers to focus only on the significant increase in inequality among \textit{major} metro areas (the \xmybox[tactic]{annotated} points).
Finally, the NYT's \xmybox[strategy]{spotlighting} of New York and Binghamton (bold font \xmybox[tactic]{encoding}) may bias readers to primarily focus on the wide variance between these two areas, while Bostock's global \xmybox[tactic]{encoding} representation emphasizes the wide variance across all metro areas as a result of significant mark overlap.

\figureselfpredict

\subsection{Example: Self-prediction}
\label{sec:example-prediction}
As demonstrated with prior examples, a central benefit to our design space is the ability to reason more precisely and descriptively about differences between augmented visualizations.
On the other hand, we can also identify similarities between charts that differ substantially in either form or presentation.
For example, the following three visualizations encourage active learning with interactive charts spanning different contexts:
(\autoref{fig:selfpredict}(A)) self-directed tool for learning the telltale signs in machine-generated images~\cite{hohman2020communicating}; (\autoref{fig:selfpredict}(C)) a visualization game for guessing binge drinking percentage rates across different US counties~\cite{diakopoulos2011playable}; and (\autoref{fig:selfpredict}(E)) an interactive visualization for predicting rates of Hispanic voters by income bracket, with other voting statistics provided as context~\cite{kim2017data}.
In \autoref{fig:selfpredict}(A), readers may hover over \xmybox[tactic]{annotated} regions to view descriptions of machine-generated artifacts; multiple examples are provided in a faceted \xmybox[tactic]{layout} at the bottom right corner for the reader to explore.
To guess health statistics in \autoref{fig:selfpredict}(C), readers can change the slider threshold to (re-) \xmybox[tactic]{encode} county data above~/~below the new slider value.
Readers (self-) \xmybox[tactic]{predict} data in \autoref{fig:selfpredict}(E) by drawing a line with their mouse.

Despite dissimilar presentation (\eg interactive features), data domains, and viewing context (\eg interactive article that includes additional text and material), our design space (\autoref{fig:selfpredict}(B,D,F)) reveals both important similarities and differences.
In particular, each of the visualizations use the same tactics and strategies to \xmybox[objective]{introspect} on the data.
In each case, the reader is encouraged to \xmybox[strategy]{reflect} by \xmybox[tactic]{predicting} something about the data that is not displayed in the default view: machine-generated artifact (\autoref{fig:selfpredict}(A)); health statistic (\autoref{fig:selfpredict}(C)); and voting trend (\autoref{fig:selfpredict}(E)).
To \xmybox[strategy]{repair} the reader's prior knowledge, including assumptions or (mis)understanding of the data, reader \xmybox[tactic]{predictions} are \xmybox[tactic]{evaluated} via feedback.
Such feedback includes \xmybox[tactic]{annotated} explanations (\autoref{fig:selfpredict}(B)), a \xmybox[tactic]{guide} that displays the running total of the reader's score (\autoref{fig:selfpredict}(D), and \xmybox[tactic]{annotated} text and arrows with red color \xmybox[tactic]{encoding} to \xmybox[strategy]{spotlight} (\autoref{fig:selfpredict}(F)) any \xmybox[tactic]{prediction} errors.

Despite sharing \xmybox[objective]{introspection} aids, the ``lower-level'' tactics and strategies which facilitate \xmybox[objective]{observation} and \xmybox[objective]{interpretation} diverge in certain respects (\eg only \autoref{fig:selfpredict}(A) \xmybox[strategy]{sequences} deliberately).
As a result, these ``lower-level'' cues may affect self-learning directly, ``guiding'' the reader through the process of understanding their mistakes.
For example, \autoref{fig:selfpredict}(A,B) utilizes \xmybox[tactic]{layout} techniques (\ie details-on-demand, which hide content behind interaction, and faceting) to provide additional \xmybox[objective]{observations} and help \xmybox[strategy]{relate} the types of machine-generated errors to facilitate learning.
However, by \xmybox[strategy]{spotlighting} these mistakes in the default view (via circle \xmybox[tactic]{annotations}), is the reader able to recognize machine-generated images in the future when augmentations are absent?
By comparison, the other two examples (\autoref{fig:selfpredict}(C,D)) and \autoref{fig:selfpredict}(E,F)) also augment with \xmybox[strategy]{context} and \xmybox[strategy]{relation} strategies, via \xmybox[tactic]{encoding} channels.
\autoref{fig:selfpredict}(C) also provides readers with a dynamic threshold (\xmybox[tactic]{transform}) that \mbox{(re-)} \xmybox[tactic]{encodes} data to \xmybox[objective]{observe} geographical correlations more readily, which could benefit learning.

Using the descriptive and systematic reasoning affordances of the design space, this analysis suggests an interesting possibility.
Do learning benefits generalize to similar \xmybox[objective]{introspection} aids used in different contexts?
For instance, self-prediction improves recall and data comprehension with graphical elicitation techniques~\cite{kim2017explaining}; does this hold for other \xmybox[tactic]{prediction} techniques (\eg details-on-demand across multiple instances in \autoref{fig:selfpredict}(A) and guessing games in \autoref{fig:selfpredict}(C))?
If not, can learning differences be attributed to divergent \xmybox[objective]{observation} and \xmybox[objective]{interpret} aids?
Future work should explore this relationship more deeply to identify the exact process of self-learning under these conditions, which we discuss in \autoref{sec:discussion}.

\section{Beyond the Chasm: Discussion \& Future Work}
\label{sec:discussion}

\subsection{Contribution \& Limitations}
Our primary contribution is a design space for augmentations in communicative visualization, which consists of \emph{tactics} (the implemented design features), \emph{strategies} (the approaches the designer selects to achieve an objective), and \emph{objectives} (how the reader is affected perceptually or cognitively).
This design space formally bridges the gap between intent and design, enabling designers to map their goals via strategies to a concrete set of actionable tactics.

\add{
We frame communicative / designer intent as explicit cognitive behaviors on behalf of the viewer.
This framing synthesizes perceptual and cognitive visualization goals from prior work to provide a useful level of abstraction for deconstructing the different roles and uses of varied augmentations.
Unlike prior objectives, our viewer-oriented goals helpfully relate an expressive range of augmentation tactics (\eg animation, annotation, self-prediction, \etc) across different contexts (\eg narrative visualization, data comics, dashboards).
As a result, our design space is able to capture more nuanced relations and dependencies (\eg concurrency and hierarchy) at multiple levels (tactics, strategies, objectives) that help analyze and evaluate communicative aims more precisely.}
Designers can reflect on the way they are using augmentations, how these decisions reinforce their high-level goals, and communicative tradeoffs between different designs.
Our design space may offer support for exploring alternatives (\autoref{fig:co2-emissions}), either with respect to different approaches to achieve the same goals, or modifying the augmentation to support alternative goals, which we aim to validate in future work.


We present the hierarchical design space as a color-coded graph and apply it to a rich set of examples to illustrate its descriptive and analytic power (\autoref{fig:co2-emissions}, \autoref{fig:income-inequality}, \autoref{fig:selfpredict}).
\add{By grounding objectives in tactics, we provide a principled framework to help generate and search over a space of augmentations (\autoref{fig:co2-emissions}). 
Designers can formulate their high-level intents as a series of viewer-oriented objectives: what do they want they viewer to pay attention to, interpret, or reflect on?
Each objective can then be mapped to a set of strategies and tactics, varied based on the data and context of use.
For example, \xmybox[tactic]{annotate} and \xmybox[tactic]{transform} tactics can both \xmybox[strategy]{spotlight} data, though the designer's choice depends on \textit{what} they wish for the viewer to \xmybox[objective]{observe} (\eg a specific point along a trendline \autoref{fig:co2-emissions}(B) vs. trend increase over time \autoref{fig:co2-emissions}(D)).
This process is further aided by analyzing structural divergences between augmentation graphs (\autoref{fig:income-inequality}), encouraging designers to reflect on potential shifts in viewer behavior and reason about communication tradeoffs.
Specifically, by adding, removing, or modifying an augmentation, what information and takeaways will the viewer likely prioritize over others?
}

\add{
Appropriate use of augmentations can also be navigated by first selecting individual tactics, then noting potential concurrent and hierarchical dependencies.
For instance, our critique of different color \xmybox[tactic]{encodings} in \autoref{fig:co2-emissions}(D,F) stems from the concurrent relationship between \xmybox[strategy]{sequence} and \xmybox[strategy]{associate}.
\autoref{fig:co2-emissions}(D) exploits the magma color \xmybox[strategy]{association} with rising temperatures, while \autoref{fig:co2-emissions}(F)'s blue gradient conflicts with these semantics.
Our design space hierarchy facilitates this line of reasoning: tactics used to first \xmybox[objective]{observe}, such as \xmybox[tactic]{encode} or \xmybox[tactic]{annotate} (\autoref{fig:income-inequality}(A)), can be matched next against concurrent strategies for \xmybox[objective]{interpret} to identify unintended effects.
In such cases, the designer can select a different implementation for the same tactic (\eg color scheme), an alternative tactic that supports the same strategy or objective, or use the proper implementation based on concurrent strategies and objectives.
}




A major advantage of our design space is that it provides a consistent lexicon for describing complex communicative visualizations that may appear very different (\eg different tactics) but have similar goals (\autoref{fig:selfpredict}).
For instance, our design space helped relate explorable explanations~\cite{victor2011explorable} with other uses of similar augmentations from visualization games~\cite{diakopoulos2011playable} or graphical self-prediction~\cite{kim2017explaining}, \add{despite significant differences in form and presentation}.
These connections could suggest that similar methodologies and assessments might be used to investigate actual learning benefits (which, for example, remains an open question for interactive articles~\cite{conlen2018idyll}).

Despite promise, some augmentations, such as self-reflection and repair (\autoref{fig:selfpredict}), are rarely used in general practice, likely due to the design and/or engineering knowledge necessary to implement them.
Future (semi-)~automated authoring systems may provide some relief on this front, leveraging our design space to evaluate or generate designs.
Going forward, we plan to also continue evaluating our design space through discussions with expert designers themselves and more tightly integrating with learning objectives.

\subsection{Scope of Augmentations}

We expect our design goals to be sufficiently high-level to capture more abstract intents, while also providing enough grounding to help designers identify actionable strategies and tactics that can facilitate design iteration and reason about readers' takeaways.
However, future work is needed to test our design space with various designers across different domains, which may reveal gaps between how they express and organize their intents and how the literature captures communicative intent.
Naturally, the intent may shift with varying communication contexts (\eg didactic, rhetorical / persuasive, affective, \etc).
In these situations, we expect that our design space may function as ``middle-ware'' in which the augmentation objectives uniquely map to the higher-level ``context'' objectives.
This connective tissue might even serve as an additional resource for mapping between ``context'' objectives (\eg how do learning objectives map to rhetorical aims via visual augmentations), or even connecting with lower-level tasks throughout perceptual~\cite{quadri2021survey} and analysis~\cite{brehmer2013multi} taxonomies.
As a result, designers may have more flexibility in generating designs that can be adapted across multiple contexts according to their intent.

Given the existence of augmentations, we expect there to be anti-augmentations (akin to anti-learning objectives as noted by Adar and Lee-Robbins~\cite{adar2020communicative}), which serve to distract or bias readers or impede learning.
For example, removing clutter (\eg gridlines, unhelpful annotations, \etc)~\cite{ajani2021declutter}.
Anti-augmentations may also entail removing particular types of content to aid observation (\eg filtering) or interpretation, which has been observed with decluttering visualizations~\cite{ajani2021declutter} and, in certain cases, animation~\cite{hullman2011benefitting}.

\subsection{Relation to Learning Objectives}
We began this work particularly inspired by Adar and Lee-Robbins' extended learning objectives taxonomy for visualization~\cite{adar2020communicative}.
We view our characterization of augmentations and the resulting design space as a step towards evaluating and reasoning about learning objectives, allowing designers to use the strategies and tactics we have surveyed to achieve different learning outcomes.
For instance, Adar \etal~\cite{adar2020communicative} operationalize learning objectives as statements of the form: \textbf{The viewer will \emph{[verb]} \emph{[noun]}}.
This construction delineates cognitive constructs (verbs) (\eg \textsc{recall}, \textsc{classify}, \textsc{implement}, \etc) and knowledge dimensions (nouns) (\eg factual knowledge like terminology or conceptual knowledge like categories).
We might leverage \xmybox[objective]{observation} mechanisms (\ie \xmybox[strategy]{spotlighting} or \xmybox[strategy]{sequencing}) to convey or focus alternative knowledge representations (nouns); \xmybox[objective]{interpretation} and \xmybox[objective]{introspection} aids can then instrument the desired learning processes (verbs).
However, the exact connection between our design space and learning objectives, as supported by empirical evaluation, remains as future work, in part due to the difficulties and training that can be required for designers to formulate their intents as learning objectives~\cite{adar2020communicative}.

\subsection{Extensions to Augmentations \& Authoring Tools}

We contribute a systematic outline of the visual augmentation space that can serve as a springboard for future empirical work and communication tools.
We hope to use our design space to ground empirical assessment and treatments of various augmentations, including their relationships with domain knowledge, other learning objectives (\eg rhetorical or affective), and augmentations in different contexts (\eg math formulas~\cite{head2022math}), to assess ecological validity.
Although trade-offs and opportunities for learning outcomes must be evaluated, visual augmentations may provide other benefits such as engagement or joy~\cite{brehmer2013multi}.

Several exciting directions for future work revolve around using our design space as the basis for a design language, potentially even a computational domain-specific language (DSL).
For instance, do the tiers of tactics, strategies, and objectives serve as useful ``building blocks'' or composable primitives for a grammar of communicative graphics?
The generative and analytic capabilities of our design space might help scaffold future authoring systems, enabling designers to create communicative visualizations more rapidly.
Such capabilities may also be used to encode knowledge in DSLs for different domains, or more generally leverage this knowledge as input to AI systems (\eg using language models to author data stories~\cite{wang2025jupybara}) to identify possible augmentations during the design process.

The space of possible augmentations is expansive, encompassing non-visual channels (\eg narration or text~\cite{mayer2005cognitive, stokes2022striking, pluto}), assumptions about readers (\eg visual literacy and background knowledge), different device types (\eg mobile vs. desktop~\cite{hoffswell2020techniques}), and context of use (collaboration, dashboards, teaching, business presentations, \etc).
Our design space examines a particular subset of augmentations (visual) within a specific context (data communication \& learning), which we hope to help scaffold future work across other modalities (\eg multimedia voice narration~\cite{mayer2005cognitive}), types of learning (\eg affective~\cite{lee2022affective} or social~\cite{kim2017data}), and contexts.


\bibliographystyle{IEEEtran}
\bibliography{bibliography}

@inproceedings{hill1991deixis,
  title={Deixis and the future of visualization excellence},
  author={Hill, William C and Hollan, James D},
  booktitle={IEEE Visualization},
  volume={91},
  pages={314--320},
  year={1991}
}

@article{tversky2002animation,
  title={Animation: can it facilitate?},
  author={Tversky, Barbara and Morrison, Julie Bauer and Betrancourt, Mireille},
  journal={International Journal of Human-Computer Studies},
  volume={57},
  number={4},
  pages={247--262},
  year={2002},
  publisher={Elsevier}
}

@inproceedings{guo2024we,
  title={What we augment when we augment visualizations: A design elicitation study of how we visually express data relationships},
  author={Guo, Grace and Stasko, John and Endert, Alex},
  booktitle={Proceedings of the 2024 International Conference on Advanced Visual Interfaces},
  pages={1--5},
  year={2024}
}

@inproceedings{quadri2024you,
  title={Do you see what i see? A qualitative study eliciting high-level visualization comprehension},
  author={Quadri, Ghulam Jilani and Wang, Arran Zeyu and Wang, Zhehao and Adorno, Jennifer and Rosen, Paul and Szafir, Danielle Albers},
  booktitle={Proceedings of the 2024 CHI Conference on Human Factors in Computing Systems},
  pages={1--26},
  year={2024}
}

@inproceedings{kim2017explaining,
  title={Explaining the gap: Visualizing one's predictions improves recall and comprehension of data},
  author={Kim, Yea-Seul and Reinecke, Katharina and Hullman, Jessica},
  booktitle={Proceedings of the 2017 CHI Conference on Human Factors in Computing Systems},
  pages={1375--1386},
  year={2017}
}

@inproceedings{conlen2019capture,
  title={Capture \& analysis of active reading behaviors for interactive articles on the web},
  author={Conlen, Matthew and Kale, Alex and Heer, Jeffrey},
  booktitle={Computer Graphics Forum},
  volume={38},
  number={3},
  pages={687--698},
  year={2019},
  organization={Wiley Online Library}
}

@article{ajani2021declutter,
  title={Declutter and focus: Empirically evaluating design guidelines for effective data communication},
  author={Ajani, Kiran and Lee, Elsie and Xiong, Cindy and Knaflic, Cole Nussbaumer and Kemper, William and Franconeri, Steven},
  journal={IEEE Transactions on Visualization and Computer Graphics},
  volume={28},
  number={10},
  pages={3351--3364},
  year={2021},
  publisher={IEEE}
}

@inproceedings{chevalier2016animations,
  title={Animations 25 years later: New roles and opportunities},
  author={Chevalier, Fanny and Riche, Nathalie Henry and Plaisant, Catherine and Chalbi, Amira and Hurter, Christophe},
  booktitle={Proceedings of the 2016 International Working Conference on Advanced Visual Interfaces},
  pages={280--287},
  year={2016}
}

@inproceedings{ge2021cast,
  title={Cast: Authoring data-driven chart animations},
  author={Ge, Tong and Lee, Bongshin and Wang, Yunhai},
  booktitle={Proceedings of the 2021 CHI Conference on Human Factors in Computing Systems},
  pages={1--15},
  year={2021}
}

@article{heer2007animated,
  title={Animated transitions in statistical data graphics},
  author={Heer, Jeffrey and Robertson, George},
  journal={IEEE Transactions on Visualization and Computer Graphics},
  volume={13},
  number={6},
  pages={1240--1247},
  year={2007},
  publisher={IEEE}
}

@article{hullman2013deeper,
  title={A deeper understanding of sequence in narrative visualization},
  author={Hullman, Jessica and Drucker, Steven and Riche, Nathalie Henry and Lee, Bongshin and Fisher, Danyel and Adar, Eytan},
  journal={IEEE Transactions on Visualization and Computer Graphics},
  volume={19},
  number={12},
  pages={2406--2415},
  year={2013},
  publisher={IEEE}
}

@inproceedings{kim2017graphscape,
  title={Graphscape: A model for automated reasoning about visualization similarity and sequencing},
  author={Kim, Younghoon and Wongsuphasawat, Kanit and Hullman, Jessica and Heer, Jeffrey},
  booktitle={Proceedings of the 2017 CHI Conference on Human Factors in Computing Systems},
  pages={2628--2638},
  year={2017}
}

@article{kim2020gemini,
  title={Gemini: A grammar and recommender system for animated transitions in statistical graphics},
  author={Kim, Younghoon and Heer, Jeffrey},
  journal={IEEE Transactions on Visualization and Computer Graphics},
  volume={27},
  number={2},
  pages={485--494},
  year={2020},
  publisher={IEEE}
}

@inproceedings{mckenna2017visual,
  title={Visual narrative flow: Exploring factors shaping data visualization story reading experiences},
  author={McKenna, Sean and Henry Riche, Nathalie and Lee, Bongshin and Boy, Jeremy and Meyer, Miriah},
  booktitle={Computer Graphics Forum},
  volume={36},
  number={3},
  pages={377--387},
  year={2017},
  organization={Wiley Online Library}
}

@book{card1999readings,
  title={Readings in information visualization: using vision to think},
  author={Card, Stuart K and Mackinlay, Jock and Shneiderman, Ben},
  year={1999},
  publisher={Morgan Kaufmann}
}

@article{ploetzner2020review,
  title={A review of learning demands in instructional animations: The educational effectiveness of animations unfolds if the features of change need to be learned},
  author={Ploetzner, Rolf and Berney, Sandra and B{\'e}trancourt, Mireille},
  journal={Journal of Computer Assisted Learning},
  volume={36},
  number={6},
  pages={838--860},
  year={2020},
  publisher={Wiley Online Library}
}

@article{plaisant2007promoting,
  title={Promoting insight-based evaluation of visualizations: From contest to benchmark repository},
  author={Plaisant, Catherine and Fekete, Jean-Daniel and Grinstein, Georges},
  journal={IEEE Transactions on Visualization and Computer Graphics},
  volume={14},
  number={1},
  pages={120--134},
  year={2007},
  publisher={IEEE}
}

@article{saraiya2005insight,
  title={An insight-based methodology for evaluating bioinformatics visualizations},
  author={Saraiya, Purvi and North, Chris and Duca, Karen},
  journal={IEEE Transactions on Visualization and Computer Graphics},
  volume={11},
  number={4},
  pages={443--456},
  year={2005},
  publisher={IEEE}
}

@article{saraiya2006insight,
  title={An insight-based longitudinal study of visual analytics},
  author={Saraiya, Purvi and North, Chris and Lam, Vy and Duca, Karen A},
  journal={IEEE Transactions on Visualization and Computer Graphics},
  volume={12},
  number={6},
  pages={1511--1522},
  year={2006},
  publisher={IEEE}
}

@inproceedings{satyanarayan2014authoring,
  title={Authoring narrative visualizations with ellipsis},
  author={Satyanarayan, Arvind and Heer, Jeffrey},
  booktitle={Computer Graphics Forum},
  volume={33},
  number={3},
  pages={361--370},
  year={2014},
  organization={Wiley Online Library}
}

@article{segel2010narrative,
  title={Narrative visualization: Telling stories with data},
  author={Segel, Edward and Heer, Jeffrey},
  journal={IEEE transactions on Visualization and Computer Graphics},
  volume={16},
  number={6},
  pages={1139--1148},
  year={2010},
  publisher={IEEE}
}

@article{victor2006magic,
  title={Magic ink: Information software and the graphical interface},
  author={Victor, Bret},
  journal={},
  volume={},
  year={2006}
}

@article{srinivasan2018augmenting,
  title={Augmenting visualizations with interactive data facts to facilitate interpretation and communication},
  author={Srinivasan, Arjun and Drucker, Steven M and Endert, Alex and Stasko, John},
  journal={IEEE Transactions on Visualization and Computer Graphics},
  volume={25},
  number={1},
  pages={672--681},
  year={2018},
  publisher={IEEE}
}

@article{heer2012interactive,
  title={Interactive dynamics for visual analysis: A taxonomy of tools that support the fluent and flexible use of visualizations},
  author={Heer, Jeffrey and Shneiderman, Ben},
  journal={Queue},
  volume={10},
  number={2},
  pages={30--55},
  year={2012},
  publisher={ACM New York, NY, USA}
}

@article{brehmer2013multi,
  title={A multi-level typology of abstract visualization tasks},
  author={Brehmer, Matthew and Munzner, Tamara},
  journal={IEEE Transactions on Visualization and Computer Graphics},
  volume={19},
  number={12},
  pages={2376--2385},
  year={2013},
  publisher={IEEE}
}

@article{ancker2024you,
  title={Do you want to promote recall, perceptions, or behavior? The best data visualization depends on the communication goal},
  author={Ancker, Jessica S and Benda, Natalie C and Zikmund-Fisher, Brian J},
  journal={Journal of the American Medical Informatics Association},
  volume={31},
  number={2},
  pages={525--530},
  year={2024},
  publisher={Oxford University Press}
}

@article{quadri2021survey,
  title={A survey of perception-based visualization studies by task},
  author={Quadri, Ghulam Jilani and Rosen, Paul},
  journal={IEEE Transactions on Visualization and Computer Graphics},
  volume={28},
  number={12},
  pages={5026--5048},
  year={2021},
  publisher={IEEE}
}

@article{hohman2020communicating,
  title={Communicating with interactive articles},
  author={Hohman, Fred and Conlen, Matthew and Heer, Jeffrey and Chau, Duen Horng Polo},
  journal={Distill},
  volume={5},
  number={9},
  pages={e28},
  year={2020}
}

@article{matuschak2019can,
  title={How can we develop transformative tools for thought?},
  author={Matuschak, Andy and Nielsen, Michael},
  journal={Numinous Productions (blog)},
  year={2019}
}

@inproceedings{crichton2021new,
  title={A new medium for communicating research on programming languages},
  author={Crichton, Will},
  booktitle={Workshop on Human Aspects of Types and Reasoning Assistants},
  year={2021}
}

@article{mackinlay1986automating,
  title={Automating the design of graphical presentations of relational information},
  author={Mackinlay, Jock},
  journal={ACM Transactions on Graphics},
  volume={5},
  number={2},
  pages={110--141},
  year={1986},
  publisher={Acm New York, NY, USA}
}

@inproceedings{bateman2010useful,
  title={Useful junk? The effects of visual embellishment on comprehension and memorability of charts},
  author={Bateman, Scott and Mandryk, Regan L and Gutwin, Carl and Genest, Aaron and McDine, David and Brooks, Christopher},
  booktitle={Proceedings of the SIGCHI Conference on Human Factors in Computing Systems},
  pages={2573--2582},
  year={2010}
}

@article{fygenson2023arrangement,
  title={The arrangement of marks impacts afforded messages: Ordering, partitioning, spacing, and coloring in bar charts},
  author={Fygenson, Racquel and Franconeri, Steven and Bertini, Enrico},
  journal={IEEE Transactions on Visualization and Computer Graphics},
  volume={30},
  number={1},
  pages={1008--1018},
  year={2023},
  publisher={IEEE}
}

@article{shah1999graphs,
  title={Graphs as aids to knowledge construction: Signaling techniques for guiding the process of graph comprehension.},
  author={Shah, Priti and Mayer, Richard E and Hegarty, Mary},
  journal={Journal of Educational Psychology},
  volume={91},
  number={4},
  pages={690},
  year={1999},
  publisher={American Psychological Association}
}

@article{YangLeni2023EWEL,
author = {Yang, Leni and Xiong, Cindy and Wong, Jason K. and Wu, Aoyu and Qu, Huamin},
address = {United States},
copyright = {Copyright The Institute of Electrical and Electronics Engineers, Inc. (IEEE) 2023},
issn = {1077-2626},
journal = {IEEE Transactions on Visualization and Computer Graphics},
language = {eng},
number = {3},
pages = {1638-1650},
publisher = {IEEE},
title = {Explaining with examples: Lessons learned from crowdsourced introductory description of information visualizations},
volume = {29},
year = {2023},
}

@article{xiong2021visual,
  title={Visual arrangements of bar charts influence comparisons in viewer takeaways},
  author={Xiong, Cindy and Setlur, Vidya and Bach, Benjamin and Koh, Eunyee and Lin, Kylie and Franconeri, Steven},
  journal={IEEE Transactions on Visualization and Computer Graphics},
  volume={28},
  number={1},
  pages={955--965},
  year={2021},
  publisher={IEEE}
}

@article{levie1982effects,
  title={Effects of text illustrations: A review of research},
  author={Levie, W Howard and Lentz, Richard},
  journal={Educational Technology Research and Development},
  volume={30},
  number={4},
  pages={195--232},
  year={1982},
  publisher={Springer}
}

@article{rahman2024qualitative,
  title={A qualitative analysis of common practices in annotations: A taxonomy and design space},
  author={Rahman, Md Dilshadur and Quadri, Ghulam Jilani and Doppalapudi, Bhavana and Szafir, Danielle Albers and Rosen, Paul},
  journal={IEEE Transactions on Visualization and Computer Graphics},
  year={2024},
  publisher={IEEE}
}

@inproceedings{bach2018design,
  title={Design patterns for data comics},
  author={Bach, Benjamin and Wang, Zezhong and Farinella, Matteo and Murray-Rust, Dave and Henry Riche, Nathalie},
  booktitle={Proceedings of the 2018 CHI Conference on Human Factors in Computing Systems},
  pages={1--12},
  year={2018}
}

@article{hullman2011benefitting,
  title={Benefitting infovis with visual difficulties},
  author={Hullman, Jessica and Adar, Eytan and Shah, Priti},
  journal={IEEE Transactions on Visualization and Computer Graphics},
  volume={17},
  number={12},
  pages={2213--2222},
  year={2011},
  publisher={IEEE}
}

@inproceedings{adar2017persalog,
  title={PersaLog: Personalization of news article content},
  author={Adar, Eytan and Gearig, Carolyn and Balasubramanian, Ayshwarya and Hullman, Jessica},
  booktitle={Proceedings of the 2017 CHI Conference on Human Factors in Computing Systems},
  pages={3188--3200},
  year={2017}
}

@article{cordova1996intrinsic,
  title={Intrinsic motivation and the process of learning: Beneficial effects of contextualization, personalization, and choice.},
  author={Cordova, Diana I and Lepper, Mark R},
  journal={Journal of Educational Psychology},
  volume={88},
  number={4},
  pages={715},
  year={1996},
  publisher={American Psychological Association}
}

@article{victor2011explorable,
  title={Explorable explanations},
  author={Victor, Bret},
  journal={},
  volume={},
  year={2011}
}

@inproceedings{conlen2018idyll,
  title={Idyll: A markup language for authoring and publishing interactive articles on the web},
  author={Conlen, Matthew and Heer, Jeffrey},
  booktitle={Proceedings of the 31st Annual ACM Symposium on User Interface Software and Technology},
  pages={977--989},
  year={2018}
}

@inproceedings{diakopoulos2011playable,
  title={Playable data: Characterizing the design space of game-y infographics},
  author={Diakopoulos, Nicholas and Kivran-Swaine, Funda and Naaman, Mor},
  booktitle={Proceedings of the SIGCHI Conference on Human Factors in Computing Systems},
  pages={1717--1726},
  year={2011}
}

@book{clark2010graphics,
  title={Graphics for learning: Proven guidelines for planning, designing, and evaluating visuals in training materials},
  author={Clark, Ruth C and Lyons, Chopeta},
  year={2010},
  publisher={John Wiley \& Sons}
}

@inproceedings{sultanum2021leveraging,
  title={Leveraging text-chart links to support authoring of data-driven articles with vizflow},
  author={Sultanum, Nicole and Chevalier, Fanny and Bylinskii, Zoya and Liu, Zhicheng},
  booktitle={Proceedings of the 2021 CHI Conference on Human Factors in Computing Systems},
  pages={1--17},
  year={2021}
}

@article{mayer2005cognitive,
  title={Cognitive theory of multimedia learning},
  author={Mayer, Richard E},
  journal={The Cambridge Handbook of Multimedia Learning},
  volume={41},
  number={1},
  pages={31--48},
  year={2005}
}

@inproceedings{boy2015storytelling,
  title={Storytelling in information visualizations: Does it engage users to explore data?},
  author={Boy, Jeremy and Detienne, Francoise and Fekete, Jean-Daniel},
  booktitle={Proceedings of the 33rd Annual ACM Conference on Human Factors in Computing Systems},
  pages={1449--1458},
  year={2015}
}

@article{victor2011scientific,
  title={Scientific communication as sequential art},
  author={Victor, Bret},
  journal={},
  volume={},
  year={2011}
}

@article{victor2011ladder,
  title={Up and down the ladder of abstraction},
  author={Victor, Bret},
  journal={},
  volume={},
  year={2011}
}

@inproceedings{cai2024watching,
  title={Watching the election sausage get made: How data journalists visualize the vote counting process in US elections},
  author={Cai, Mandi and Kay, Matthew},
  booktitle={Proceedings of the 2024 CHI Conference on Human Factors in Computing Systems},
  pages={1--19},
  year={2024}
}

@inproceedings{ge2024v,
  title={V-FRAMER: Visualization framework for mitigating reasoning errors in public policy},
  author={Ge, Lily W and Easterday, Matthew and Kay, Matthew and Dimara, Evanthia and Cheng, Peter and Franconeri, Steven L},
  booktitle={Proceedings of the 2024 CHI Conference on Human Factors in Computing Systems},
  pages={1--15},
  year={2024}
}

@inproceedings{shi2021communicating,
  title={Communicating with motion: A design space for animated visual narratives in data videos},
  author={Shi, Yang and Lan, Xingyu and Li, Jingwen and Li, Zhaorui and Cao, Nan},
  booktitle={Proceedings of the 2021 CHI Conference on Human Factors in Computing Systems},
  pages={1--13},
  year={2021}
}

@incollection{padilla2020uncertainty,
    author={Padilla, Lace and Kay, Matthew and Hullman, Jessica},
    title={Uncertainty visualization},
    booktitle={Balakrishnan N., Colton T., Everitt B., Piegorsch W., Ruggeri F., Teugels J. L. (Eds.)},
    publisher={Wiley},
    year={2022} 
}

@book{tufte1997visual,
  title={Visual explanations},
  author={Tufte, Edward R and Robins, David},
  year={1997},
  publisher={Graphics Cheshire, CT}
}

@book{alma99126810900001452,
    title={Understanding comics},
    publisher = {Kitchen Sink Press, Inc.},
    year={1993},
    author={McCloud, Scott}
}

@article{adar2020communicative,
  title={Communicative visualizations as a learning problem},
  author={Adar, Eytan and Lee, Elsie},
  journal={IEEE Transactions on Visualization and Computer Graphics},
  volume={27},
  number={2},
  pages={946--956},
  year={2020},
  publisher={IEEE}
}

@article{lee2021learning,
  title={Learning objectives, insights, and assessments: How specification formats impact design},
  author={Lee-Robbins, Elsie and He, Shiqing and Adar, Eytan},
  journal={IEEE Transactions on Visualization and Computer Graphics},
  volume={28},
  number={1},
  pages={676--685},
  year={2021},
  publisher={IEEE}
}

@incollection{chi2013self,
  title={Self-explaining expository texts: The dual processes of generating inferences and repairing mental models},
  author={Chi, Michelene TH},
  booktitle={Advances in Instructional Psychology, Volume 5},
  pages={161--238},
  year={2013},
  publisher={Routledge}
}

@article{pinker1990theory,
  title={A theory of graph comprehension.},
  author={Pinker, Steven},
  year={1990},
  publisher={Lawrence Erlbaum Associates, Inc}
}

@inproceedings{joshi2024constrained,
  title={Constrained Highlighting in a Document Reader can Improve Reading Comprehension},
  author={Joshi, Nikhita and Vogel, Daniel},
  booktitle={Proceedings of the 2024 CHI Conference on Human Factors in Computing Systems},
  pages={1--10},
  year={2024}
}

@article{satyanarayan2016vega,
  title={Vega-lite: A grammar of interactive graphics},
  author={Satyanarayan, Arvind and Moritz, Dominik and Wongsuphasawat, Kanit and Heer, Jeffrey},
  journal={IEEE transactions on visualization and computer graphics},
  volume={23},
  number={1},
  pages={341--350},
  year={2016},
  publisher={IEEE}
}

@article{bostock2011d3,
  title={D$^3$ data-driven documents},
  author={Bostock, Michael and Ogievetsky, Vadim and Heer, Jeffrey},
  journal={IEEE transactions on visualization and computer graphics},
  volume={17},
  number={12},
  pages={2301--2309},
  year={2011},
  publisher={IEEE}
}

@misc{d3incomeinequality,
  key = {Inequality in American cities.},
  title = {Inequality in American cities},
  howpublished = {\url{https://observablehq.com/@d3/inequality-in-american-cities}},
  year = {}
}

@misc{nytincomeinequality,
  key = {Watch 4 Decades of Inequality Drive American Cities Apart.},
  title = {Watch 4 decades of inequality drive american cities apart},
  howpublished = {\url{https://www.nytimes.com/2019/12/02/upshot/wealth-poverty-divide-american-cities.html}},
  year = {2019}
}

@misc{nytco2,
  key = {Carbon Dioxide Levels Have Passed a New Milestone.},
  title = {Carbon dioxide levels have passed a new milestone},
  howpublished = {\url{https://www.nytimes.com/interactive/2024/04/20/upshot/carbon-dioxide-growth.html}},
  year = {2024}
}

@misc{economist,
  key = {The Phillips curve may be broken for good.},
  title = {The phillips curve may be broken for good},
  howpublished = {\url{https://www.economist.com/graphic-detail/2017/11/01/the-phillips-curve-may-be-broken-for-good}},
  year = {2017}
}

@misc{ft,
  key = {Trump broke the Democrats' thermostat.},
  title = {Trump broke the democrats' thermostat},
  howpublished = {\url{https://www.ft.com/content/73a1836d-0faa-4c84-b973-554e2ca3a227}},
  year = {2024}
}

@misc{aisch,
  key = {Six Decades of Carbon Dioxide Concentration in the Atmosphere},
  title = {Six decades of carbon dioxide concentration in the atmosphere},
  howpublished = {\url{https://www.datawrapper.de/_/OHgEm/}}
}

@misc{nytcovid,
  key = {How America Reached One Million Covid Deaths},
  title = {How america reached one million covid deaths},
  howpublished = {\url{https://www.nytimes.com/interactive/2022/05/13/us/covid-deaths-us-one-million.html}}
}

@misc{nytcopenhagen,
  key = {Copenhagen: Emissions, Treaties and Impacts},
  title = {Copenhagen: Emissions, treaties and impacts},
  howpublished = {\url{https://archive.nytimes.com/www.nytimes.com/interactive/2009/12/05/world/climate-graphic-background.html}}
}

@misc{idyllex,
  key = {Kernel Density Estimation},
  title = {Kernel density estimation},
  howpublished = {\url{https://mathisonian.github.io/kde/}}
}

@article{hullman2011visualization,
  title={Visualization rhetoric: Framing effects in narrative visualization},
  author={Hullman, Jessica and Diakopoulos, Nick},
  journal={IEEE Transactions on Visualization and Computer Graphics},
  volume={17},
  number={12},
  pages={2231--2240},
  year={2011},
  publisher={IEEE}
}

@article{lee2022affective,
  title={Affective learning objectives for communicative visualizations},
  author={Lee-Robbins, Elsie and Adar, Eytan},
  journal={IEEE Transactions on Visualization and Computer Graphics},
  volume={29},
  number={1},
  pages={1--11},
  year={2022},
  publisher={IEEE}
}

@article{kim2017data,
  title={Data through others' eyes: The impact of visualizing others' expectations on visualization interpretation},
  author={Kim, Yea-Seul and Reinecke, Katharina and Hullman, Jessica},
  journal={IEEE Transactions on Visualization and Computer Graphics},
  volume={24},
  number={1},
  pages={760--769},
  year={2017},
  publisher={IEEE}
}

@inproceedings{head2022math,
  title={Math augmentation: How authors enhance the readability of formulas using novel visual design practices},
  author={Head, Andrew and Xie, Amber and Hearst, Marti A},
  booktitle={Proceedings of the 2022 CHI Conference on Human Factors in Computing Systems},
  pages={1--18},
  year={2022}
}

@book{10.5555/1095597,
author = {Bertin, Jacques},
title = {Semiology of graphics},
year = {1983},
isbn = {0299090604},
publisher = {University of Wisconsin Press}
}

@book{playfair1801commercial,
  title={The commercial and political atlas: Representing, by means of stained copper-plate charts, the progress of the commerce, revenues, expenditure and debts of england during the whole of the eighteenth century},
  author={Playfair, W.},
  lccn={05021912},
  url={https://books.google.com/books?id=dgRdAAAAcAAJ},
  year={1801},
  publisher={T. Burton}
}

@article{kosslyn1989understanding,
  title={Understanding charts and graphs},
  author={Kosslyn, Stephen M},
  journal={Applied Cognitive Psychology},
  volume={3},
  number={3},
  pages={185--225},
  year={1989},
  publisher={Wiley Online Library}
}

@article{cleveland1984graphical,
  title={Graphical perception: Theory, experimentation, and application to the development of graphical methods},
  author={Cleveland, William S and McGill, Robert},
  journal={Journal of the American Statistical Association},
  volume={79},
  number={387},
  pages={531--554},
  year={1984},
  publisher={Taylor \& Francis}
}

@inproceedings{heer2023living,
  title={Living papers: A language toolkit for augmented scholarly communication},
  author={Heer, Jeffrey and Conlen, Matthew and Devireddy, Vishal and Nguyen, Tu and Horowitz, Joshua},
  booktitle={Proceedings of the 36th Annual ACM Symposium on User Interface Software and Technology},
  pages={1--13},
  year={2023}
}

@article{sarikaya2018we,
  title={What do we talk about when we talk about dashboards?},
  author={Sarikaya, Alper and Correll, Michael and Bartram, Lyn and Tory, Melanie and Fisher, Danyel},
  journal={IEEE Transactions on Visualization and Computer Graphics},
  volume={25},
  number={1},
  pages={682--692},
  year={2018},
  publisher={IEEE}
}

@article{shannon1948mathematical,
  title={A mathematical theory of communication},
  author={Shannon, Claude E},
  journal={The Bell System Technical Journal},
  volume={27},
  number={3},
  pages={379--423},
  year={1948},
  publisher={Nokia Bell Labs}
}

@article{adams2015bloom,
  title={Bloom’s taxonomy of cognitive learning objectives},
  author={Adams, Nancy E},
  journal={Journal of the Medical Library Association},
  volume={103},
  number={3},
  pages={152},
  year={2015}
}

@article{chandra2024watchat,
  title={Watchat: Explaining perplexing programs by debugging mental models},
  author={Chandra, Kartik and Collins, Katherine M and Crichton, Will and Chen, Tony and Li, Tzu-Mao and Weller, Adrian and Nigam, Rachit and Tenenbaum, Joshua and Ragan-Kelley, Jonathan},
  journal={arXiv preprint arXiv:2403.05334},
  year={2024}
}

@article{koffka1922perception,
  title={Perception: an introduction to the gestalt-theorie},
  author={Koffka, Kurt},
  journal={Psychological Bulletin},
  volume={19},
  number={10},
  pages={531},
  year={1922},
  publisher={Psychological Review Company}
}

@inproceedings{heer2007voyagers,
  title={Voyagers and voyeurs: supporting asynchronous collaborative information visualization},
  author={Heer, Jeffrey and Vi{\'e}gas, Fernanda B and Wattenberg, Martin},
  booktitle={Proceedings of the SIGCHI Conference on Human Factors in Computing Systems},
  pages={1029--1038},
  year={2007}
}

@inproceedings{nobarany2012facilitating,
  title={Facilitating the reuse process in distributed collaboration: a distributed cognition approach},
  author={Nobarany, Syavash and Haraty, Mona and Fisher, Brian},
  booktitle={Proceedings of the ACM 2012 Conference on Computer Supported Cooperative Work},
  pages={1223--1232},
  year={2012}
}

@article{tufte1990data,
  title={Data-ink maximization and graphical design},
  author={Tufte, ER},
  journal={Oikos},
  pages={130--144},
  year={1990},
  publisher={JSTOR}
}

@inproceedings{hoffswell2020techniques,
  title={Techniques for flexible responsive visualization design},
  author={Hoffswell, Jane and Li, Wilmot and Liu, Zhicheng},
  booktitle={Proceedings of the 2020 CHI Conference on Human Factors in Computing Systems},
  pages={1--13},
  year={2020}
}

@article{stokes2023role,
  title={The role of text in visualizations: How annotations shape perceptions of bias and influence predictions},
  author={Stokes, Chase and Bearfield, Cindy Xiong and Hearst, Marti A},
  journal={IEEE Transactions on Visualization and Computer Graphics},
  volume={30},
  number={10},
  pages={6787--6800},
  year={2023},
  publisher={IEEE}
}

@article{stokes2022striking,
  title={Striking a balance: Reader takeaways and preferences when integrating text and charts},
  author={Stokes, Chase and Setlur, Vidya and Cogley, Bridget and Satyanarayan, Arvind and Hearst, Marti A},
  journal={IEEE Transactions on Visualization and Computer Graphics},
  volume={29},
  number={1},
  pages={1233--1243},
  year={2022},
  publisher={IEEE}
}

@inproceedings{pluto,
author = {Srinivasan, Arjun and Setlur, Vidya and Satyanarayan, Arvind},
title = {Pluto: Authoring semantically aligned text and charts for data-driven communication},
year = {2025},
publisher = {Association for Computing Machinery},
booktitle = {Proceedings of the 30th International Conference on Intelligent User Interfaces},
pages = {1123–1140},
}

@inproceedings{wang2025jupybara,
  title={Jupybara: Operationalizing a design space for actionable data analysis and storytelling with LLMs},
  author={Wang, Huichen Will and Birnbaum, Larry and Setlur, Vidya},
  booktitle={Proceedings of the 2025 CHI Conference on Human Factors in Computing Systems},
  pages={1--24},
  year={2025}
}

@article{maceachren2012visual,
  title={Visual semiotics \& uncertainty visualization: An empirical study},
  author={MacEachren, Alan M and Roth, Robert E and O'Brien, James and Li, Bonan and Swingley, Derek and Gahegan, Mark},
  journal={IEEE Transactions on Visualization and Computer Graphics},
  volume={18},
  number={12},
  pages={2496--2505},
  year={2012},
  publisher={IEEE}
}

@inproceedings{gunturu2024augmented,
  title={Augmented Physics: Creating Interactive and Embedded Physics Simulations from Static Textbook Diagrams},
  author={Gunturu, Aditya and Wen, Yi and Zhang, Nandi and Thundathil, Jarin and Kazi, Rubaiat Habib and Suzuki, Ryo},
  booktitle={Proceedings of the 37th Annual ACM Symposium on User Interface Software and Technology},
  pages={1--12},
  year={2024}
}

@book{haladyna2004developing,
  title={Developing and validating multiple-choice test items},
  author={Haladyna, Thomas M},
  year={2004},
  publisher={Routledge}
}

@article{watts1998collective,
  title={Collective dynamics of ‘small-world’networks},
  author={Watts, Duncan J and Strogatz, Steven H},
  journal={Nature},
  volume={393},
  number={6684},
  pages={440--442},
  year={1998},
  publisher={Nature Publishing Group}
}




\appendix

\subsection{Open Coding Procedure}
One author analyzed the filtered corpus of 52 papers, extracting \textit{why} a visual augmentation or feature described in the paper was used (\textit{objectives}), \textit{what} purpose it served (\textit{strategies}), and \textit{how} it was realized (\textit{tactics}).
\add{Codes for the extracted data were generated and iteratively refined, organized, and categorized through multiple rounds of discussion between two of the authors, with all authors reviewing the final design space.
Our ideas and categories evolved significantly throughout this process (\eg observing shared strategies across different tactics); we revisited prior examples to re-categorize when appropriate, doing our best to maintain consistent evaluation.
}

Other design space dimensions were considered, such as data semantics (\eg type or distribution of data), usage context (\eg news article, mobile, supporting text or tables, \etc), and chart characteristics (\eg static vs. interactive, use of titles or captions, data representation, \etc).
Overall, we found that many of these dimensions did not help to meaningfully partition the space of augmentations, as most were used to implement different variations of the same strategy (\eg faceting vs. interactive steppers can both \textit{sequence} a chart), but did not otherwise affect higher-level design objectives.

\subsection{Augmentation Categorization of Paper Corpus}
\add{
To assess the completeness of our design space in representing the existing literature, we revisited our paper corpus to categorize each paper along our design space dimensions. 
The outcome of this process is shown in
}
\autoref{table:augmentation-map}, though we do not delineate how these dimensions relate to each other (\eg which tactics support which strategies).
We provide seven additional visualizations from this table in the following sections, showcasing how our design space can be used to describe and analyze their visual augmentations.

Dimensions in~\autoref{table:augmentation-map} are only activated based on the manner they are used or discussed in a given paper, \add{including results from user studies, system features and design, or features of a design space / taxonomy}.
\add{
Further, we only activate dimensions that are used to communicate information beyond denotative graphical decoding (as defined in section II of the main paper).
For example, Mackinlay's APT system~\cite{mackinlay1986automating} presents a standard set of encoding choices (c.f., Fig. 15), though they are not discussed in the context of our higher-level design strategies (\eg using \xmybox[tactic]{encodings} to \xmybox[strategy]{spotlight} particular information).
If this were the case, then any discussion of chart types or \textit{denotative} encoding choices (nearly all papers in our corpus) would qualify as observation and / or interpretation aids since they inherently affect chart comprehension~\cite{pinker1990theory, kosslyn1989understanding}.
However, Mackinlay does discuss faceting (a \xmybox[tactic]{layout} tactic) in the context of \xmybox[strategy]{sequencing} (\ie arranging content in a particular order), which can serve as an \xmybox[objective]{observation} aid for identifying salient trends less obvious with standard encodings.
We used this line of reasoning to identify strategies and objectives in a consistent manner for all tactics described in a given paper.
}

\tableaugmentationmap

\newpage

\subsection{Example 1}
\begin{figure}[!h]
    \centering
    \resizebox{1\linewidth}{!}{\includegraphics{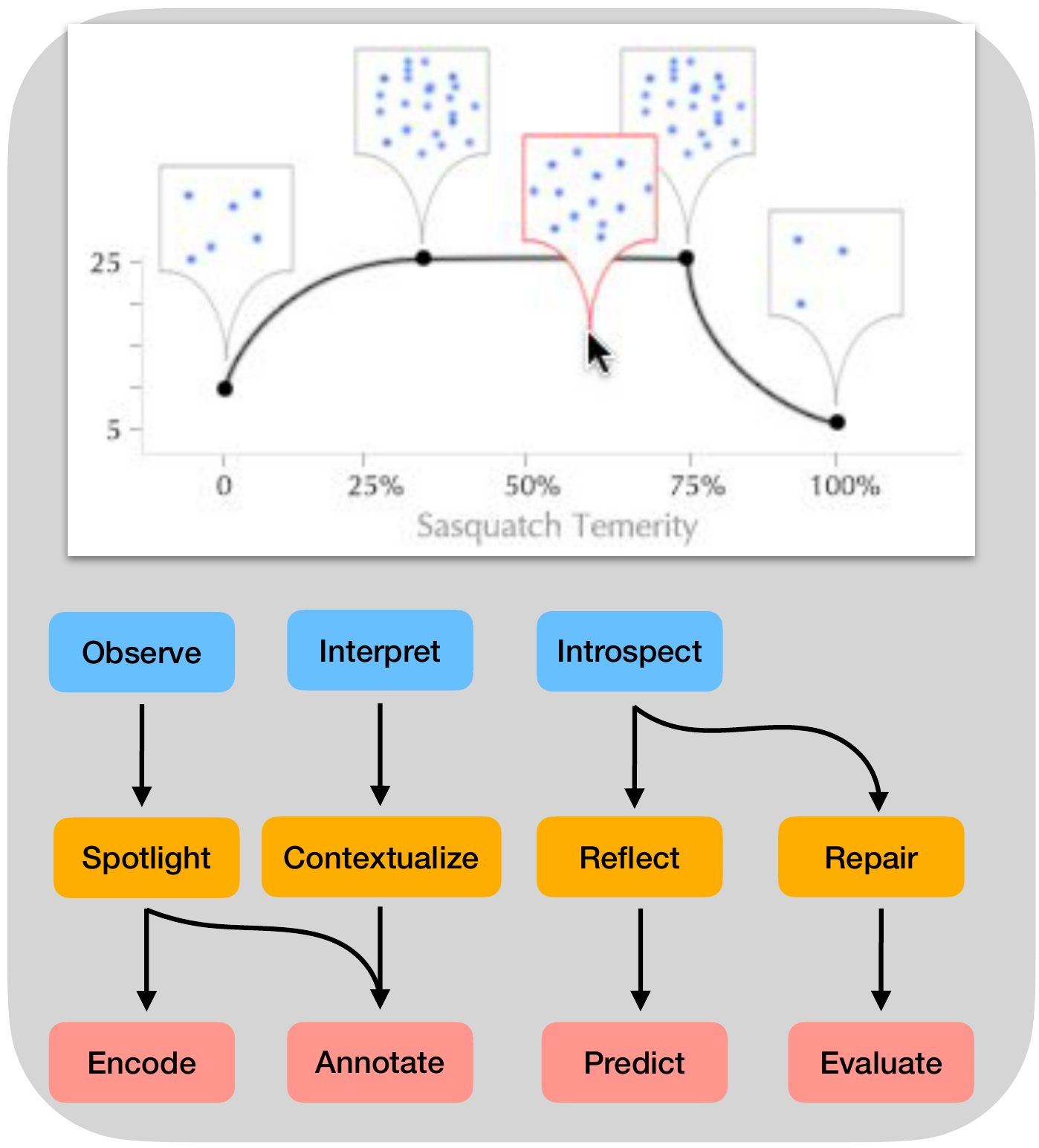}}
\end{figure}
This visualization was used by Victor~\cite{victor2006magic} to demonstrate complementary interactions between concrete examples (annotated pop-ups) and abstract representations (the underlying curve).
Specifically, \xmybox[tactic]{annotated} cut-outs \xmybox[strategy]{spotlight} particular points along the curve's path, with example illustrations for \xmybox[strategy]{context}.
Further, readers can click anywhere on the plane to see an example at that point.
This feature may encourage readers to \xmybox[objective]{introspect} and assess their understanding in which they \xmybox[tactic]{predict} examples (prompting \xmybox[strategy]{reflection}), then clicking to \xmybox[tactic]{evaluate} their prediction and \xmybox[strategy]{repair} incorrect understanding.
Readers' interactions are \xmybox[tactic]{encoded} with a red hue for \xmybox[strategy]{spotlighting}.

\newpage
\subsection{Example 2}
\begin{figure}[!h]
    \centering
    \resizebox{1\linewidth}{!}{\includegraphics{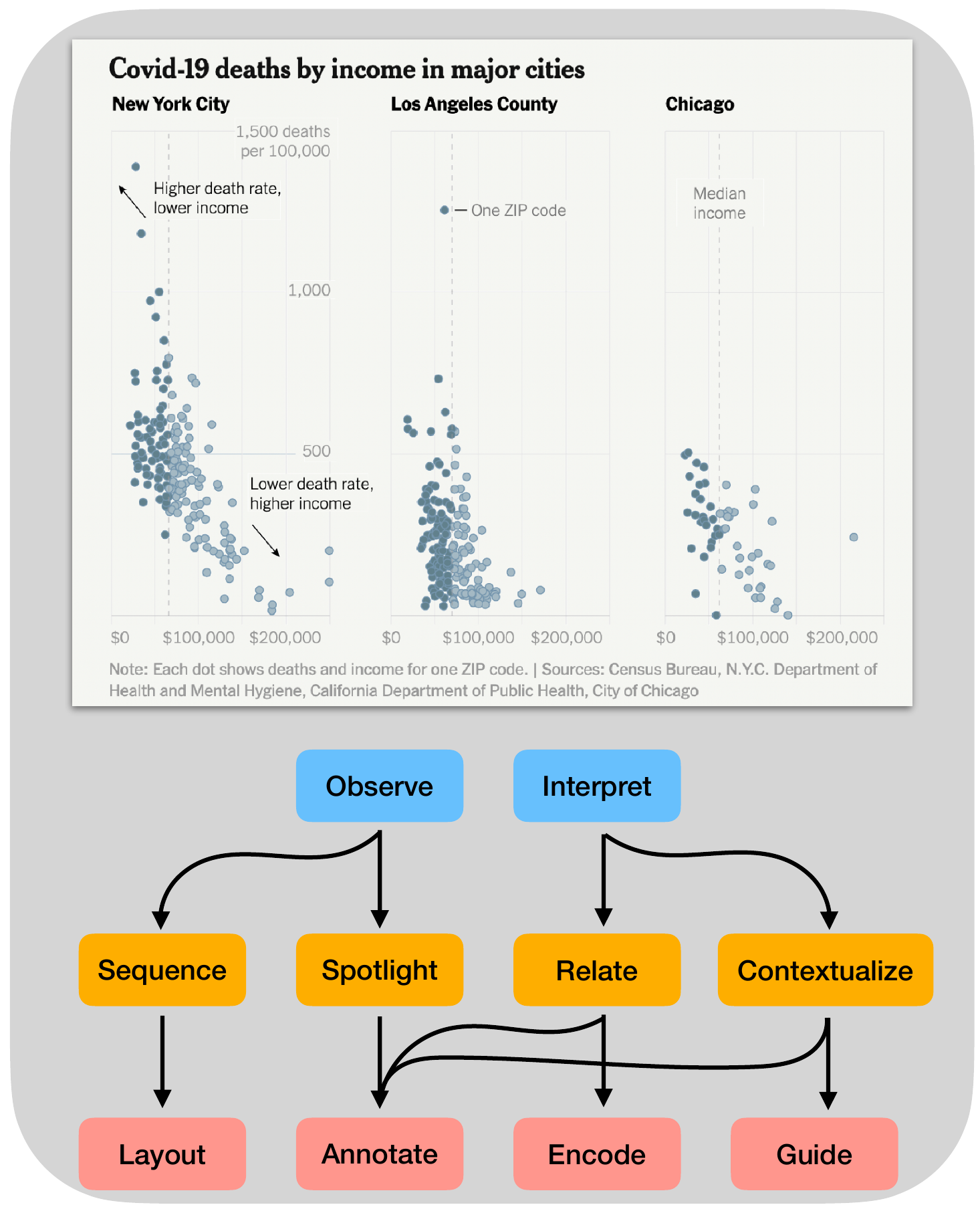}}
\end{figure}
This visualization from The New York Times~\cite{nytcovid} shows COVID-19 deaths stratified by three major cities (New York, Los Angeles, and Chicago) and income.
Quadri \etal~\cite{quadri2024you} used this chart in a pilot study to explore alignment between readers' takeaways and designers' intent.

\xmybox[tactic]{Layout} (faceting) \xmybox[strategy]{sequences} the different cities to \xmybox[objective]{observe} differences in COVID-19 deaths.
A dotted line \xmybox[tactic]{annotation} serves as a \xmybox[strategy]{spotlight} for the median income; points are color \xmybox[tactic]{encoded} to \xmybox[strategy]{relate} their position with the median divider.
Text \xmybox[tactic]{annotations} and \xmybox[tactic]{guides} (title and footnote) provide important \xmybox[strategy]{context}.
Finally, arrow \xmybox[tactic]{annotations} help \xmybox[strategy]{spotlight} outliers in particular regions (top left and bottom right) for the reader to \xmybox[objective]{observe}.

\newpage
\subsection{Example 3}
\begin{figure}[!h]
    \centering
    \resizebox{1\linewidth}{!}{\includegraphics{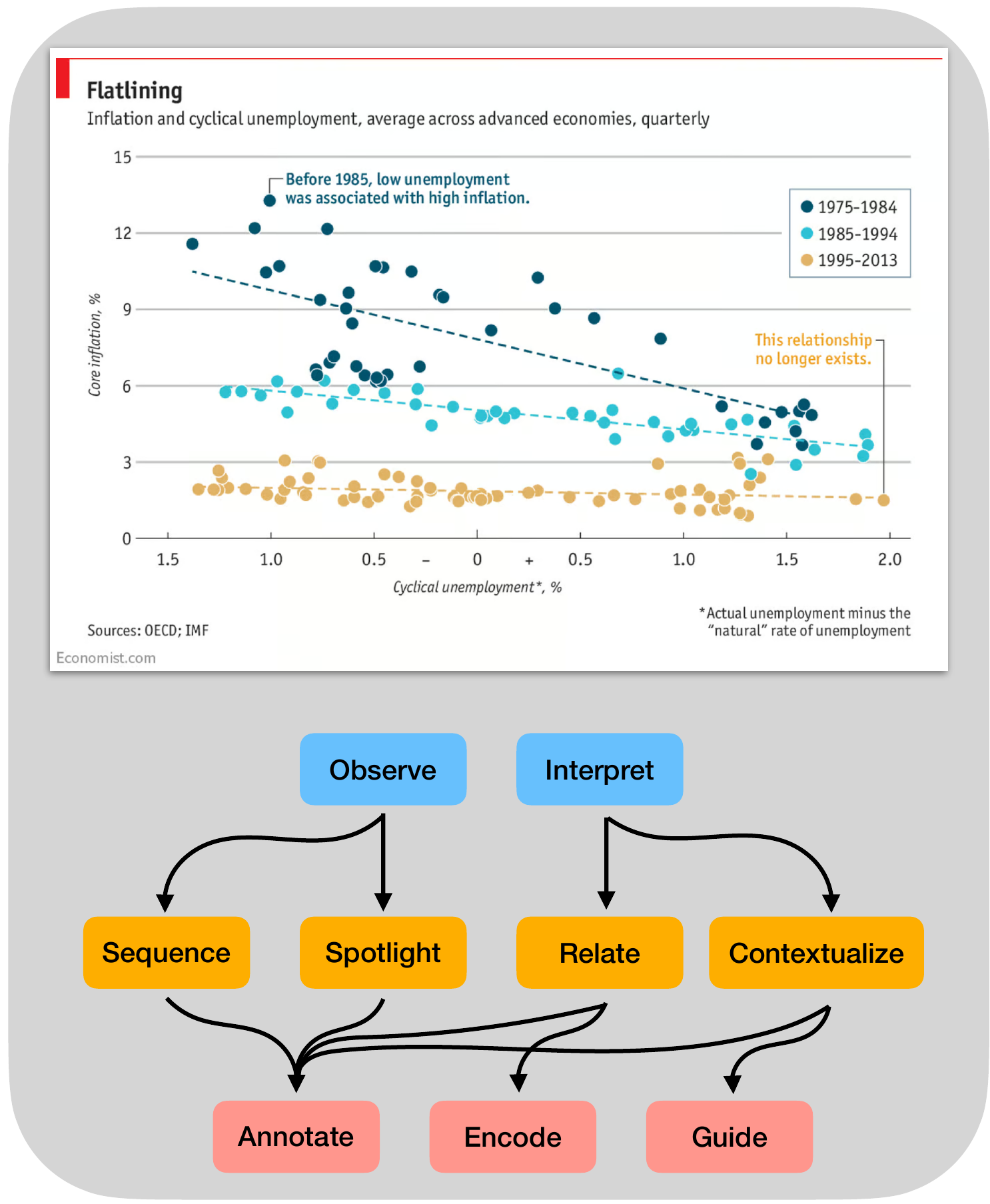}}
\end{figure}

This visualization displays the weakening correlation between unemployment and inflation across decades.
Originally from The Economist~\cite{economist}, this graph was used by Adar and Lee-Robbins~\cite{adar2020communicative} to illustrate learning objectives. 

\xmybox[tactic]{Annotations} accomplish four goals: \xmybox[strategy]{Sequencing} and \xmybox[strategy]{spotlighting} with verbal phrasing and left-to-right text placement to \xmybox[objective]{observe} the weakening correlation over time; adding \xmybox[strategy]{context} to help \xmybox[objective]{interpret} the relationship between unemployment and inflation; and \xmybox[strategy]{relating} the annotated text with their respective time periods.
By also using \xmybox[tactic]{encoding} to \xmybox[strategy]{relate} these items, we can \xmybox[objective]{interpret} the annotation as referring to the entire time period rather than the single point it connects with.
Finally, footnotes, title and subtitle/caption serve as \xmybox[tactic]{guides} for additional \xmybox[strategy]{context} about the chart's message and data sources.
           
\newpage
\subsection{Example 4}
\begin{figure}[!h]
    \centering
    \resizebox{1\linewidth}{!}{\includegraphics{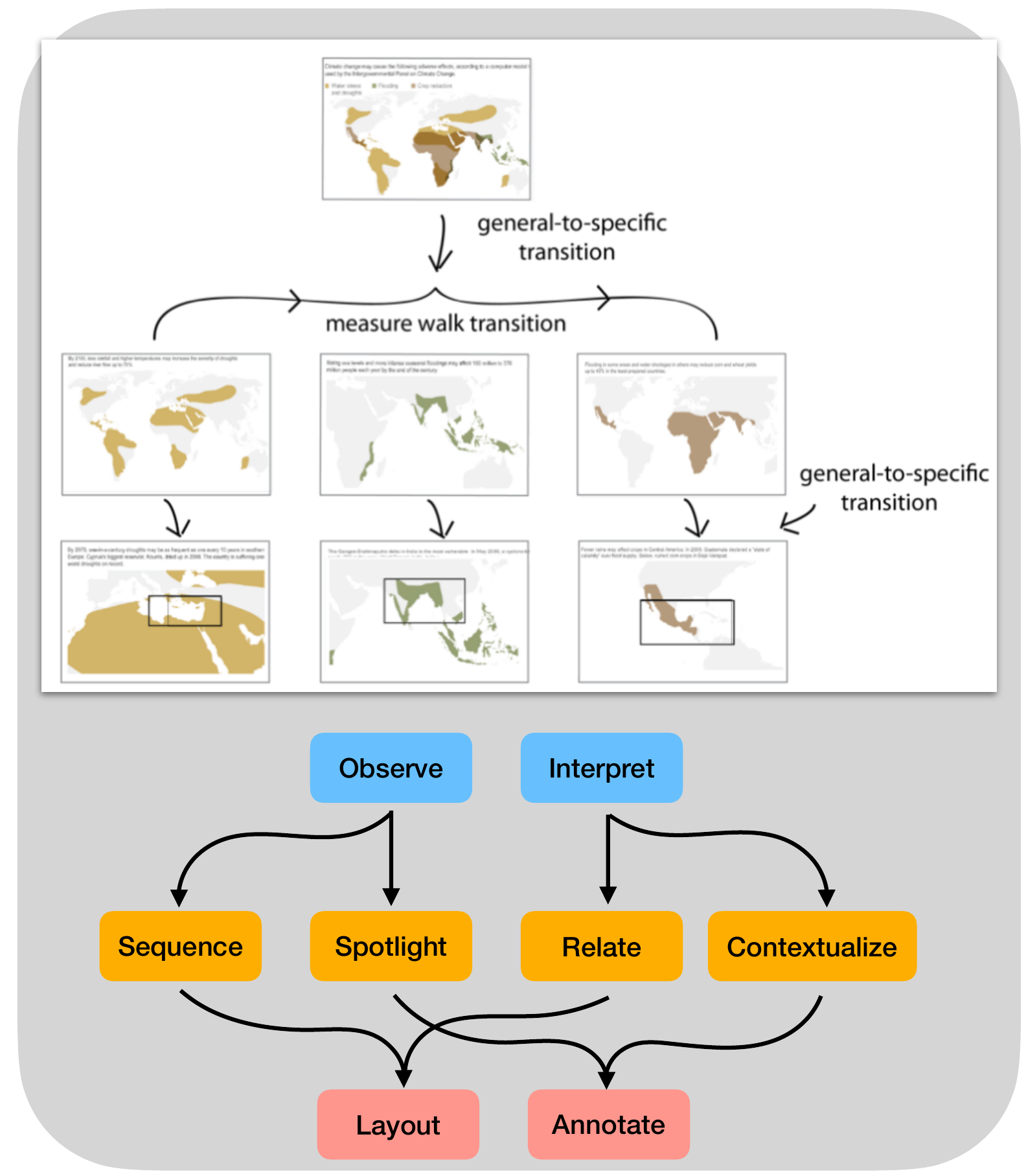}}
\end{figure}
This visualization was used by Hullman \etal~\cite{hullman2013deeper} (originally from The New York Times~\cite{nytcopenhagen}) to analyze forms of sequencing in data journalism. 
In this example, \xmybox[tactic]{layout} is used to not only \xmybox[strategy]{sequence} a set of insights for the reader to \xmybox[objective]{observe}, but also \xmybox[strategy]{relate} these insights and better \xmybox[objective]{interpret} the overall narrative through parallelism (\ie walking through general-to-specific transitions for each region).
\xmybox[tactic]{Annotations} are used in each frame to provide \xmybox[strategy]{context} about the current view, as well as \xmybox[strategy]{spotlight} specific countries near the end of a general-to-specific transition (black cut-out).

\newpage
\subsection{Example 5}
\begin{figure}[!h]
    \centering
    \resizebox{1\linewidth}{!}{\includegraphics{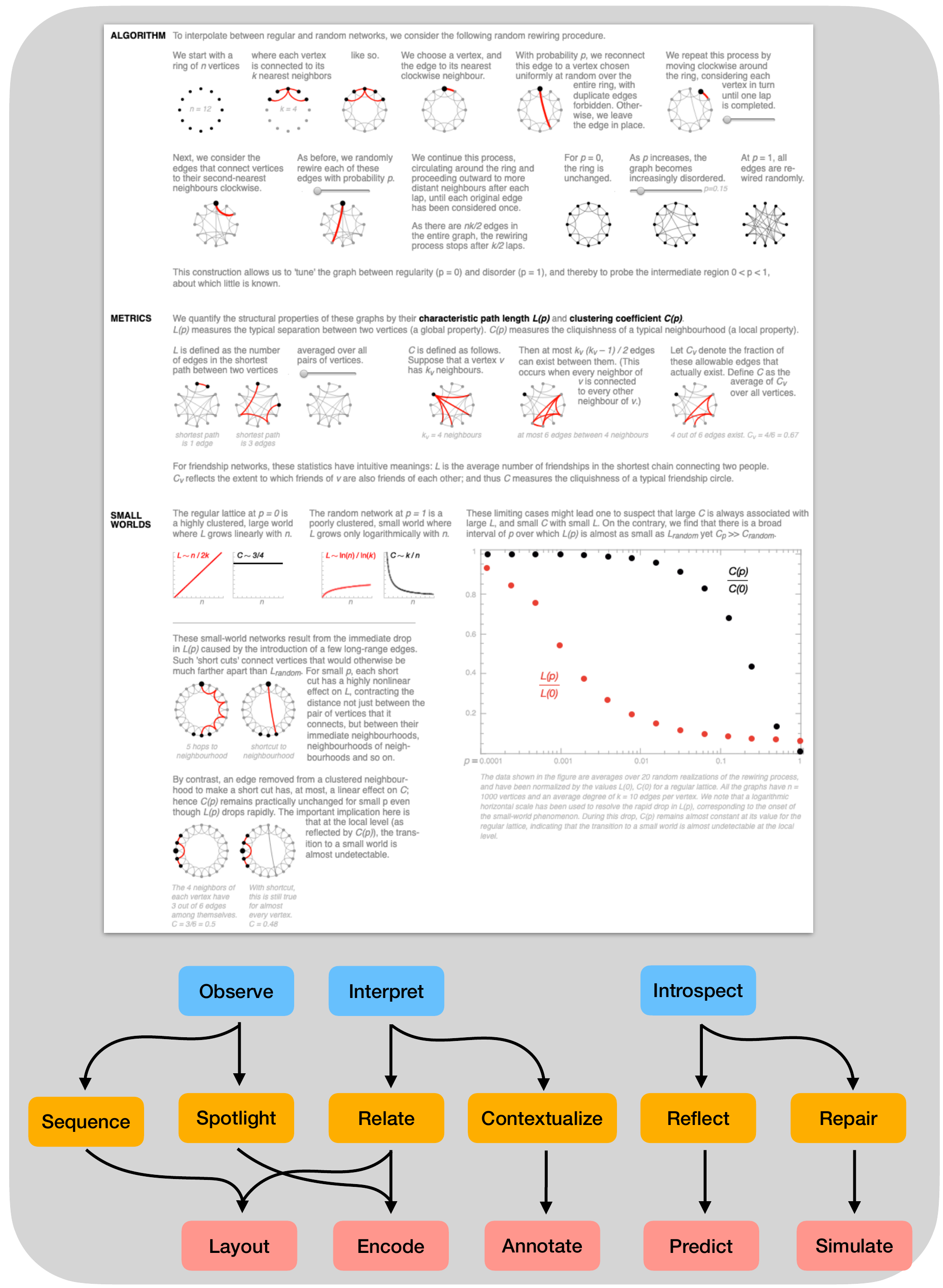}}
\end{figure}
This illustration was authored by Victor~\cite{victor2011scientific} to demonstrate the benefits of sequential art for communicating `small-world' networks~\cite{watts1998collective}.
\xmybox[tactic]{Layout} helps \xmybox[strategy]{sequence} sections of content (algorithm, metrics, small worlds) and to help \xmybox[objective]{observe} the overall organization. This \xmybox[tactic]{layout} also \xmybox[strategy]{relates} the ordering of algorithmic steps to \xmybox[objective]{interpret} its process.
Red hue and bold font \xmybox[tactic]{encoding} \xmybox[strategy]{spotlights} structural graph properties and \xmybox[strategy]{relates} their meaning (red hue to represent paths and bold font to represent clustering coefficients).
\xmybox[tactic]{Annotations} (in light gray color) are also placed below some of the visuals to provide helpful \xmybox[strategy]{context}.
Finally, interactive sliders \xmybox[tactic]{simulate} certain steps of the algorithm (to \xmybox[strategy]{repair} understanding), enabling the reader to \xmybox[objective]{introspect} and \xmybox[strategy]{reflect} by making tacit \xmybox[tactic]{predictions} about the algorithm's functionality.

\newpage
\subsection{Example 6}
\begin{figure}[!h]
    \centering
    \resizebox{1\linewidth}{!}{\includegraphics{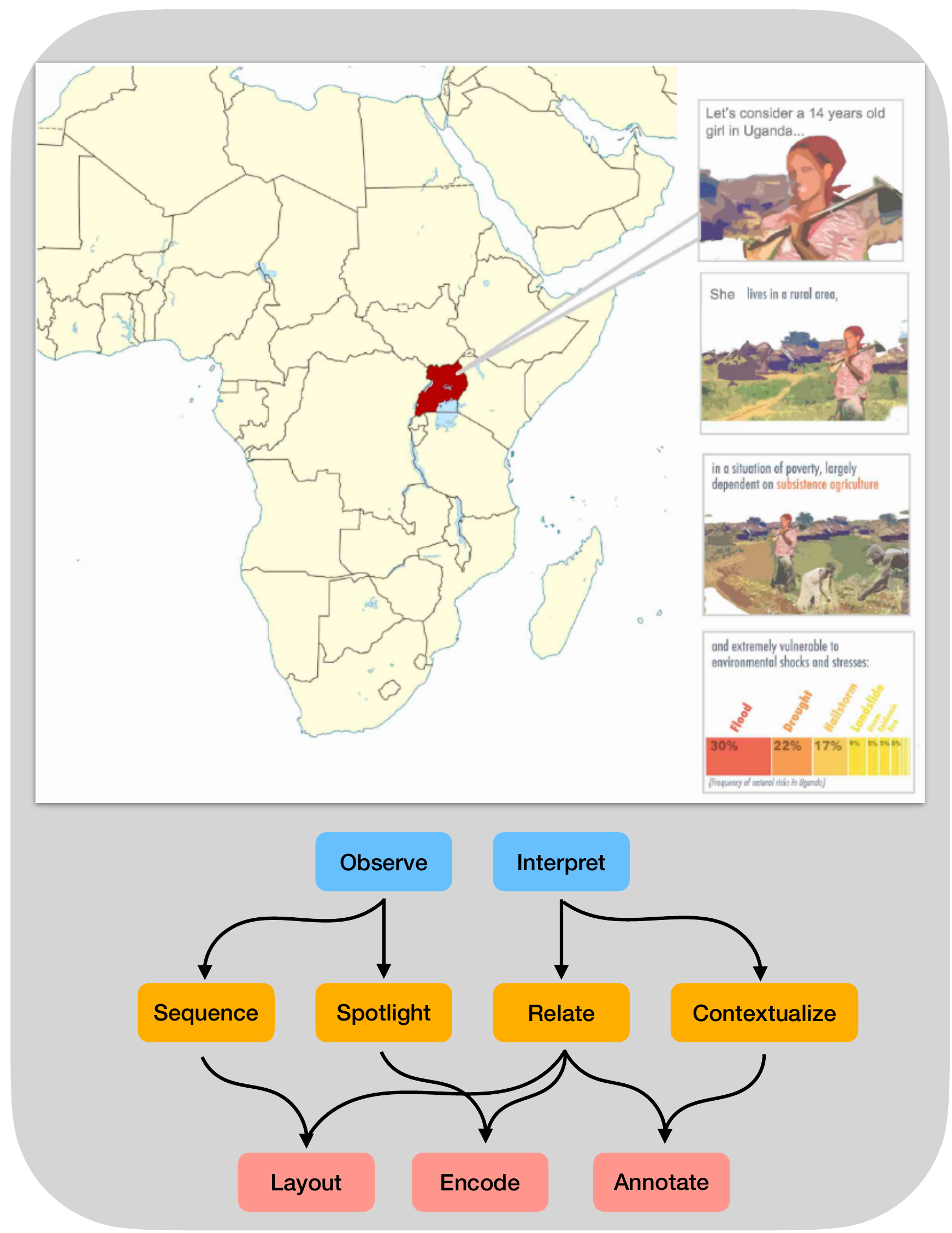}}
\end{figure}
This visualization was used by Bach \etal~\cite{bach2018design} to demonstrate the use of data comics, narrating conditions of poverty in Uganda. 
A \xmybox[tactic]{layout} of four frames on the right \xmybox[strategy]{sequence} a narrative for \xmybox[objective]{observation}, helping \xmybox[strategy]{relate} events of the story and data (last frame at the bottom).
A bright red hue \xmybox[tactic]{encoding} \xmybox[strategy]{spotlights} Uganda on the map, with a tapered \xmybox[tactic]{annotation} that \xmybox[strategy]{relates} it to the narrative sequence. \xmybox[tactic]{Annotations} are also provided in each of the narrative frames for additional \xmybox[strategy]{context}.
Finally, orange hue \xmybox[tactic]{encoding} \xmybox[strategy]{relates} the third and fourth frames, indicating that the dependency on ``subsistence agriculture'' is affected by droughts.

\newpage
\subsection{Example 7}
\begin{figure}[!h]
    \centering
    \resizebox{1\linewidth}{!}{\includegraphics{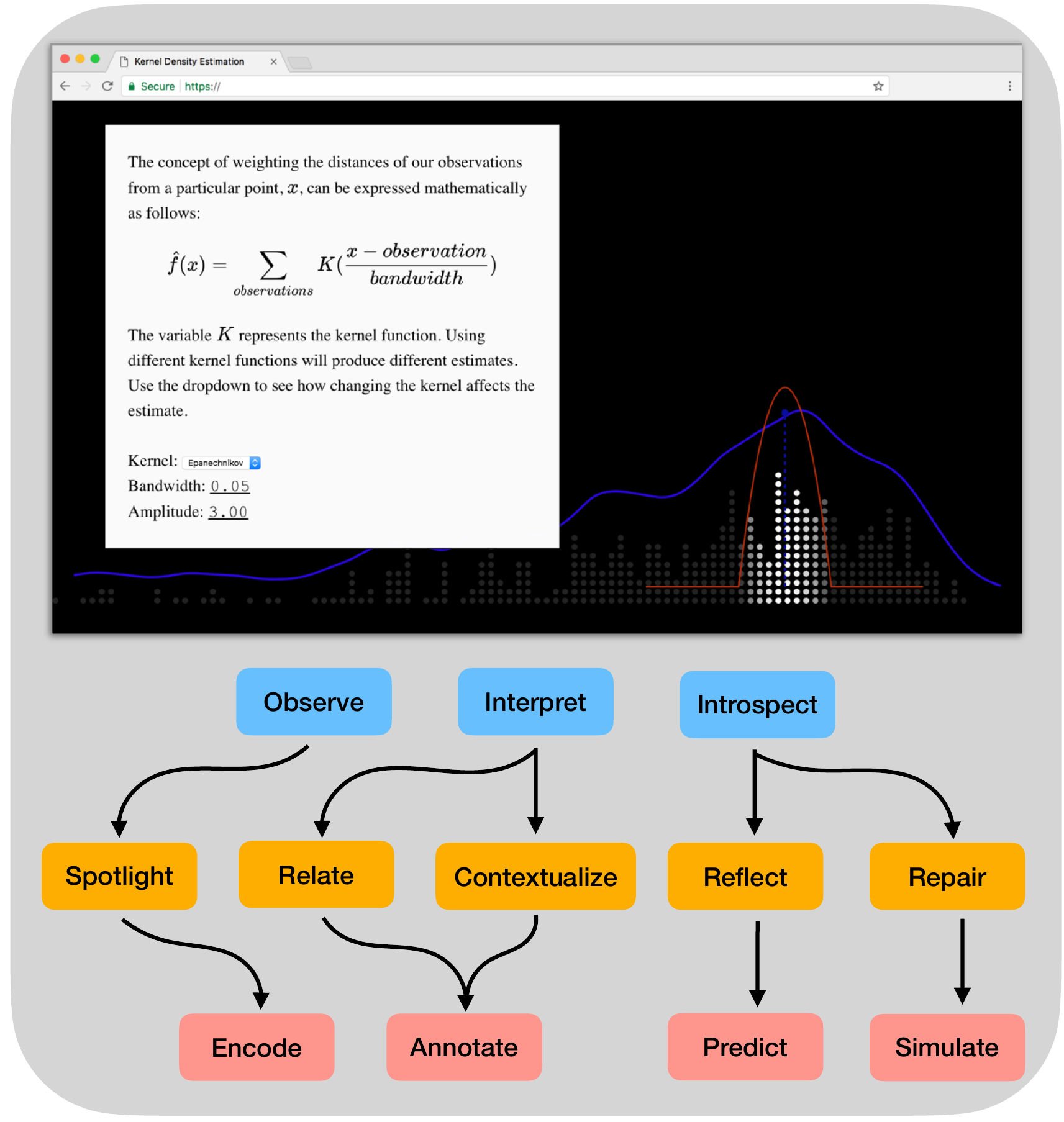}}
\end{figure}
This visualization~\cite{idyllex} was created by Matt Conlen using Idyll~\cite{conlen2018idyll} to explain kernel desnity estimation (KDE).
\xmybox[tactic]{Annotated} text in the white box provides important \xmybox[strategy]{context} to help the reader \xmybox[objective]{interpret} KDE parameters and their effect on the background visualization.
Underlined text \xmybox[tactic]{annotations} (bandwidth and amplitude) serves to \xmybox[strategy]{relate} interactive parameters throughout the entire document.
Based on the selected parameters, data inside the KDE bandwidth is \xmybox[strategy]{spotlit} with opacity \xmybox[tactic]{encoding}.
Finally, readers can \xmybox[strategy]{reflect} via \xmybox[tactic]{predictions} about the effect of different parameters on the background visualization by changing their values directly, which then \xmybox[tactic]{simulate} the KDE algorithm to provide visual feedback and \xmybox[strategy]{repair} comprehension. 

\newpage

\end{document}